\documentclass[prd,preprintnumbers,floatfix,
nofootinbib,superscriptaddress]{revtex4}

\usepackage{float}
\usepackage{nicefrac}
\usepackage{mathtools}
\usepackage{amsfonts} 
\usepackage{amssymb} 
\usepackage{amsmath} 
\usepackage{graphicx} 
\usepackage{subfigure} 
\usepackage{array} 
\usepackage{dcolumn} 
\usepackage{bm} 
\usepackage{latexsym} 
\usepackage{longtable} 
\usepackage{hyperref} 
\usepackage{verbatim}
\usepackage{epsfig}
\usepackage{slashed}
\usepackage{color}
\usepackage{tikz}
\DeclareRobustCommand\sampleline[1]{%
  \tikz\draw[#1] (0,0) (0,\the\dimexpr\fontdimen22\textfont2\relax)
  -- (2em,\the\dimexpr\fontdimen22\textfont2\relax);%
}

\newcommand{\cB}[0]{\mathcal B}
\newcommand{\cD}[0]{\mathcal D}
\newcommand{\cF}[0]{\mathcal F}
\newcommand{\cK}[0]{\mathcal K}
\newcommand{\cL}[0]{\mathcal L}
\newcommand{\cM}[0]{\mathcal M}
\newcommand{\cR}[0]{\mathcal R}
\newcommand{\cY}[0]{\mathcal Y}
\newcommand{\cU}[0]{\mathcal U}
\newcommand{\cS}[0]{\mathcal S}
\newcommand{\cT}[0]{\mathcal T}
\newcommand{\cZ}[0]{\mathcal Z}

\newcommand{\df}[0]{\mathrm{df}}
\newcommand{\K}[0]{\mathcal K}
\newcommand{\Kt}[0]{\widetilde{\K}}

\newcommand{\Y}[0]{\mathcal Y}

\newcommand{\Ki}[1]{ \mathcal K_{2#1}}
\newcommand{\Kti}[1]{ \widetilde{\mathcal K}_{2#1}}
\newcommand{\ttil}[0]{{\widetilde 2}}
\newcommand{\PV}[0]{\widetilde{\mathrm{PV}}}
\renewcommand{\u}[0]{{(u)}}
\newcommand{\s}[0]{{(s)}}
\newcommand{\off}[0]{{{\rm off}}}

\newcommand{\CLF}[2]{C_{L, #1 F}^{(#2)}}

\newcommand{\f}[0]{\mathrm{off}}
\newcommand{\on}[0]{\mathrm{on}}
\newcommand{\uu}[0]{{{(u,u)}}}

\newcommand{\bKdf}[0]{\mathbf K_{\df}}

\newcommand{\iKdfMat}[0]{
 \begin{pmatrix}   
i \K_{\df, \ttil \ttil}     &   i \K_{\df,  \ttil 3} \\  
i \K_{\df, 3 \ttil } &   i \K_{\df, 33} 
\end{pmatrix}
}
\newcommand{\bKdfMat}[0]{
 \begin{pmatrix}   
 \mathbf K_{\df, \ttil \ttil}     &   \mathbf K_{\df,  \ttil 3} \\  
\mathbf K_{\df, 3 \ttil } &   \mathbf K_{\df, 33} 
\end{pmatrix}
}
\newcommand{\iFMat}[0]{
\begin{pmatrix}
i F_{\ttil \ttil} & i F_{ \ttil 3} \\ 
i F_{ 3 \ttil } & i F_{33}
\end{pmatrix}
}
\newcommand{\bFMat}[0]{
\begin{pmatrix}
\bF_{\ttil \ttil} & \bF_{ \ttil 3} \\ 
\bF_{ 3 \ttil } & \bF_{33}
\end{pmatrix}
}
\newcommand{\iApMat}[0]{
\begin{pmatrix} i A'_{\widetilde 2} & i A'_3 \end{pmatrix}
}
\newcommand{\iAMat}[0]{
 \begin{pmatrix}  iA_{\widetilde 2} \\  iA_3  \end{pmatrix} 
}
\newcommand{\bApMat}[0]{
\begin{pmatrix} \bA'_{\widetilde 2} & \bA'_3 \end{pmatrix}
}
\newcommand{\bAMat}[0]{
 \begin{pmatrix}  \bA_{\widetilde 2} \\  \bA_3  \end{pmatrix} 
}

\newcommand{\bF}{{\mathbf F}}
\newcommand{\bG}{{\mathbf G}}
\newcommand{\bFrp}{{\mathbf F_{\! \rho \pi}}}
\newcommand{\bGr}{{\mathbf G_{ \rho }}}
\newcommand{\bGrD}{{\overline {\mathbf G}_{ \rho }}}
\newcommand{\bGam}{{\pmb \Gamma}}
\newcommand{\bK}{{\mathbf K_2}}

\newcommand{\bKLth}{\mathbf K_{L,33}}
\newcommand{\bcFrp}{{\pmb {\mathcal F}_{\! \rho \pi}}}
\newcommand{\sbKLth}{\slashed{\mathbf K}_{L,33}}
\newcommand{\bKLtwth}{\mathbf K_{L,\ttil 3}}
\newcommand{\sbKLtwth}{\slashed {\mathbf K}_{L,\ttil 3}}
\newcommand{\sbKLthtw}{\slashed {\mathbf K}_{L, 3 \ttil}}
\newcommand{\sbKLtw}{\slashed {\mathbf K}_{L, \ttil \ttil}}
\newcommand{\bKdftwth}{\mathbf K_{\mathrm{df},\ttil 3}}
\newcommand{\bKdfthtw}{\mathbf K_{\mathrm{df}, 3 \ttil}}
\newcommand{\bKdfth}{\mathbf K_{\mathrm{df}, 3 3}}
\newcommand{\bKdftw}{\mathbf K_{\mathrm{df}, \ttil \ttil}}
\newcommand{\sbKL}{\slashed {\mathbf K}_{L}}
\newcommand{\bKi}{\mathbf K}
\newcommand{\bcGK}{\pmb {\mathcal G}_{\mathbf K}}
\newcommand{\bKLtw}{\mathbf K_{L,\ttil \ttil}}
\newcommand{\bKLthtw}{\mathbf K_{L, 3 \ttil}}
\newcommand{\bV}{\mathbf V}
\newcommand{\bVD}{\overline {\mathbf V}}
\newcommand{\bKL}{\mathbf K_{L}}
\newcommand{\bGK}{{\, \mathbf G_{\mathbf K} \,}}
\newcommand{\bSig}{{\pmb \sigma^*}}
\newcommand{\bSigD}{{\pmb \sigma^{\dagger*}}}
\newcommand{\bA}{{\mathbf A}}
\newcommand{\bX}{{\mathbf X}}
\newcommand{\bT}{{\mathbf T}}
\newcommand{\bEL}{{\pmb {\mathcal E}_L}}
\newcommand{\bER}{{\pmb {\mathcal E}_R}}
\newcommand{\bZ}{{\mathbf Z}}
\newcommand{\bY}{{\mathbf Y}}
\newcommand{\bB}{\mathbf B_3}
\newcommand{\bFL}{{\mathbf F_{\mathbf L}}}
\newcommand{\bFR}{{\mathbf F_{\mathbf R}}}
\newcommand{\bSL}{{\pmb {\mathcal S}_{\mathbf L}}}
\newcommand{\bSR}{{\pmb {\mathcal S}_{\mathbf R}}}
\newcommand{\IFL}{\mathcal I_{\bf F \bf L}}
\newcommand{\IFR}{\mathcal I_{\bf F \bf R}}
\newcommand{\IXY}{\mathcal I_{\textbf X \textbf Y}}
\newcommand{\DC}{\mathcal D_{C}}
\newcommand{\DA}{\mathcal D_{\!\bf A}}
\newcommand{\DAp}{\mathcal D_{\!\bf A'}}

\newcommand{\BHSQC}[0]{Briceno:2017tce}
\newcommand{\BHSnum}[0]{Briceno:2018mlh}
\newcommand{\HSQCa}[0]{Hansen:2014eka}
\newcommand{\HSQCb}[0]{Hansen:2015zga}

\newcommand{\nn}[0]{\nonumber} 
\newcommand{\pNCA}[0]{{\em ($\checkmark$NCA\,$\checkmark$\!)}}

\begin{document}

\preprint{\vbox{\hbox{JLAB-THY-18-2819} }}
\preprint{\vbox{\hbox{CERN-TH-2018-216} }}

\title{Three-particle systems with resonant subprocesses in a finite volume}

\author{Ra\'ul A. Brice\~no}
\email[e-mail: ]{rbriceno@jlab.org}
\affiliation{Thomas Jefferson National Accelerator Facility, 12000 Jefferson Avenue, Newport News, VA 23606, USA}
\affiliation{ Department of Physics, Old Dominion University, Norfolk, Virginia 23529, USA}

\author{Maxwell T. Hansen}
\email[e-mail: ]{maxwell.hansen@cern.ch}
\affiliation{Theoretical Physics Department, CERN, 1211 Geneva 23, Switzerland}

\author{Stephen R. Sharpe}
\email[e-mail: ]{srsharpe@uw.edu}
\affiliation{Physics Department, University of Washington, Seattle, WA 98195-1560, USA}

\date{\today}

\begin{abstract}
In previous work, we have developed a relativistic, model-independent three-particle quantization
condition, but only under the assumption that no poles are present in the two-particle K matrices that appear as scattering subprocesses~\cite{\HSQCa,\HSQCb,\BHSQC}.
Here we lift this restriction, by deriving the quantization condition for identical scalar
particles with a G-parity symmetry, in the case that the two-particle K matrix has a pole
in the kinematic regime of interest.
As in earlier work, our result involves intermediate infinite-volume 
quantities with no direct physical interpretation, and we show how these are related to the physical three-to-three scattering
amplitude by integral equations.
This work opens the door to study processes such as $a_2 \to\rho\pi \to \pi \pi \pi$, in which the
$\rho $ is rigorously treated as a resonance state.
 \end{abstract}


\nopagebreak

\maketitle


\section{Introduction}


Studies of hadronic resonances using lattice QCD (LQCD) have progressed rapidly in recent 
years.\footnote{%
 For recent reviews, see Refs.~\cite{Wilson:2016rid,Briceno:2017max,Mohler:2018lvr}. }
The present frontier of this effort involves resonances that have significant branching ratios into
channels with three (or more) particles.
Here the results from lattice calculations are, in some cases, more advanced than the theoretical developments
needed to interpret them. In particular, 
energy levels above three-particle thresholds are already being calculated, 
using three-particle operators~\cite{Cheung:2017tnt}.
Thus a fully developed theoretical formalism to interpret LQCD quantities in this sector is of great importance. In recent years significant progress has been made, using a variety of approaches~\cite{Polejaeva:2012ut,Briceno:2012rv,Aoki:2013cra,\HSQCa,\HSQCb,Hammer:2017uqm,Hammer:2017kms,Guo:2017ism,Mai:2017bge,\BHSQC,Doring:2018xxx,\BHSnum, Mai:2018djl}. 
In this work we consider the relativistic model-independent framework of
Refs.~\cite{\HSQCa,\HSQCb,\BHSQC},
and remove the last major theoretical restriction on this formalism.

LQCD studies of resonances proceed in two basic steps. 
First, one uses numerical LQCD to determine
the energy levels in a finite volume for a given range of total energy.
Second, these levels are related to infinite-volume scattering parameters by solving a
quantization condition.\footnote{%
In practice, this requires truncation of the quantization condition by assuming that higher partial
waves are negligible. Such truncation schemes for the three-particle case have
been discussed in Refs.~\cite{\HSQCa,Hammer:2017uqm,Hammer:2017kms,Mai:2017bge,\BHSQC,Doring:2018xxx,\BHSnum}. We do not consider these further in the present work.}
In the case of a single channel of identical scalar particles, the relation between finite-volume energies and the scattering amplitude was first derived by L\"uscher~\cite{Luscher:1986n2,Luscher:1991n1}. This has since been extended to describe all possible, multi-channel two-particle systems \cite{Rummukainen:1995vs, Kim:2005gf, He:2005ey, Hansen:2012tf, Briceno:2012yi, Briceno:2014oea}
and by now there is a large body of work extracting energy levels above multiple
open thresholds and relating these to the different components of the coupled-channel
scattering amplitudes~\cite{Dudek:2014qha,Briceno:2017qmb,Moir:2016srx,Dudek:2016cru,Wilson:2015dqa,Woss:2018irj}. For resonances with three-particle decay
channels, a further step is required, in which intermediate infinite-volume quantities
are related to the 
scattering amplitudes.  This step also requires knowledge of the
scattering amplitudes in each of the two-particle subsystems.

The approach we follow here was originally derived in Refs.~\cite{\HSQCa,\HSQCb}
under two major assumptions: first, that a G-parity-like symmetry forbids 
$2\leftrightarrow 3$ transitions and, second, that two-particle subsystems are  nonresonant
within the kinematic range of interest (or, more precisely, that the two-particle
K matrices have no poles).\footnote{%
In addition the particles were taken to be
identical and spinless. Based on experience with the two-particle
case, we expect the extensions to multiple channels of non-identical and non-degenerate particles, as well as particles with intrinsic spin, will be relatively straightforward.}
We removed the former restriction in Ref.~\cite{\BHSQC}, and it is the purpose of
the present paper to lift the second restriction, i.e. to allow arbitrary interactions in the
two-particle subsystems.
This removes the last major theoretical obstacle to general implementation of the
formalism.

Removing the restriction on sub-channel K matrices is necessary
for the application of the formalism to many interesting three-particle systems.
Consider, for example, the three-pion system in the isospin-symmetric limit.
Only for the maximal isospin channel, $I=3$, are all
two-pion subchannels nonresonant (since they all have $I_{\rm sub}=2$).
For $I_{\rm tot}<3$, however, the subchannels can have $I_{\rm sub}=1$ or $0$,
and thus contain either the $\rho$ or $\sigma$ resonance, respectively.
For example, the $a_2(1320)$ ($I^G=1^-$, $J^{PC}=2^{++}$) decays predominantly
to three pions with the $\rho$ and $f_2(1270)$ resonances in two-pion subchannels.
Another example where sub-channel resonances must be included is the Roper resonance, which has
a significant branch to the $\Delta\pi\to p\pi\pi$ channel.

\bigskip

It is useful to recall the reason why the analysis in Refs.~\cite{\HSQCa,\BHSQC} 
had to assume the absence of poles in the two-particle K matrix, $\K_2$.
These works study finite-volume correlation functions and determine the spectrum
from the position of the poles in these functions.
The correlation functions are considered to all orders in perturbation theory in an
arbitrary effective field theory.
The core step in the analysis is the replacement of the 3-momentum
sums that appear in finite volume with the corresponding infinite-volume integrals, 
together with a volume-dependent residue.
If the summand is smooth,
this residue is exponentially suppressed (i.e. suppressed by $e^{-m L}$ where $m$
is the particle mass and $L$ the box size),
and such exponentially suppressed corrections are assumed negligible.
However, if the summand is singular then the residue falls only like inverse powers of $L$,
and must be kept. Such singularities occur either when intermediate states
can go on shell or when intermediate infinite-volume quantities are themselves singular. In our approach the latter class of singularities arise as K-matrix poles. In Refs.~\cite{\HSQCa,\BHSQC} we did not include the finite-volume effects associated with these and thus the formalism derived in those works only applies if they are absent.

Poles in $\K_2$ do not correspond to physical particles. If the theory has a narrow resonance in the two-particle subsystem the two-body scattering amplitude, $\mathcal M_2$, will have a complex-valued pole close to the real axis. For this scenario, $\K_2$ will have a real-valued pole close to that of $\mathcal M_2$. Therefore, the $\K_2$ pole is approximately equal to the mass of the resonance. Away from the very-narrow limit of a resonance, however, $\K_2$ poles do not have a direct physical interpretation. Nevertheless, at intermediate stages of the analysis of Ref.~\cite{\HSQCa}, terms
appear whose summands contain such singularities.  
These lead to additional power-law finite-volume dependence, and this
must be accounted for, as it ultimately impacts the form of the quantization condition.
The analysis presented here incorporates all such contributions.

It is worth noting that one may envision taking a different approach than that proposed in Refs.~\cite{\HSQCa,\BHSQC} in which two-particle subprocesses are encoded via $\mathcal M_2$ instead of $\K_2$, and thus the scattering amplitude appears inside of the summand. The original reason for preferring $\mathcal K_2$ is that $\mathcal M_2$ has a cusp at the two-particle threshold and one must then include finite-volume effects associated with this singularity. Furthermore, if the system contains a narrow resonance, of width $\Gamma$, then $\mathcal M_2$ will vary rapidly as a function of energy and this will induce neglected $e^{- \Gamma L}$ volume effects, if the contribution is not explicitly incorporated. In addition, in QCD many resonances lie close to thresholds, leading to dynamically enhanced cusp effects. In short, one would have to develop a framework to address finite-volume effects associated with all possible scenarios. With these considerations in mind, we find it preferable to work with $\K_2$ and properly treat its poles in the kinematic window of interest.

In order to keep track of these singularities, we find it convenient to
express the problem in terms of two effective channels: one containing the physical three-particle state, and a second built from a particle and a pseudoparticle arising from the $\K_2$ pole, which we refer
to as the ``$\rho\pi$ channel". The quantization condition turns out to take a relatively simple form in this presentation, one that
is similar to that in the multichannel two-particle 
problem~\cite{He:2005ey,Briceno:2012yi,Hansen:2012tf}. 
An important consistency check is that our final expression for
the finite-volume correlator does not contain $\rho\pi$ poles, despite their
appearance at intermediate stages. 

\bigskip

The addition of an unphysical channel appears at first as a negative feature
of our approach. We have explored various alternatives that do not require
this artifact, but have not yet been able to use them to derive a useful alternative formalism.
There is, however, one reason to view the appearance of this channel as natural.
To explain this, we return to the example of the $\rho$ resonance, and imagine 
continuously increasing
the quark masses, starting from their physical values. 
As is well known, as this is done, the $\rho$ becomes narrower,
eventually becoming a bound-state at threshold, and, beyond that, a physical particle
lying below the two-pion threshold. If the masses are chosen such that the $\rho$ is deeply bound, then the pole in $\K_2$ moves far below threshold and becomes irrelevant to our formalism. Thus, in this case, the unphysical channel is no longer needed. However, the presence of the $\rho$
particle implies that one should use the $2+3$ particle formalism of Ref.~\cite{\BHSQC},
including a physical $\rho\pi$ channel. Given that the stable $\rho$ can be described in terms of a new open channel, it is natural that this is continuously connected to the effective two-particle channel for narrow resonances that arises in this work.

As was the case in Refs.~\cite{\HSQCa,\HSQCb,\BHSQC},
the derivation of the quantization condition is rather lengthy, despite the fact that we have found
ways to shorten and simplify certain steps compared to the earlier works.
To make this paper more accessible, we have focused in the main text on the logic
and key steps of the derivation, pushing most of the details into appendices.
In addition, we have provided a {\em Mathematica} notebook as supplementary
material in which the package {\em The NCAlgebra Suite} is used to check the key results by algebraically manipulating matrices of unspecified size as generic non-commuting objects \cite{NCA}.

\bigskip

This article is organized as follows. We begin in Sec.~\ref{sec:summary}
 by presenting the final result and defining all of the objects appearing in it. 
This section is meant to stand alone so that the lattice practitioner does not 
need to look elsewhere in order to make use of the result. 
In Sec.~\ref{sec:derivation} we present the derivation of the quantization condition,
with technical details given in Appendix~\ref{app:derapp}. 
The quantization condition is written in terms of the three-body K matrix, 
which we relate to the physical scattering amplitude in Sec.~\ref{sec:KtoM}. 
We summarize, compare to previous work, and give an outlook in Sec.~\ref{sec:conclusion}. 

The framework presented here relies heavily on two facts: First, that the off-shell version of $\K_2$ has the same poles as its on-shell limit and second, that at the residues of the poles of the off-shell $\K_2$ can be written as a product of functions separately describing the incoming and outgoing two-particle states. In Appendix~\ref{app:factorize} we 
demonstrate these two results using constraints from unitarity and all-orders perturbation theory.

\section{Summary of the final result \label{sec:summary}}


The main result of this article is a quantization condition with solutions equal to the energies of finite-volume three-particle states in a generic, relativistic quantum field theory. In contrast to earlier work, this result also holds for systems with a two-particle resonant subchannel.
The particles are assumed to be identical, of physical mass $m$, and to have a G-parity-like symmetry that restricts interactions to those involving an even number of fields.  

We assume that $\K_2$ diverges only for a single angular momentum, denoted $J$,
in the energy range of interest, specified below.
We further assume that there is only one pole in $\K_2^{(J)}$ in this energy range,
occurring when the two-particle center of mass (c.m.)~energy equals $M$.\footnote{%
In the following we refer to this energy as the ``resonance mass", which is a convenient
label despite the fact that the correct definition of the resonance mass differs for all but
a very narrow resonance. As noted in the introduction, we also refer to the resonance
channel as the $\rho$.}
These assumptions simplify the discussion and derivation. The extension to completely general K matrices, achieved by promoting certain quantities introduced here to matrices, will be described in a future publication.

The result presented in this work holds for fields restricted to a cubic spatial volume of side length $L$, with periodic boundary conditions.
Following the pattern that is by now well established from previous work \cite{\HSQCa,\HSQCb,\BHSQC}, we find that for a given total momentum, $\vec P = 2 \pi \vec n_P /L$, 
the discrete finite-volume spectrum
is given by all solutions in $E$ to the condition
 \begin{equation}
 \label{eq:det}
 \mathrm{det} \big [1 + \K_\df(E^*)  \mathcal F(E, \vec P, L) \big ] = 0 \,,
 \end{equation}
where $E^* = \sqrt{E^2 - \vec P^2}$ is the total energy in the c.m.~frame. Here both $\K_\df(E^*)$ and $\mathcal F(E, \vec P, L)$ are matrices on a two-channel space
\begin{align}
i \K_\df &\equiv \iKdfMat \,,
\label{eq:Kmat_def}
\\[5pt]
 i \mathcal F &\equiv \iFMat \,,
\label{eq:Fmat_def}
\end{align}
where the index $3$ denotes the three-particle channel while $\tilde 2$ denotes an effective two-particle channel containing the two-particle resonance with the third non-resonating particle. This result holds up to neglected corrections of the form $e^{- m L}$, with $m$ the physical mass of the stable particle, and applies only in the region $m < E^* < 5m$.

In the remainder of this section we provide the definitions of the quantities $\K_\df$ and
$\cF$ appearing in the quantization condition. We only note here that $\K_\df$ is a real, infinite-volume quantity that is related to the three-to-three 
scattering amplitude, while $\cF$ has volume dependence but can be expressed in terms of known geometric functions together with
the two-particle scattering amplitude, including parameters describing the K matrix pole.

We discuss strategies for the practical implementation of the quantization condition,
the generalization to multiple K matrix poles,
and the relation of this result to earlier work, in Sec.~\ref{sec:conclusion}.


\subsection{Kinematics}

In this subsection we introduce the kinematic variables used throughout the paper
to describe two- and three-particle states, and the index space implicit in the matrices
appearing in the quantization condition, Eq.~(\ref{eq:det}).
These results are summarized in Table~\ref{tab:kin}, which we hope will provide a useful 
reference for the reader. Many of the results are self-explanatory; for the others we provide
further explanation in the following.

\renewcommand{\arraystretch}{2}

\begin{table}
\begin{center}
\begin{tabular}{c | c | l}
\ \ Quantity \ \  & \ \  Definition/Key relation  \ \ & \ \ \ Description \\ \hline \hline
\multicolumn{3}{c}{\em Basic kinematics used throughout} \\
$\vec k$ & $(k_x, k_y, k_z) = 2 \pi \vec n/L$ & \ \ \ 3-momentum (often of the spectator particle)\\
$\omega_k$ & $\sqrt{\vec k^2 + m^2}$ & \ \ \  on-shell time component of 4-vector $k^\mu$ (with physical mass $m$) \\
$\ell m$ &  indices on $Y_{\ell m}$  & \ \ \  angular-momentum indices (e.g.~of the non-spectator pair)\\
$M_J$ & $M_J = - J, -J +1 , \cdots, J$ & \ \ \  azimuthal component of total angular-momentum $J$ \\ \hline \hline
\multicolumn{3}{c}{\em Multi-particle energies and momenta} \\
 $(E, \vec P)$ & $\vec P = 2 \pi \vec n_P/L$& \ \ \ total energy and momentum of the three-particle state \\ 
$P_{2,k}$ & $(E - \omega_k, \vec P - \vec k)$ & \ \ \ 4-momentum of the non-spectator pair 
or of the resonance \\
$E^*_{2,k}$ &  $  \sqrt{(E - \omega_k)^2 - (\vec P - \vec k)^2}$ & \ \ \  energy of the non-spectator pair (two-particle c.m.~frame) \\
$q^*_{2,k} $ & $ \sqrt{E^{*2}_{2,k}/4 - m^2}$ & \ \ \ on-shell momentum of a non-spectator (two-particle c.m.~frame) \\ \hline \hline
\multicolumn{3}{c}{\em Individual particles within the three-particle state} \\
$\vec a$, $\vec b_{ka}$ & $\vec b_{ka} \equiv \vec P - \vec k - \vec a$ & \ \ \ individual 3-momenta of the non-spectators (finite-volume frame) \\
$(\omega_{a}, \vec a )$ & & \ \ \ 4-momentum of the $a$-momentum particle (finite-volume frame) \\
$(\omega^*_{a; 2,k}, \vec a^*_{2, k})$ && \ \ \ 4-momentum of the $a$-momentum particle (two-particle c.m.~frame) \\ 
$(E - \omega_k - \omega_a, \vec b_{ka})$ & & \ \ \ 4-momentum of the $b$-momentum particle (finite-volume frame) \\
$ (E^*_{2,k} - \omega_{a; 2,k}^* , \vec b^*_{ka; 2, k})$ & $ \vec b^*_{ka; 2, k} = - \vec a^*_{2, k}$ &  \ \ \  4-momentum of the $b$-momentum particle (two-particle c.m.~frame) \\  
$ \omega_{Pka}$ & $ \sqrt{m^2 + (\vec P - \vec k - \vec a)^2 }$   & \ \ \  on-shell time component of the $b$-momentum particle  \\ \hline \hline
\multicolumn{3}{c}{\em Individual particles within the $\ttil$-state} \\
$M$ & $\lim_{E_{2}^* \to M} \mathcal K_2(E_2^*) =  \infty$ & \ \ \ position of the $\mathcal K_2$ pole \\ \cline{1-2}
\multicolumn{2}{c |}{  $q^*_\rho \ \ \ \text{defined via} \ \ \  E^* = \sqrt{m^2 + q^{*2}_\rho} + \sqrt{M^2 + q^{*2}_\rho}$ }  & \ \ \ on-shell momentum of the $\ttil$ spectator (c.m.~frame) \\ \cline{1-2}
$\omega_{\rho,  k}$ & $ \sqrt{M^2 + (\vec P - \vec k)^2}$ & \ \ \ on-shell time component of the resonance   \\
$(\omega_{k}, \vec k)$ && \ \ \ 4-momentum of the $\ttil$ spectator (finite-volume frame) \\
$(E-\omega_{k}, \vec P -\vec k)$  &  & \ \ \ 4-momentum of the resonance (finite-volume frame) \\
$(\omega^*_{k},\vec k^*)$ && \ \ \  4-momentum of the $\ttil$ spectator (c.m.~frame)   \\
$(E^*-\omega_{k}^*, -\vec k^*)$  &   & \ \ \ 4-momentum of the resonance (c.m.~frame) \\
\hline \hline 
\multicolumn{3}{c}{\em On-shell conditions and index spaces} \\
\multicolumn{2}{c |}{ $E - \omega_k - \omega_a = \omega_{Pka} \ \Leftrightarrow \ E^*_{2,k} = 2 \omega_{a; 2,k}^* \ \Leftrightarrow \   a^*_{2,k} =  q^*_{2,k} $}   & \ \ \ equivalent on-shell conditions for the 3-particle state \\
\multicolumn{2}{c |}{ $E - \omega_{k} = \omega_{\rho,k}  
\ \Leftrightarrow \ E^* - \omega_{k}^*= \sqrt{M^2+k^{*\,2} }
\ \Leftrightarrow  \ k^* = q^*_\rho $  } & \ \ \ equivalent on-shell conditions for the $\ttil$-state \\ \hline 
\multicolumn{2}{c |}{ $ k \ell  m = \vec k, \ell , m = k_x, k_y, k_z, \ell, m$   \ \ where  \ \ $k_i = 2 \pi n_i/L$  } & \ \ \  index-space for an on-shell $3$-state (implicit with $3$ subscript)  \\
\multicolumn{2}{c |}{  $M_J, \ell, m$ } & \ \ \ index-space for an on-shell $\ttil$-state (implicit with $\ttil$ subscript)
\end{tabular}
\caption{Summary of kinematics used throughout the paper. \label{tab:kin}}
\end{center}
\end{table}

\renewcommand{\arraystretch}{1.0}

Each entry in the two-by-two matrices $\mathcal K_{\mathrm{df}}$ and $ \mathcal F$ is itself a matrix in a space that describes the on-shell degrees of freedom, either for three particles or for the resonance together with the spectator. In particular, $3$, when used as an index, is shorthand for $3; k \ell m = 3; k_x k_y k_z \ell m$ and $\ttil$ is shorthand for $ \ttil; M_J \ell m$. 
We use $\ttil$ rather than $2$ to emphasize that the K matrix pole does not correspond to a physical particle, and so the $\ttil$ channel is not a physical two-particle channel.

In the three-particle state, one of the three particles, referred to as the spectator, carries the 3-momentum $\vec k = (k_x, k_y, k_z)$. In infinite volume this momentum can take on a continuous range of values (within the range allowed by total energy and momentum conservation), but in our quantization condition it is restricted to discrete values: $\vec k = 2 \pi \vec n/L$ where $\vec n$ is a 3-vector of integers. Within the three-particle state, $\ell m$ describes the angular momentum of the non-spectator pair. 

In the $\ttil$ state, $M_J$ labels the different azimuthal components for a K-matrix pole with angular momentum $J$. Roughly speaking, it plays the role of a channel index, labeling different degrees of freedom rather than different momentum configurations. For  a given value of $M_J$, $\ell m$ describes the angular momentum of the 
spectator-resonance pair.\footnote{%
We stress that the index pair $\ell m$ plays a very different role in the $3$ and $\ttil$ states.
This causes no problems, however, as these two sets of indices are never contracted.
}

\bigskip

The kinematics used for the on-shell three-particle state are described in detail in Refs.~\cite{\HSQCa,\HSQCb}. For completeness, and to introduce new notation, we summarize the discussion here. For a given total energy and momentum, $(E, \vec P)$, we label one of the three particles (the spectator) with on-shell 4-momentum $(\omega_k, \vec k)$, where
$\omega_k = \sqrt{\vec k^2 + m^2} $.
The 4-momentum of the remaining two particles is then $P_{2,k} \equiv (E - \omega_k, \vec P - \vec k)$ and their two-particle c.m.~energy is 
\begin{equation}
E^*_{2,k} = \sqrt{P_{2,k}^2} = \sqrt{(E - \omega_k)^2 - (\vec P - \vec k)^2}\,.
\end{equation}
We denote the individual 3-momenta of these two particles in the finite-volume frame by $\vec a$ and $\vec b_{ka} = \vec P - \vec k - \vec a$. 

Often we must consider the case were the $\vec a$ particle is on shell with 4-momentum $(\omega_a, \vec a)$ whereas the $\vec b$ particle is not necessarily on shell, and carries 4-momentum $(E - \omega_k - \omega_a, \vec b_{ka})$. Boosting the 4-vectors corresponding to $\vec a$ and 
$\vec b$ to the two-particle c.m.~frame then gives, respectively,
\begin{equation}
(\omega^*_{a; 2,k}, \vec a^*_{2, k}) \,, 
\ \ \ \ \ 
(E^*_{2,k} - \omega_{a; 2,k}^* , \vec b^*_{ka; 2, k}) = (E^*_{2,k}- \omega_{a; 2,k}^* , - \vec a^*_{2, k}) \,.
\end{equation} 
Here the notation is somewhat involved as we must label both the momenta and the frame. Finally, we need to know the conditions on the kinematic variables such that the 
$\vec b$ particle is also on shell; these are given towards the bottom of Table~\ref{tab:kin}. 
The upshot is that, for three on-shell particles with total energy and momentum $(E, \vec P)$, the remaining degrees of freedom are the spectator momentum, $\vec k$, and 
the direction of the $\vec a$ particle in the non-spectator-pair c.m.~frame, $\hat a^*_{2,k}$. Decomposing the latter in spherical harmonics leads to the indices $\vec k, \ell, m$, 
which we abbreviate to $k \ell m$.\footnote{%
As mentioned above, the quantization condition depends only on the allowed finite-volume spectator momenta, $\vec k = 2 \pi \vec n/L$ with $\vec n$ a 3-vector of integers.}

We turn now to the $\ttil$ state, built from a particle of mass $m$
and the resonance of mass $M$.
In the overall c.m.~frame, each of these has a 3-momentum with a magnitude that we denote by $q^*_\rho$,  given by solving
\begin{equation}
E^* = \sqrt{m^2 + q^{*2}_\rho} + \sqrt{M^2 + q^{*2}_\rho} \,.
\label{eq:krho}
\end{equation}
In the finite-volume frame, if the particle has momentum $\vec k$ and is on shell, 
then the resonance has 4-momentum $(E-\omega_{k}, \vec P-\vec k)$.
Boosting these to the overall c.m.~frame gives 
$(\omega_{k}^*, \vec k^*)$ and $(E^* - \omega^*_{k},- \vec k^*)$. 
The second particle is then on-shell when any of the 
three equivalent conditions listed in Table \ref{tab:kin} are satisfied.

Thus, for fixed $(E, \vec P)$, the two on-shell particles have 
remaining degree of freedom $\hat k^*$ and decomposing this
in spherical harmonics gives the indices $\ell, m$. 
Combining this with the azimuthal angular momentum of the resonance
 gives the full index set, $M_J \ell m$.


\subsection{K-matrix poles \label{sec:Kpoles}}

The central aim of this paper is to include the finite-volume effects from poles in $\K_2$.
In order to complete the definitions of the quantities entering the quantization condition,
we need to understand the properties of these poles. 
This is nontrivial, because, unlike poles in the scattering amplitude, poles in $\K_2$
do not correspond to propagation of physical degrees of freedom.
Nevertheless, as we show in this subsection and the accompanying
 Appendix~\ref{app:factorize}, two key results do carry over from poles in $\cM_2$:
the off-shell K matrix has the same poles as the on-shell version, 
and the residues of the poles factorize.
Both results play an important role in the subsequent derivation.

We begin by recalling that the $\ell$th angular-momentum component of the two-to-two on-shell scattering amplitude satisfies a unitarity constraint, relating it to the scattering phase-shift via
\begin{equation}
\mathcal M^{(\ell)}_2(P_{2,k}^2)   =   \frac{16 \pi E_{2,k}^*}{q_{2,k}^*} \frac{1}{\cot \delta_{\ell}(q_{2,k}^*)  - i } \,,
\label{eq:MvsK}
\end{equation}
or equivalently
\begin{equation}
\label{eq:MtotanD}
\mathcal M^{(\ell)}_2(P_{2,k}^2)^{-1}  - \bigg [   \frac{16 \pi E_{2,k}^*}{q_{2,k}^*\cot \delta_{\ell}(q_{2,k}^*)}    \bigg ]^{-1} =  - i    \frac{q_{2,k}^*} {16 \pi E_{2,k}^*}    \,.
\end{equation}
In anticipation of three-particle scattering, 
we have taken the squared c.m.~energy in the two-to-two amplitude as 
$P_{2,k}^2 = E_{2,k}^{*2}$,
where we recall that $P_{2,k} \equiv (E - \omega_k, \vec P - \vec k)$ is 
our notation for the 4-momentum of the non-spectator pair (see Table~\ref{tab:kin}).
Thus the spectator momentum $\vec k$ serves a proxy for the two-particle c.m~frame energy.
We are assuming in Eqs.~(\ref{eq:MvsK}) and (\ref{eq:MtotanD}) that the scattering is
above threshold and in the region where only two-particle states can propagate, $2 m \leq E_{2,k}^* < 4 m$.

The quantity appearing in square braces in Eq.~(\ref{eq:MtotanD}) 
defines the conventional K matrix when working above threshold.
It is a real function containing all dynamical information about the two-particle scattering.
We will also need the continuation below threshold, and here,
following Ref.~\cite{\HSQCa}, we use a nonstandard choice that is convenient for
the derivation of the quantization condition.
Our K matrix is given by
\begin{equation}
\label{eq:ourK}
\mathcal M_2^{(\ell)}(P_{2,k}^2)^{-1}  - \mathcal K_2^{(\ell)}(P_{2,k}^2)^{-1}   \equiv H(\vec k) \tilde \rho(P_{2,k}^2) \,,
\end{equation}
where $\tilde \rho$ is the standard phase-space factor, including below-threshold analytic continuation,
\begin{align}
\tilde\rho(P_{2,k}^2) &\equiv \frac{1}{16 \pi  \sqrt{P_{2,k}^2}} \times
\begin{cases} 
-  i \sqrt{P_{2,k}^2/4-m^2} & (2m)^2< P_{2,k}^{2} \,, 
\\[10pt]
\Big \vert \sqrt{P_{2,k}^2/4-m^2} \Big \vert &   0<P_{2,k}^{2} \leq (2m)^2 \,,
\end{cases}
\label{eq:rhotdef}
\end{align}
and $H(\vec k)$ is a smooth, real cut-off function that equals $1$ when $E_{2,k}^*\ge 2m$
(so that the particles in the nonspectator pair can propagate on-shell) 
and then smoothly interpolates to 0 in the sub-threshold region. 
Our choice of $\K_2$ differs from the analytic continuation of the above-threshold
K matrix once $H$ differs from unity. 

Although we do not need to make a choice of $H$ for the derivation, it is useful to
have one in mind as an example. The choice suggested in Ref.~\cite{\HSQCa},
and used in our recent numerical investigation~\cite{\BHSnum}, is
\begin{equation}
\label{eq:Hdef}
H(\vec k) \equiv J(P_{2,k}^{2}/[4m^2]) \,,
\end{equation}
with
\begin{equation}
\label{eq:Jdef}
J(x) \equiv
\begin{cases}
0 \,, & x \le 0 \,; 
\\ 
\exp \left( - \frac{1}{x} \exp \left [-\frac{1}{1-x} \right] \right ) \,, 
& 0<x \le 1 \,; 
\\ 
1 \,, & 1<x \,.
\end{cases}
\end{equation}
With this definition, $H$ vanishes for $E_{2,k}^{*2} \leq 0$.

Up to this point, we have considered only the on-shell K matrix, including the analytic continuation to
sub-threshold momenta. However, in our derivation we also require its off-shell extension,
in which the individual particle momenta take on values of $p^2$ differing from $m^2$. 
Although not necessary for the implementation of the main result of this work, namely
Eq.~(\ref{eq:det}), we find it informative to discuss the off-shell extension of the K matrix.
Off-shell extensions are not uniquely defined,
as they depend on the choice of single-particle interpolator. 
In our all-orders diagrammatic analysis, based in a generic effective field theory, we
define the fully off-shell scattering amplitude $\cM_{2,\f,\f}$
by amputating the corresponding four-point correlation function.
The presence of two ``off"s indicates that both initial and final state particles are off shell.
This corresponds to choosing the interpolator to be
the fundamental field in the theory, renormalized so that
it couples to an on-shell particle with unit amplitude. 
This is a natural choice in perturbation theory.

In the diagrammatic framework, this definition is naturally extended to the K matrix.
To go from the off-shell $\cM_2$ to the off-shell $\cK_2$,
one considers the same amputated correlation function,
but replaces the $i\epsilon$ prescription for integrals over poles
with the principal value (PV) prescription modified by multiplication by $H(\vec k)$---as
described in Ref.~\cite{\HSQCa}.
For our kinematic range, $0 < E_{2,k}^* < 4 m$, this only impacts two-particle
intermediate states, and the difference between the prescriptions occurs only
when the intermediate state is on shell.
This allows one to write the fully off-shell K matrix in terms of the fully on-shell K matrix:
\begin{equation}
i\K_{2,\f,\f}^{(\ell)}
=
i\cM_{2,\f,\f}^{(\ell)}
-
i\cM_{2,\f,\on}^{(\ell)} i H \tilde\rho\, \K_{2,\on,\on}^{(\ell)} 
\cM_{2,\on,\on}^{(\ell)\ -1} i\cM_{2,\on,\f}^{(\ell)}
\,,
\label{eq:Kofffromon}
\end{equation}
where momentum arguments are suppressed for the sake of brevity.
We derive this result in Appendix~\ref{app:factorize}.
As also discussed in the Appendix, 
it follows from Eq.~(\ref{eq:Kofffromon}) that the off- and on-shell K matrices
have poles at the same positions, the first of the key results mentioned in the introduction.

We now turn to case of interest in which $\cK_{2,\on,\on}^{(\ell)}$ has a pole for $\ell=J$.
Above threshold, this happens when $\cot\delta_J$ vanishes,
i.e. when the phase shift passes through $\pi/2 + n \pi$ for any integer $n$.
If the phase shift is increasing this corresponds to a nearby resonance, but
we stress that we must also consider the situation in which $\delta_J$ decreases through
$\pi/2+n\pi$, which does not correspond to a resonance but still leads to
power-law finite-volume effects.
$\cK_2$ can also have a pole below threshold, 
when $\cM_2^{(\ell) -1}=-H \tilde \rho$ [see Eq.~(\ref{eq:ourK})].
This is not directly associated with anything physical, e.g.~a bound state, but nevertheless
also contributes finite-volume effects.
In all cases, near the pole the on-shell K matrix has the form
 \begin{equation}
\K_{2,\on,\on}^{(J)}(P_{2,k}^2) = \frac{R}{P_{2,k}^2 - M^2} + \textrm{non-pole}
\,,
\label{eq:poleinK2}
\end{equation}
with $M$ the pole position and $R$ a real constant. 
The pole must have a Lorentz-invariant form
as $\K_2$ is relativistically invariant.
Inserting Eq.~(\ref{eq:poleinK2}) into Eq.~(\ref{eq:Kofffromon}), it follows from
the structure of the second term on the right-hand side of the latter equation
that the off-shell momentum dependence factorizes, as discussed in 
Appendix~\ref{app:factorize}. This allows us to write
 \begin{multline}
 \label{eq:K2off_v2}
i \Ki{, \f, \f}^{(J)}(P^2_{2,k}; p^2,b^2,a'^2,b'^2) =  
 (p^*_{2,k})^J  i \Gamma_{J}(M^2, p^2, b^2) \frac{i \eta_{J}}{P_{2,k}^2 - M^2} 
i \Gamma_J(M^2, a'^2, b'^2) (a'^*_{2,k})^{J} \\ 
+ i \Kti{, \f, \f}^{(J)}(P_{2,k}^2; p^2,b^2,a'^2,b'^2) \,.
 \end{multline}
Here we have made the momentum arguments explicit:
$a'$ and $b'$ are the incoming four-momenta, while $p$ and $b$ are the outgoing. 
Factorization manifests itself as the dependence on $a'^2$ and $b'^2$
being independent of that on $p^2$ and $b^2$. These dependences arise,
respectively, from the factors of $\cM_{2,\on,\off}$ and $\cM_{2,\off,\on}$,
in Eq.~(\ref{eq:Kofffromon}). Since they are related by time reversal, 
the residue functions $\Gamma_J$ that carry the off-shell dependence are the
same for initial and final momenta. 
These residue functions are real.

The remaining factors in Eq.~(\ref{eq:K2off_v2}) can be understood as follows:
$\eta_J=\pm 1$ encodes the sign of the residue,
with both values allowed since this is not a physical pole.
The $\Kti{}$ term is the non-pole residue and is a smooth function of its arguments.
Finally, the ``barrier factors" $(p^*_{2,k})^J $ and $(a'^*_{2,k})^{J}$ 
have been pulled out\footnote{%
The quantities $a'^*_{2,k}$ and $p^*_{2,k}$ are similar to the $a_{2,k}^*$ defined
in Table~\ref{tab:kin}. They are obtained by boosting the four momenta
$a'$ and $p$, respectively, into the two particle c.m.~frame.
They differ from $a_{2,k}^*$ slightly because $a'$ and $p$ are not, in general, on shell
4-vectors, while, in Table~\ref{tab:kin}, $a$ is on shell.}
so that when $\cK_{2,\f,\f}^{(J)}$ is multiplied by
spherical harmonics to reconstruct the full $\cK_{2,\f,\f}$ there are no nonanalyticities
when $p^*_{2,k}$ and $a'^*_{2,k}$ vanish.\footnote{%
The key point here is that $a^\ell Y_{\ell m}(\hat a)$ is a polynomial in 
the components of $\vec a$, while $Y_{\ell m}(\hat a)$ is nonanalytic at $\vec a=0$.}

We choose in Eq.~(\ref{eq:K2off_v2}) to set the first argument in $\Gamma_J$
(which, in general, is $P_{2,k}^{*\, 2}$) to its value at the pole, $M^2$.
This choice is convenient for the derivations.
It is allowed as the difference cancels the pole and leads to a term
that can be absorbed in $\Kti{}$.
We stress, however, that we do not evaluate the barrier factors at the pole,
since this would reintroduce the nonanalyticities that these factors remove.

 Taking the on-shell limit, i.e.~sending $p^2,b^2,a'^2,b'^2 \longrightarrow m^2$, we reach
 \begin{equation}
 \label{eq:K2poles}
 i   {\K}_2^{(J)}(P^2_{2,k}) = (q^*_{2,k})^J i \Gamma_{J} \frac{i \eta_{J}}{P_{2,k}^2 - M^2} i \Gamma_J (q^*_{2,k})^{J} + i \widetilde {\K}_2^{(J)}(P_{2,k}^2) \,,
 \end{equation}
  where we have introduced the following shorthand for the fully on-shell residue function
 \begin{equation}
\Gamma_J \equiv  \Gamma_J(M^2, m^2, m^2)   \,,
\label{eq:GammaJ_def}
 \end{equation}
 together with analogous notation for ${\K}_2^{(J)}$ and $\widetilde {\K}_2^{(J)}$. As is shown explicitly in the following subsection, the quantities ${\K}_2^{(\ell)}$, $\Gamma_J$ and $M^2$ all enter the definition of the finite-volume matrix $\mathcal F$.


\subsection{Definition of $\mathcal F$}

We now have the ingredients necessary to define the entries in the matrix
$\cF$, Eq.~(\ref{eq:Fmat_def}).
We begin with the $33$ component, $F_{33} = F_{33; k' \ell' m'; k \ell m}$.
This is defined by
\begin{equation}
\label{eq:F3def}
i F_{33} \equiv \frac{1}{2 \omega L^3} \bigg [
 \frac{iF}{3} + iF   i \mathcal T_L i F \bigg ] \,,
\end{equation}
where
\begin{equation}
i \mathcal T_L \equiv \frac1{1-i\K_2 (iF+iG)} i\K_2\,,
  \label{eq:TL}
\end{equation}
and
\begin{align}
\left[ \frac{1}{2 \omega L^3} \right]_{k',\ell',m';k,\ell,m} 
& \equiv
\delta_{k',k} \delta_{\ell',\ell} \delta_{m',m} \frac{1}{2 \omega_k L^3} 
\label{eq:Domegadef}
\,,\\ 
\label{eq:Gdef}
i G_{p, \ell', m' ; k, \ell, m} 
& \equiv \mathcal Y_{3, \ell' m'}(\vec k_{2,p}^*) 
 i \mathbf S^{0}_3(\vec p, \vec k )
   \mathcal Y^*_{3, \ell m}(\vec p_{2,k}^{\,*}) 
\frac{1}{2 \omega_k L^3} \,, 
\\
\label{eq:Fdef1}
i F_{k', \ell',m';k,\ell,m} 
& \equiv 
\delta_{k',k} i F_{\ell',m';\ell,m}(\vec k)\,,
\\
\label{eq:Fdef2}
i F_{\ell',m';\ell,m}(\vec k)
&\equiv
i F^{i\epsilon}_{\ell',m';\ell,m}(\vec k)
+
i \rho_{\ell', m'; \ell, m}(\vec k) \,,
\\
\label{eq:Fdef3}
i F^{i\epsilon}_{\ell',m';\ell,m}(\vec k)
&\equiv
\frac12 \left[\frac{1}{L^3} \sum_{\vec a} - \int_{\vec a} \right]  \frac{1}{2 \omega_a} \mathcal Y_{3, \ell' m'}(\vec a_{2,k}^*) 
 i \mathbf S^{i \epsilon}_3(\vec p, \vec k )
   \mathcal Y^*_{3, \ell m}(\vec a_{2,k}^{\,*}) \,, \\
   i \mathcal K_{2; k' \ell' m'; k \ell m } & \equiv \delta_{k', k} \delta_{\ell', \ell} \delta_{m', m} i \mathcal K^{(\ell)}_2(P_{2,k}^2) \,,
\end{align}
with $\int_{\vec a} \equiv \int d^3 a/(2 \pi)^3$ and $\sum_{\vec a} = \sum_{\vec n \in \mathbb Z^3,\ \vec a = 2 \pi \vec n/L}$. 
We have introduced a compact notation for poles and harmonic polynomials
\begin{gather}
  \mathcal Y_{3, \ell m}(\vec k_{2,p}^*) \equiv
 \sqrt{4 \pi}  \left ( \frac{k_{2,p}^*}{q^*_{2,p}}  \right )^{\ell} Y_{\ell m}(\hat k^*_{2,p})
   \,, \ \ \ 
\ i \mathbf S^{i \epsilon}_3(\vec p, \vec k ) \equiv 
\frac{i H_3(\vec p, \vec k)}{ b_{pk}^2 - m^2 + i \epsilon} \,.
\label{eq:harmonic3}
\end{gather}
In Eq.~(\ref{eq:Fdef2}), $\rho(\vec k)$ is a phase-space factor defined by
\begin{align}
\label{eq:rhodef}
\rho_{\ell',m';\ell,m}(\vec k)& \equiv 
\delta_{\ell',\ell} \delta_{m',m} H(\vec k) \tilde\rho(P_{2,k})\,,
\end{align}
where $\tilde \rho(P_{2,k}^2)$ and $H(\vec k)$ are defined in Eqs.~(\ref{eq:rhotdef}) and (\ref{eq:Hdef}) above. 
Finally, $H_3(\vec p, \vec k)$ is a symmetric product of the smooth cutoff function, $H(\vec k)$,
\begin{equation}
H_3(\vec p, \vec k) = H(\vec k) H(\vec p) H(\vec b_{pk}) \,.
\label{eq:H3_def}
\end{equation}

This definition is nearly the same as that used in Refs.~\cite{\HSQCa,\HSQCb}.
There are two differences.
The first is that $F$ and $G$ are expressed here in a manifestly Lorentz covariant way---the
pole term in $\mathbf S_3$ involves the square of a four-vector rather than the 
energies in the finite-volume frame. This changes $F$ only by exponentially suppressed
contributions, but for $G$ the modification is significant. In particular, using the definition above
leads to $\K_\df$ being a Lorentz scalar, as noted in Ref.~\cite{\BHSQC}.
The second change is that, in Refs.~\cite{\HSQCa,\HSQCb}, $G$ is defined
with $H_3 \to H(\vec k) H(\vec p)$ rather than the form with three $H$ functions
given in Eq.~(\ref{eq:H3_def}).
The present definition is that which appears in the case of no $\mathbb Z_2$ symmetry,
as shown in Ref.~\cite{\BHSnum}. Thus, although it is not mandatory here,
this choice of $H_3$ seems more likely to lead to a formalism that
smoothly goes over to the result when the resonance becomes stable.
In any case, it is one possible choice.

The other three entries of $\mathcal F$ are new to this work, and are all brought about
by the presence of the pole in $\K_2$. They are defined as
\begin{align}
i F_{\ttil\ttil} &\equiv i F_{\rho\pi} 
+ i G_\rho i\Gamma_J \frac{1}{2 \omega L^3} (iF+iG) \frac1{1- i\K_2(iF+iG)} i\Gamma_J i G_\rho^\dagger\,,
\label{eq:F22def}
\\
i F_{\ttil 3} &\equiv iG_\rho i\Gamma_J \frac{1}{2 \omega L^3} \frac1{1- (iF+iG) i\K_2} iF\,,
\label{eq:F23def}
\\
i F_{3\ttil} &\equiv \frac{1}{2 \omega L^3}  iF \frac1{1-i\K_2(iF+iG)} i\Gamma_J iG_\rho^\dagger \,, 
\label{eq:F32def}
\end{align}
where $\Gamma_J$ is the on-shell residue defined in Eq.~(\ref{eq:GammaJ_def}),
and we have introduced two new kinematic functions, needed to describe the 
finite-volume dependence arising from the K-matrix pole:
\begin{align}
i G_{\rho; M_J' \ell' m'; k \ell m} & \equiv
{\mathcal Y_{\ttil, \ell' m'}(\vec k^*)} \   i \mathbf S_{\ttil}(\vec k) \  \delta_{J, \ell} \delta_{M_J', m} (q^*_{2,k})^J  \, , \\
 i F_{ \rho \pi; M_J' \ell' m'; M_J \ell m} &\equiv   \delta_{M_J' M_J}   \frac{1}{L^3} \sum_{\vec k}   \frac{1}{2 \omega_k}   {              \mathcal Y_{\ttil, \ell' m'}(\vec k^*)} \  i \mathbf S_{\ttil}(\vec k) \   
 {\mathcal   Y^*_{\ttil, \ell m}(\vec k^*)} \,,
 \label{eq:Frhopi_def}
\\
\mathcal Y_{\ttil, \ell m}(\vec k^*) &\equiv 
\sqrt{4 \pi}  \left ( \frac{k^*}{q^*_\rho}  \right )^{\ell} Y_{\ell m}(\hat k^*)  \,, 
\label{eq:harmonic2}
\\
i \mathbf S_{\ttil}(\vec k) &\equiv     \frac{i \eta_J H_\rho(\vec k\,)}{P_{2,k}^2 - M^2}   \,,
\end{align}
where $\eta_J = \pm 1$ encodes the sign of the residue of the $\K_2$ pole and 
is defined in Eq.~(\ref{eq:K2off_v2}).

Here we require an additional cutoff function, $H_\rho(\vec k)$, 
the role of which is to provide a ultraviolet cutoff 
for the sum in Eq.~(\ref{eq:Frhopi_def}).\footnote{%
 $H_\rho(\vec k)$ also appears in $G_\rho$ through the pole factor, $\textbf S_\ttil$. In fact, here the cutoff function has no effect because $G_\rho$ is always accompanied by $F$ or $G$ and thus $H_{\rho}(\vec k)$ is always multiplied by $H(\vec k)$. From the definitions of the cutoff function it trivially follows that $H_\rho(\vec k) H(\vec k) = H(\vec k)$. We nonetheless find it convenient to keep the cutoff within $G_\rho$ as written. 
}
The range of possible $\K_2$ pole masses that we need to accommodate is $0 < M <  4 m$,
with the lower limit set by the value of $P_{2,k}^2$ for which $H(\vec k)$ vanishes,
and the upper limit set by the opening of the five-particle threshold with respect to $E^*$.
For any choice of $\vec k$ such that $P_{2,k}^2$ lies in this range, we need $H_\rho(\vec k)=1$,
so as not to distort the pole. However, as $P_{2,k}^2$ drops
below zero, the cutoff function should smoothly drop to zero.
The detailed choice is not important, but we display one example
for illustration,
\begin{equation}
H_\rho(\vec k) = J\left( \frac{P_{2,k}^{*\,2} + 4 m^2}{4m^2}\right)\,.
\end{equation}
This is chosen so that $H_\rho$ vanishes when $P_{2,k}^{*\,2}\le - 4 m^2$.

\bigskip
One of the important properties of $F_{33}$, stressed in Ref.~\cite{\HSQCa,\HSQCb},
is that it is fully determined if one knows $\K_2$ in the relevant kinematic range.
Thus a separate study using the two-particle quantization condition can, in principle,
determine the finite-volume function. We stress here that the same is true for all four components
of $\cF$. The only difference is that we must pull out the pole contribution from $\K_2$,
and use this in the determination of $F_{3\ttil}$, $F_{\ttil3}$ and $F_{\ttil\ttil}$.
The added complexity when there is a pole in ${\K}_2^{(J)}$ manifests only in the way that information about two-particle interactions appears in the finite-volume functions.

\subsection{Definition of $\K_\df$}

We close this section with some brief comments on $\K_\df$, whose components
are given by Eq.~(\ref{eq:Kmat_def}).
These four entries ($\K_{\df, \ttil \ttil}$,~$\K_{\df, \ttil 3}$,~$\K_{\df, 3 \ttil}$~and~$\K_{\df, 3 3}$) are each infinite-volume quantities, characterizing scattering
in the indicated channels. They are, themselves, matrices with indices matching those of the corresponding components of $\cF$, Eq.~(\ref{eq:Fmat_def}).
When multiplied by the appropriate spherical harmonics, and summed over angular
momentum indices, they become real, Lorentz-invariant functions of the on-shell
kinematic variables.

Another key property is that, in each of the four components, 
all kinematic singularities and possible $\K_2$ poles have been removed
from $\K_\df$. Thus, these can be viewed in position space as quasi-local vertices connecting
the various channels. 
This analogy is not perfect, however, since the components of $\K_\df$ are not
physical, as they depend on the details of the cutoff functions
described above.
It is also possible, just as for $\K_2$, that there are dynamical singularities in $\K_\df$.  

The derivation presented in the next section provides a complicated and implicit  definition of
the components of $\K_\df$. This turns out to be sufficient, however,
because what really matters is how these components are related to 
the physically measurable three-to-three scattering amplitude.
This relation can be derived based solely on how $\K_\df$ enters the final result. This is presented in Sec.~\ref{sec:KtoM}, following the approach of Ref.~\cite{\HSQCb}.

\section{Derivation \label{sec:derivation}}


We now derive the result described in the previous section. 
Begin by defining a finite-volume correlation function
\begin{equation}
\label{eq:CLintermsofOO}
C_L(E, \vec P) \equiv - \int_L d^4 x \, e^{- i E t + i \vec P \cdot \vec x} \, \langle \Omega \vert \text{T} \mathcal O(x) \mathcal O^\dagger(0) \vert \Omega \rangle \,,
\end{equation}
where $\mathcal O^\dagger(0)$ is any operator with the quantum numbers of the 
three-particle states that we are after.\footnote{%
The overall minus sign included in this definition should be understood as a factor of $i^2$. 
We choose to accompany each operator with a factor of $i$ for reasons explained below.
}
Inserting a complete set of states, one can show that this object has poles in $E$ 
at the finite-volume energies. 
Our aim is thus to derive an equation---the quantization condition---for 
the locations of these poles.

In the following subsections we show that the correlator can be written as
\begin{equation}
\label{eq:CL}
C_L(E, \vec P) = C_\infty(E, \vec P) + i A' i \mathcal F \frac{1}{1 - i \mathcal K_{\mathrm{df}} i \mathcal F }    i A \,,
\end{equation}
up to exponentially suppressed corrections.
Here $C_\infty(E, \vec P)$ and 
\begin{equation}
i A' \equiv \iApMat \,, \ \ \ \ \ i A \equiv \iAMat \,,
\label{eq:AApdef}
\end{equation}
are infinite-volume quantities, defined in the course of the following subsections. 
Note that the second term in Eq.~(\ref{eq:CL}) is a product of a row vector, 
a matrix, and a column vector, with all indices contracted. 
As the infinite-volume quantities contain no finite-volume poles, 
the poles in $C_L(E, \vec P)$ correspond to divergent eigenvalues in the matrix 
between $A'$ and $A$. 
This is equivalent to the inverse of the matrix having a vanishing determinant,
and thus to the quantization condition given in Eq.~(\ref{eq:det}) above.


\subsection{Compact notation for the derivation}

In order to make the following derivation more readable, 
in this section we introduce a compact notation 
for the various quantities introduce above. 
Our aim is to minimize explicit factors of $i$ and of $2\omega L^3$.
We thus define
\begin{gather}
{\bf G}   \equiv  \frac{1}{2 \omega L^3} i G \,, \ \ \ \ \ {\bf F}  \equiv \frac{1}{2 \omega L^3} i F \,, \ \ \ \ \  {\bf K}_2    \equiv 2 \omega_k L^3 i \K_2   \,, \\ 
\bGam \equiv i \Gamma_J\,, \ \ \ \ \  \bGr   \equiv i G_\rho \,, \ \ \ \ \    \bGrD \equiv i G_\rho^\dagger \,, \ \ \ \ \    \bFrp    \equiv i F_{\rho \pi} \,.
\label{eq:newnotation}
\end{gather}
One advantage of these definitions is that ${\bf G}$ is now given by
\begin{align}
{\bf G}_{p, \ell', m'; k, \ell, m} & \equiv  \frac{1}{2 \omega L^3} i G_{p, \ell', m' ; k, \ell, m} 
 \equiv \frac{1}{2 \omega_p L^3} \mathcal Y_{3, \ell' m'}(\vec k_{2,p}^*) 
 i \mathbf S^{0}_3(\vec p, \vec k )
   \mathcal Y^*_{3, \ell m}(\vec p_{2,k}^{\,*}) 
\frac{1}{2 \omega_k L^3} \,, 
\end{align}
and is therefore anti-hermitian (due to the factor of $i$ in the definition). 
This avoids the need to define the separate object $[2 \omega L^3]^{-1} G [2 \omega L^3]$
that would otherwise appear in the derivation. 

In this new notation, the quantization condition becomes
\begin{equation}
 \mathrm{det} \big [1 - \mathbf {K}_\df(E^*) \pmb { \mathcal F}(E, \vec P, L) \big ] = 0 \,,
 \end{equation}
where
\begin{equation}
i \K_\df \equiv   {\bf K}_\df \equiv \bKdfMat \,, \ \ \ \ \ \ \ \ i \mathcal F \equiv \pmb {\mathcal F} \equiv \bFMat \,,
\label{eq:bKdfbFMatdef}
\end{equation}
and 
\begin{align}
\mathbf F_{33}&  =  \frac13 \mathbf F  + \mathbf F   \mathbf T_L    \mathbf F   \,, \ \ \ \ \ \text{with} \ \ \ \ \ \mathbf T_L   \equiv \frac1{1- {\bf K}_2 (\bf F + \bf G)}  {\bf K}_2 \,, \label{eq:bFandbTdef} \\
\bF_{\ttil\ttil} &= \bFrp  + \bGr \bGam  (\bF + \bG) \frac1{1- \bK (\bF + \bG)} \bGam \bGrD \,,
\label{eq:bF22def}
\\
\bF_{\ttil 3} &= \bGr \bGam  \frac1{1- (\bF+\bG) \bK} \bF\,,
\label{eq:bF23def}
\\
\bF_{3\ttil} &=  \bF \frac1{1- \bK (\bF+\bG)} \bGam \bGrD \,.
\label{eq:bF32def}
\end{align}


\subsection{Definition and decomposition of $C_L^{[B_2]}$}
\label{sec:CLB2initaldecom}

We begin by following the same steps as taken by Ref.~\cite{\HSQCa} 
in the derivation of the quantization condition in the absence of $\K_2$ poles.
This will allow us to reuse a fair amount of work from that reference.
The derivation begins with an all-orders skeleton expansion in which $C_L$ is 
defined diagrammatically in terms of two- and three-particle Bethe-Salpeter kernels, 
denoted $iB_2$ and $iB_3$ respectively, as well as fully dressed propagators.
Examples are shown in Fig.~4 of Ref.~\cite{\HSQCa}.
The skeleton expansion is designed to make all power-law finite-volume effects explicit. 
Such effects arise from on-shell intermediate states in Feynman diagrams and, 
since we constrain the overall c.m.~energy to the range $m < E^* < 5 m$, 
this amounts to keeping track of three-particle states. 
The restriction to a finite, periodic spatial volume is effected by summing the 
spatial components of all loop momenta over $\vec p = (2 \pi /L) \vec n$ where 
$\vec n \in \mathbb Z^3$ runs over all 3-vectors of integers.

As in Ref.~\cite{\HSQCa}, the challenging part of the derivation is that involving the
kernels $B_2$. Thus it is useful to begin by analyzing a reduced correlator, denoted $C_L^{[B_2]}$,
defined by the same skeleton expansion except that
all three-particle kernels are set to zero ($B_3 \to 0$).
Adding back in the effects of $B_3$ is relatively straightforward and will be done at a later
stage.
To decompose $C_L^{[B_2]}$ we can  piggyback on Ref.~\cite{\HSQCa} 
by directly taking over Eq.~(174) of that work,
since this equation was derived without assuming smoothness of $\cK_2$ 
as a function of the two-particle center-of-mass energy.
Written in our present notation, the result is
\begin{equation}
C^{[B_2]}_{L}  = 
 C_{L,0F} 
- \frac23 \bSig  \bF  \bSigD
+ 
\bA'^{(u)}_{L,3} 
\bF_{33}^{(0)}
\sum_{n=0}^\infty 
\left( \bKLth^{(u,u)}  
       \bF_{33}^{(0)} \right)^n  
\bA^{(u)}_{L,3} \,,
\label{eq:Cno3MLres}
\end{equation}
where the quantity called $[\mathcal A]$ in Ref.~\cite{\HSQCa} is here denoted $\bF^{(0)}_{33}$,
and is given by
\begin{align}
\bF^{(0)}_{33} & \equiv \bF \frac{1}{1 - \bK \bF} 
\,. 
\label{eq:F33zerodef}
\end{align}
The other quantities in Eq.~(\ref{eq:Cno3MLres}) will be explained shortly.

What has been achieved in Eq.~(\ref{eq:Cno3MLres}) is to make explicit a subset of
the finite-volume effects due to three-particle intermediate states.
We recall from Eqs.~(\ref{eq:Fdef3}) and (\ref{eq:newnotation})
that $\bF$ is defined by a sum-integral difference of a quantity with a three-particle pole. This object therefore has power-law finite-volume dependence, and also sets the
quantities multiplying it on either side to be on shell.
Thus $\bF^{(0)}_{33}$ collects such dependence from a sequence of three-particle ``cuts''
separated by two-to-two interactions occuring between the same pair.
The subscript ``$33$'' is included here to distinguish this object from similar
quantities involving $\K_2$ poles that arise below.

\bigskip

We now turn to the definitions of the remaining quantities in Eq.~(\ref{eq:Cno3MLres}).
With the exception of $\bSig$ and $\bSigD$, these are finite-volume quantities,
involving some loops in which momenta are summed rather than integrated.\footnote{%
Here and in the following a loop momentum being ``summed" is shorthand for a sum 
over spatial components and an integral over the temporal component.}
This is indicated by the subscripts $L$.

We begin with $C_{L,0F}$. This is the contribution to $C_L^{[B_2]}$ containing no factors of $\bF$.
It can be expanded according to the number of factors of $\cK_2$ that it contains
\begin{equation}
C_{L,0F} = \sum_{n=0}^\infty C_{L,0F}^{(n)}\,.
\label{eq:allordersCL0F}
\end{equation}
The objects on the right-hand side are identical to those with the same names appearing in
Ref.~\cite{\HSQCa}. They are defined in
Eqs.~(114), (154), (169), (173) and (176) of that work,
and shown diagrammatically there in Figs. 11(c), 15(b) and 17(c).
We repeat the diagrammatic representation in Fig.~\ref{fig:CnL0F}(a) below.

The quantity $\bKLth^{(u,u)}$ involves three-to-three transitions that are built
from $\K_2$ interactions alternating between different pairs.
It is closely related to $\K_{3,L}^{(n,u,u)}$, defined in Ref.~\cite{\HSQCa}:
\begin{equation}
\bKLth^{(u,u)} \equiv \sum_{n=2}^\infty \bKLth^{(n,u,u)}\,,\quad
\bKLth^{(n,u,u)} \equiv i \K_{3,L}^{(n,u,u)}
\,.
\label{eq:BKLdef}
\end{equation}
Here $\bKLth^{(n,u,u)}$ is the contribution containing $n$ factors of $\cK_2$, where $n\ge 2$.
As above, we have amended the subscripts to facilitate the addition of the $\rho\pi$ channel,
and also absorbed a factor of $i$.
The definition of $\mathcal K_{3,L}^{(n,u,u)}$ is given in Ref.~\cite{\HSQCa}
by Eqs.~(155) and (171)  and Figs.~11(a), 15(c) and 17(c).
We repeat the diagrammatic representation of $\bKLth^{(n,u,u)}$ in Fig.~\ref{fig:KL}(a) below.

The remaining quantities in Eq.~(\ref{eq:Cno3MLres}) are endcaps.
$\bA'^{(u)}_{L,3}$ and $\bA^{(u)}_{L,3}$ can be expanded as above according to the number of factors of $\cK_2$, with $\bSig$ and $\bSigD$ being the zeroth order terms in these expansions:
\begin{gather}
  \bA'^{(u)}_{L,3}   = \sum_{n=0}^\infty \bA'^{(n,u)}_{L,3} \,,  \ \ \ \ \ 
\bA^{(u)}_{L,3}   =   \sum_{n=0}^\infty \bA^{(n,u)}_{L,3} \,,\ \ \ \ \
\bSig    \equiv   \bA'^{(0,u)}_{L,3} \,, \ \ \ \ \ 
  \bSigD   \equiv  \bA^{(0,u)}_{L,3} \,.
\label{eq:endcaps}
\end{gather}
The relation to the corresponding quantities from Ref.~\cite{\HSQCa} simply involves
a change of subscripts to allow for future $\cK_2$ pole contributions
\begin{equation}
\bA'^{(n,u)}_{L,3} \equiv i A'^{(n,u)}_L\,,\quad
\bA^{(n,u)}_{L,3} \equiv i A^{(n,u)}_L\,,
\label{eq:endcaps2}
\end{equation}
where the quantities on the right-hand side are those appearing in Ref.~\cite{\HSQCa}.

The expressions for these quantities
are given in Eqs.~(60), (84), (85), (113), (153) and (170) of Ref.~\cite{\HSQCa}, 
and illustrated in Figs.~9(c), 11(b), and 17(a) of that work.
We repeat the diagrammatic representation of the left endcap
$\bA'^{(n,u)}_{L,3}$ in Fig.~\ref{fig:CnL0F}(c) below.
Note that it can be obtained from that for $C^{(n)}_{L,0F}$ by removing
the $\bSigD$ at the right end.
The representation of the corresponding right endcap, $\bA^{(u)}_{L,3}$, 
is given simply by a horizontal reflection of that for $\bA'^{(u)}_{L,3}$,
or equivalently by removing $\bSig$ from the left end of $C^{(n)}_{L,0F}$.

In the following subsections, our aim is to make explicit the full volume dependence of 
$C_{L,0F}$, $\bA'^{(u)}_{L,3}$, $\bA^{(u)}_{L,3}$ and $\bKLth^{(u,u)}$.


\subsection{Decomposition of $C_{L,0F}$}

We first consider the quantity $C_{L,0F}$, and show in this subsection that it can be
decomposed as
\begin{align}
C_{L,0F}
&=
{C}^{[B_2]}_{\infty }
+
\left( 2 \bA'^{(s)}_3  {\bF} +
\bA'_{\ttil}  \bGr \bGam  \bG \right)
(\bA_{L,3}^{(u)} - \bSigD)
+
\bA'_{\ttil} \bFrp  \slashed \bA_{L, \ttil} \,.
\label{eq:CL0Ffinal}
\end{align}
This is a partial decomposition, involving both finite- and infinite-volume quantities 
(the latter having subscripts including $L$)
separated by ``cuts".
In deriving this result, we must, for the first time, account for the poles in $\cK_2$,
as shown by the presence of factors of $\bGr$ and $\bFrp$, which set the $\rho\pi$ states
on either side on shell.
If these two quantities are set to zero we reproduce the result in Eq.~(189) of
Ref.~\cite{\HSQCa}.

\begin{figure}
\begin{center}
\includegraphics[width=.8\textwidth]{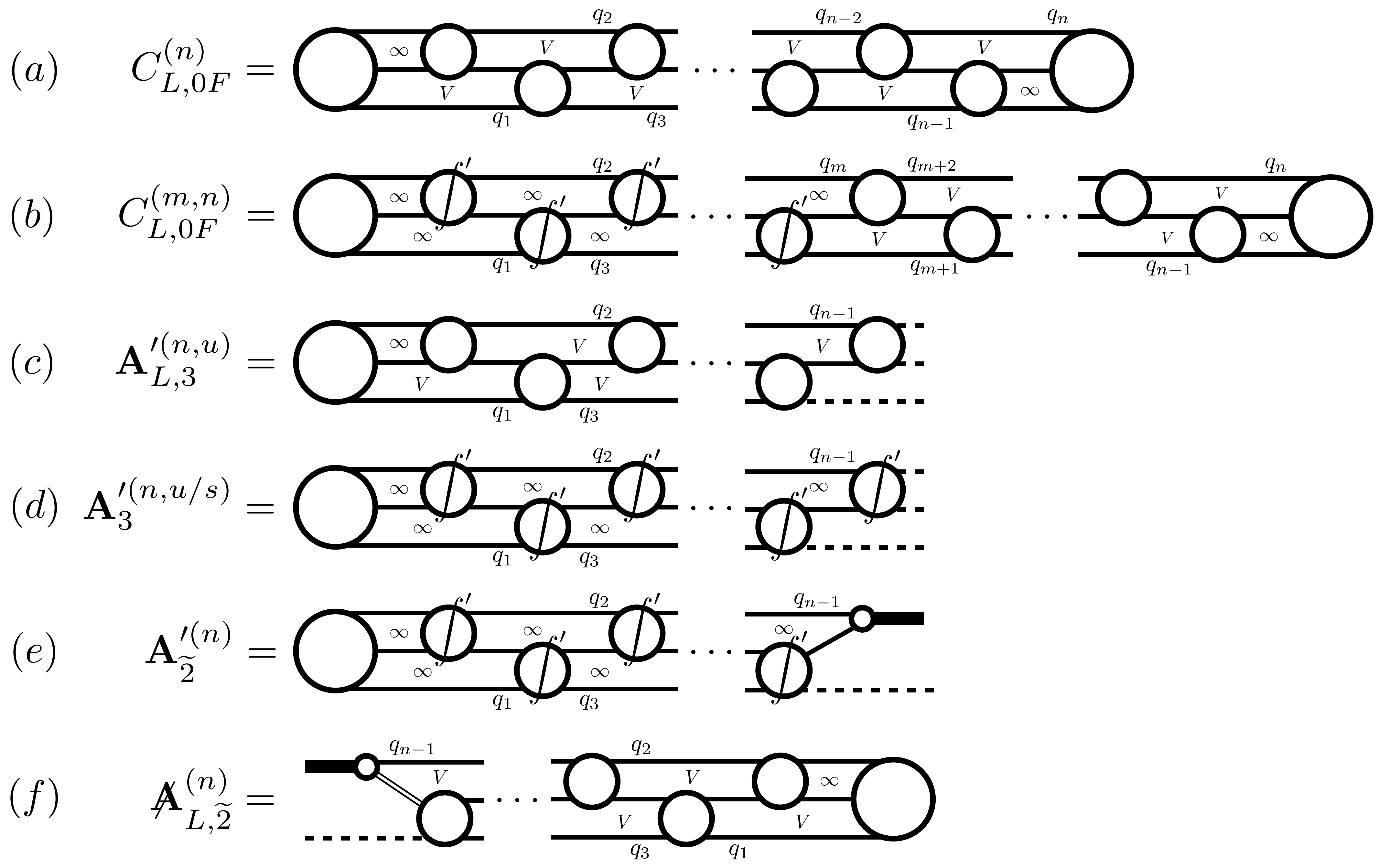}
\caption{
Diagrammatic definitions of quantities entering
the derivation of Eq.~(\ref{eq:CL0Ffinal}). 
Open circles on the left and right ends represent $\bSig$ and $\bSigD$, respectively.
Open circles in the middle represent the full off-shell $i\K_2$, 
while the circles with an integral sign indicate that only the smooth component,
$i\widetilde \K'_2$, is included. 
Loops that are summed contain a ``$V$'',
while those that are integrated contain an ``$\infty$''.
The superscript $n$ indicates the number of factors of $\cK_2$ or its smooth counterpart.
Thin lines are fully-dressed propagators, with
unit residue at the one-particle pole. Thick lines [present in (e) and (f)] represent the resonance,
corresponding to the subscript $\ttil$.
Double thin lines [present only in (f)] indicate that only the smooth
part of the exchanged particle propagator is kept.
In (c), the superscript ``$u$" indicates that the index $k$ corresponds to the momentum carried by 
the spectator propagator at the right-hand end.
In (d), the superscript ``$u/s$" indicates that the diagram serves to define
the quantity with both superscripts. If the superscript is $u$, then the momentum $k$ is assigned to the spectator propagator, while if it is $s$, then $k$ is assigned to the upper
propagator.
Further details are given in the text.  
}
\label{fig:CnL0F}
\end{center}
\end{figure}

As can be seen from Fig.~\ref{fig:CnL0F}(a),
$C_{L,0F}$ is defined by the sum over all pairwise scatterings 
in which the interaction switches to a different pair with each new insertion of $i\K_2$. 
By construction, this quantity has $n$ summed loop momenta
plus two additional loops with integrated momenta.
It is convenient to extend this notation by defining
$C_{L,0F}^{(m,n)}$ to be the same quantity as $\CLF 0 n$ but with
the leftmost $m$ momentum sums converted to integrals
(with the single-particle poles integrated using the $\PV$ prescription of Ref.~\cite{\HSQCa}),
and with the integrated $i\K_2$ factors replaced by their smooth parts, 
$i\widetilde \K'_2$ [{see Fig.~\ref{fig:CnL0F}(b)}].\footnote{%
As explained in Appendix~\ref{app:decomCLmn},
several terms combine to give the smooth part,
with $i\widetilde \K_2$ from Eq.~(\ref{eq:modK2poleDecom}) being
just one contribution.
This is illustrated in Fig.~\ref{fig:CL0Frecursion_v2}(f). It is for this reason that we require the prime to denote, $i \widetilde{\cK}_2'$, the quantity entering integrated loops of $C_{L,0F}^{(m,n)}$.
}
This requires $0\le m \le n$, with $m=0$ leading to
$C_{L,0F}^{(0,n)}\equiv \CLF 0 n$, and $m=n$ to
$C_{L,0F}^{(n,n)}\equiv C_\infty^{(n)}$, a fully-integrated, infinite-volume quantity. 

To decompose $C_{L,0F}^{(m,n)}$ we consider the leftmost sum, i.e.~that directly adjacent to the $m$ integrated loops. 
Finite-volume effects arise in this sum due to both
 the pole in $\cK_2$ and 
 the intermediate on-shell three-particle state.
In Appendix \ref{app:decomCLmn} we explain the procedure for converting a given summed loop (with the full $\cK_2$) to an integrated loop (with the smooth part only). In other words we derive a system for converting $C_{L,0F}^{(m,n)}$ to $C_{L,0F}^{(m+1,n)}$, plus finite-volume correction terms. This leads to the following recursion relation 
\begin{equation}
C_{L,0F}^{(m,n)} = \begin{cases}
C_{L,0F}^{(m + 1,n)} + 
\left(2 \bA'^{(m + 1,s)}_3 \bF 
+ \bA'^{(m + 1)}_2 \bGr \bGam \bG \right) 
\bA_{L,3}^{(n-m-1,u)}
+ \bA'^{(m + 1)}_2 \bFrp {   \slashed{\bA}_{L,\ttil}^{(n - m)}}
 & 0 \le m < n - 1 \,,
 \\
 C_{L,0F}^{(n,n)}
+ \bA'^{(n)}_2 \bFrp {  \slashed{\bA}_{L,\ttil}^{(1)}}
 & 0 \le m =n-1  \,,
 \\
C_{\infty}^{(n)} & 0 \le m=n\,.
 \end{cases}
\label{eq:CL0Fmaster}
\end{equation}
The pole in $\cK_2$ leads to the terms involving $\bGr$ and $\bFrp$.
These equations contain three new quantities,
$\bA'^{(n,s)}_3$, $\bA^{(n)}_\ttil$ and $\slashed{\bA}_{L,\ttil}^{(n)}$,
in addition to the right endcap $\bA^{(n,u)}_{L,3}$ introduced in the previous subsection.

The infinite-volume left endcap $\bA'^{(n,s)}_3$ is defined 
diagrammatically in Fig.~\ref{fig:CnL0F}(d).
It contains $n$ factors of $i \widetilde \K'_2$, with all loop momenta integrated.
The superscript $s$ indicates the manner in which the on-shell external three-particle state
is projected into spherical harmonics, as explained in Ref.~\cite{\HSQCa}.

The second new quantity is the infinite-volume left endcap $\bA'^{(n+1)}_\ttil$, 
defined diagrammatically in Fig.~\ref{fig:CnL0F}(e).
Here the ``on-shell" external state consists of the K-matrix pole plus the spectator, 
which we refer to as the $\rho\pi$ state.
$\bA'^{(n+1)}_\ttil$ contains $n$ factors of $\widetilde {\K}'_2$, $n$ loop integrals, and one factor of $\bGam$ in the loop adjacent to the external state.
We will later need the analogous right endcap, denoted $\bA^{(n+1)}_\ttil$.

The final new quantity, $\slashed{\bA}_{L,\ttil}^{(n+1)}$, is defined diagrammatically in Fig.~\ref{fig:CnL0F}(f). 
It is closely related to $\bA^{(n+1)}_{L,2}$, the reflection of $\bA'^{(n+1)}_{L,2}$,
which consists of $n$ factors of $i\K_2$, $n$ summed loops, 
and one factor of $\bGam$ adjacent to the external $\rho\pi$ state.
The slashed version differs in that the leftmost three-particle intermediate state is replaced
by the smooth difference that remains when the $\bG$ singularity is subtracted. This subtraction in indicated in the figure with a double line.
The lowest value of $n$, $n=1$, is a special case, for which there is
no summed loop and $\slashed{\bA}_{L,\ttil}^{(1)} = \bA_{\ttil}^{(1)}$. 
For further discussion, see Appendix~\ref{app:decomCLmn}.

Iterating Eq.~(\ref{eq:CL0Fmaster}) leads to an expression for 
$C_{L,0F}^{(0,n)} = C_{L,0F}^{(n)}$ in terms of $C_\infty^{(n)}$. 
Summing over $n$ then gives the desired result, Eq.~(\ref{eq:CL0Ffinal}),
where we define
\begin{equation}
{C}^{[B_2]}_{\infty } = \sum_{n=0}^\infty C_\infty^{(n)}\,,
\quad
\bA'^{(s)}_3 = \sum_{n=1}^\infty \bA'^{(n,s)}_{3}\,,
\quad
\bA'_{\ttil} = \sum_{n=1}^\infty \bA'^{(n)}_{\ttil}\,,
\quad
\slashed{\bA}_{L, \ttil} = \sum_{n=1}^\infty \slashed{\bA}^{(n)}_{L,\ttil}\,.
\label{eq:allorders2}
\end{equation}
Note that the last three sums begin at $n=1$, 
in contrast to the sum for $\bA_{L,3}^{(u)}$, Eq.~(\ref{eq:endcaps}), 
which begins at $n=0$.
It is because of this that $\bSigD$ must be
subtracted from $\bA_{L,3}^{(u)}$ in the final result, Eq.~(\ref{eq:CL0Ffinal}).


\subsection{Decompositions of $\bA'^{(u)}_{L,3}$, $\bA^{(u)}_{L,3}$ 
and $\slashed{\bA}_{L,\ttil} $}

\begin{figure}
\begin{center}
\includegraphics[width=.7\textwidth]{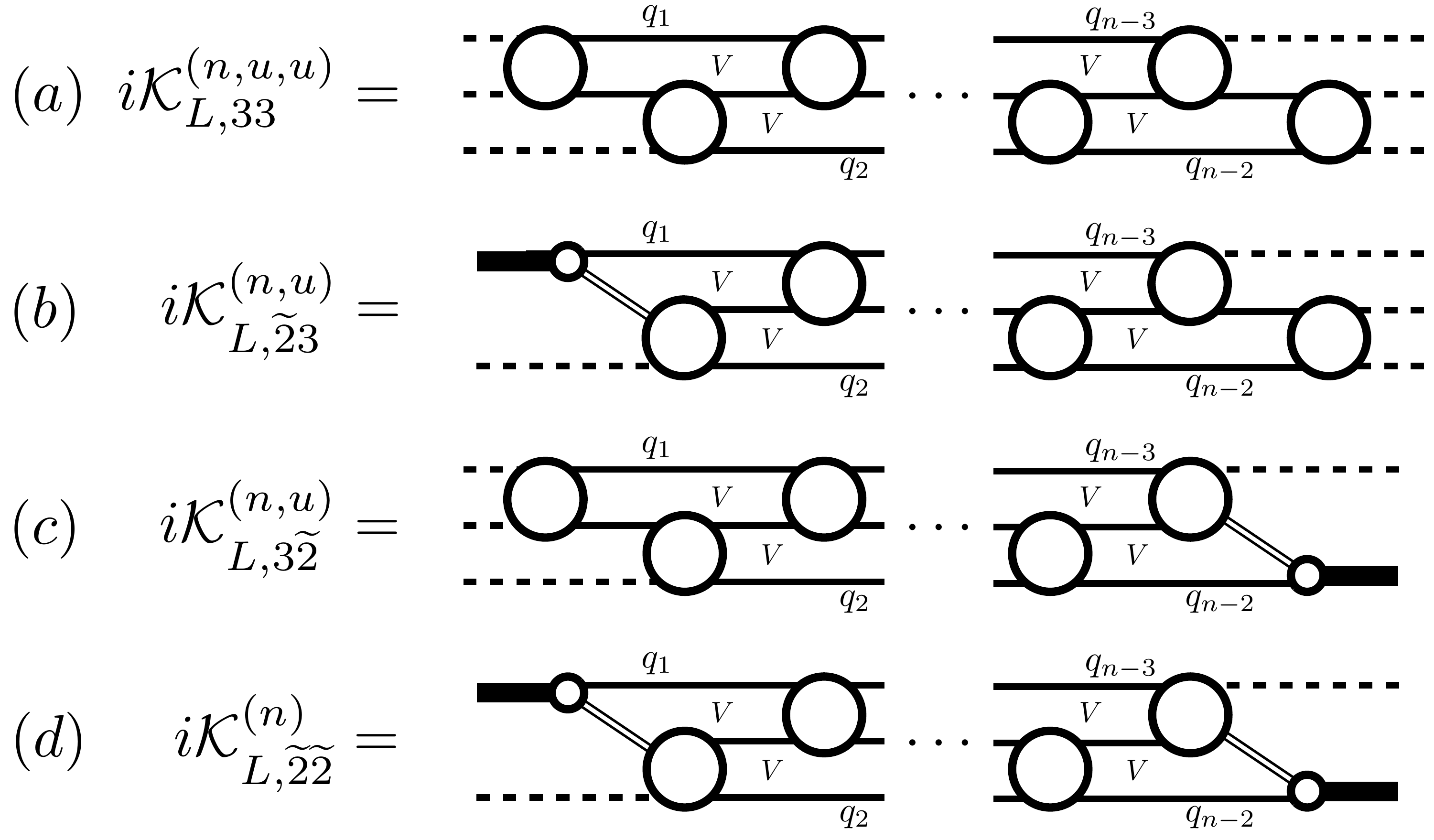}
\caption{
Diagrammatic definitions of the elements of the finite-volume K matrices
involving three-particle or $\rho\pi$ external states.
The notation is as in Fig.~\ref{fig:CnL0F}.
}
\label{fig:KL}
\end{center}
\end{figure}

In this subsection, we continue the decomposition of the quantities entering
$C_L^{[B_2]}$, Eq.~(\ref{eq:Cno3MLres}), by considering the finite-volume
endcaps $\bA'^{(u)}_{L,3}$ and $\bA^{(u)}_{L,3}$.
In addition, we decompose the related quantity $\slashed{\bA}_{L,\ttil}$ that
appears in Eq.~(\ref{eq:CL0Ffinal}).
The results we obtain are
\begin{align}
\bA'^{(u)}_{L,3}
& =
\bA'^{(u)}_3
+
\left( 2 \bA'^{(s)}_3  \bF
+  \bA'_{\ttil}   \bGr \bGam \bG \right) 
\left(  \bKLth^{(u,u)}+ \bK \right)
+
\bA'_{\ttil} \left(\bFrp \bKLtwth^{(u)} + \bGr \bGam \right)
\,, 
\label{eq:ApL3final} \\
\bA^{(u)}_{L,3}
& =
\bA^{(u)}_3
+
\left( \bKLth^{(u,u)} + \bK \right)
\left( \bF 2 \bA^{(s)}_3 +  \bG \bGam \bGrD \bA_{\ttil}\right)
+ 
\left( \bKLthtw^{(u)} \bFrp + \bGam \bGrD \right) \bA_{\ttil}
\,,
\label{eq:AL3final} \\
  \slashed{\bA}_{L,\ttil} & =  \slashed{\bA}_{\ttil}
+
 \bKLtwth^{(u)}
\left( \bF \, 2 \bA^{(s)}_3 +  \bG \bGam \bGrD \bA_{\ttil}\right)
+ 
\bKLtw \bFrp \bA_{\ttil}
\,.
\label{eq:slashedALfinal}
\end{align}
Here we have introduced three new finite-volume K matrices,
\begin{equation}
\bKLtw = \sum_{n=2}^\infty \bKLtw^{(n)}\,, \qquad \bKLtwth^{(u)} \equiv \sum_{n=2}^\infty \bKLtwth^{(n,u)} \,, \qquad \bKLthtw^{(u)} \equiv \sum_{n=2}^\infty \bKLthtw^{(n,u)}\,,
\label{eq:KLnsums}
\end{equation}
all closely related to $\bKLth^{(u,u)}$.
Specifically, $\bKLtw^{(n)}$ is obtained from $\bKLth^{(n,u,u)}$ by replacing
the $\K_2$s on both ends with factors of $\bGam$, and connecting these $\bGam$s
to the adjacent $\K_2$s with the smooth ($\bG$-subtracted) part of the neighboring exchange propagator. 
This is shown in Fig.~\ref{fig:KL}(d).
$\bKLtwth^{(n,u)} $ and $\bKLthtw^{(n,u)}$ are obtained by performing these steps 
on only one side of $\bKLth^{(n,u,u)}$ while leaving the other side unchanged,
as shown in Figs.~\ref{fig:KL}(b) and (c), respectively.
Also new in Eqs.~(\ref{eq:ApL3final})-(\ref{eq:slashedALfinal}) are the
infinite-volume endcaps $\bA'^{(u)}_3$, $\bA^{(u)}_3$, $ \slashed{\bA}_{\ttil}$. and $\bA_\ttil$.
These are defined below.

Intuitively, Eqs.~(\ref{eq:ApL3final})-(\ref{eq:slashedALfinal})
summarize the various ways that
finite-volume effects enter the endcap functions. For example,
Eq.~(\ref{eq:ApL3final}) splits $\bA'^{(u)}_{L,3}$ into its
infinite-volume counterpart, $\bA'^{(u)}_3$, plus six types of
finite-volume corrections (counted by multiplying out the products of
binomials in the middle term). Each term is characterized by a
different type of cut-factor, e.g.~$\bF$ encodes the finite-volume
effects associated with two of the three particles propagating between
adjacent pairwise re-scatterings and $\bG$ describes the volume
effects of an exchanged particle. Similarly, $\bGr$ and $\bFrp$
correspond to different types of volume effects associated with the
K-matrix poles.  $\bA^{(u)}_{L,3}$, in Eq.~(\ref{eq:AL3final}), is
simply a mirror image of $\bA'^{(u)}_{L,3}$, while
$\slashed{\bA}_{L,\ttil} $, in Eq.~(\ref{eq:slashedALfinal}), is given
by replacing the right-most state with a $\ttil$ and dropping terms
that do not arise with this type of external state.

\bigskip

To derive these results we use a similar method to that of the previous subsection.
We first consider $\bA'^{(u)}_{L,3}$, and introduce $\bA'^{(m,n,u)}_{L,3}$,
which contains $n$ two-particle kernels, $m$ integrated loop momenta and
$n-m$ summed loop momenta. As for $C^{(m,n)}_{L,0F}$, the integrated K matrices
are $i\widetilde \K'_2$s, while the summed ones are the full $i\K_2$s. 
We need this quantity for $n \ge m \ge 1$ together with the special case $n=m=0$, giving $\bA'^{(0,0,u)}_{L,3}=\bA'^{(0,u)}_{L,3}=\bSig$.
A second special case is
$\bA'^{(1,n,u)}_{L,3}=\bA'^{(n,u)}_{L,3}$ 
(since there is always one integrated loop for $n\ge 1$). 
Finally, we note that the fully integrated version is an infinite-volume quantity,
$\bA'^{(n,n,u)}_{L,3} = \bA'^{(n,u)}_3$.
This quantity is shown diagrammatically in Fig.~\ref{fig:CnL0F}(d), 
and differs from the quantity
$\bA'^{(n,s)}$ encountered above only by the choice of spectator propagator.

The steps detailed in Appendix~\ref{app:decomCLmn} apply also here, except that the
right endcaps $\bSigD$ are replaced with an on-shell three-particle state.
We find that the resulting recursion equations are 
\begin{equation}
\bA'^{(m+1,n,u)}_{L,3} = \begin{cases}
\bA'^{(m+2,n,u)}_{L,3}  +   
\left(2  \bA'^{(m+1,s)}_3  \bF 
\!+\! \bA'^{(m+1)}_\ttil \bGr \bGam \bG \right) 
\bKLth^{(n-m-1,u,u)}
+  \bA'^{(m+1)}_\ttil \bFrp  \bKLtwth^{(n-m,u)}
  & 0 \le m < n-2 \,,
\\[5pt]
\bA'^{(n,n,u)}_{L,3} + 
\left(2 \bA'^{(n-1,s)}_3 \bF 
\!+\! \bA'^{(n-1)}_\ttil \bGr \bGam \bG \right) 
 \bK
+ \bA'^{(n-1)}_\ttil \bFrp \bKLtwth^{(2,u)}
  & 0 \le m = n-2 \,,
 \\[5pt]
 \bA'^{(n,u)}_{3}
+ \bA'^{(n)}_\ttil \bGr \bGam
 &0\le m=n-1
\,.
 \end{cases} 
\label{eq:Apmaster}
\end{equation}
We stress that all quantities to the left of the cuts are, by construction, 
identical to those appearing in Eq.~(\ref{eq:CL0Fmaster}). 
The quantities appearing to the right, however, have changed:
$\bA_{L,3}^{(n,u)}$ has been replaced by $\bKLth^{(n,u,u)}$
and $\slashed{\bA}^{(n)}_{L,\ttil}$ has been replaced by $\bKLtwth^{(n,u)}$.

Solving the recursion relation for $\bA'^{(n,u)}_{L,3}$ and summing over $n$ using
the definition
\begin{equation}
 \bA'^{(u)}_{3} \equiv \sum_{n=0}^\infty \bA'^{(n,u)}_3 \,,\qquad
\label{eq:Apallorders}
\end{equation}
yields Eq.~(\ref{eq:ApL3final}).
We observe that the combination $\bKLth^{(u,u)}+\bK$ appears.
This arises because the sum over $n$ for $\bKLth^{(u,u)}$ begins at $n=2$,
since at least two factors of $i\K_2$ are
needed for a connected scattering of three particles.
The $n=1$ term then becomes simply $\bK$.
Similarly, the $n=1$ term is absent in the definition of $\bFrp \bKLtwth^{(u)}$
and this leads to the additional contribution containing $\bGr\bGam$.
Note that, if $\bGr$ and $\bFrp$ are set to zero, then we recover the
result given in Eq.~(186) of Ref.~\cite{\HSQCa}.

The horizontal reflection
of Eq.~(\ref{eq:ApL3final}) gives the decomposition of the other endcap, Eq.~(\ref{eq:AL3final}).

Finally, we need to decompose $ \slashed \bA_{L, \ttil}$.
We recall that this is the finite-volume right endcap, 
defined diagrammatically in Fig.~\ref{fig:CnL0F}(f). 
It thus differs from $\bA_{L,3}$ only in its final state, in which a factor of $\bGam$
combines with the smooth part of the exchange propagator.
This means that we can adapt the result from that for $\bA_{L,3}$ by replacing the
three-particle external state with a two-particle one, and dropping the
contribution from the $\bK$ factor on the end (since this is replaced by smooth quantities).
The result is given in Eq.~(\ref{eq:slashedALfinal}).


\subsection{Decomposition of $\bKLtw$, $\bKLtwth^{(u)}$, $\bKLthtw^{(u)}$ and $\bKLth^{(u,u)}$ \label{sec:KLdecom}}

In this subsection we complete the decomposition of the quantities entering
$C_L^{[B_2]}$ into infinite-volume objects and finite-volume cuts,
with some technical details relegated to Appendix~\ref{app:KLdecom}.

What remains is to decompose the four finite-volume 
K matrices whose components are shown in Fig.~\ref{fig:KL}.
They are conveniently packaged into a two-by-two matrix
\begin{equation}
  \bKL^{(u)}   \equiv  \begin{pmatrix}  
   \bKLtw & \bKLtwth^{(u)} \\
   \bKLthtw^{(u)} & \bKLth^{(u,u)}
   \end{pmatrix}
    \,.
\label{eq:KLmatrix}
\end{equation}
The result we will derive in the following can be written compactly as
\begin{equation}
 \bKL^{(u)}  
= 
\begin{pmatrix}
0 & 0 \\ 0 &  \bKLth^{(0)}
\end{pmatrix}
+
\bEL   \bVD   \bKdf^{(u)}  \frac1{1 - \bX  \bKdf^{(u)} } \bV \bER \,,
\label{eq:KLmatrix3}
\end{equation}
where
\begin{gather} 
\label{eq:KL330}
\bKLth^{(0)}  \equiv \bK \bGK \bK  =  \frac1{1  - \bK \bG } \bK \bG \bK
\,, \\[5pt]
\bX  \equiv 
\begin{pmatrix}
\bFrp + \bGr  \bGam \bGK \bGam \bGrD \       &      \ \bGr \bGam \bGK  \\[5pt]
\bGK \bGam \bGrD       &       \bGK
\end{pmatrix}
\label{eq:Xdef} \,, \\[5pt]
\bV  \equiv
\begin{pmatrix}
1 & \bGr \bGam \\ 0 & 1
\end{pmatrix} 
\,, \qquad
 \bVD  \equiv 
\begin{pmatrix}
1 & 0 \\   \bGam  \bGrD & 1 
\end{pmatrix}
\,, \\[5pt]
 \bEL  \equiv
\begin{pmatrix}
1 & 0 \\ 0 & 1+ \bT \bG
\end{pmatrix}
\,, \qquad
\bER  \equiv
\begin{pmatrix}
1 & 0 \\ 0 & 1 + \bG \bT
\end{pmatrix}
\,, \\[5pt]
\bGK  \equiv   \frac1{1- \bG \bK} \bG 
\,, \qquad
\bT  \equiv \bK \frac1{1- \bG \bK} 
\,.
\label{eq:bT}
\end{gather}
The final new quantity is $\bKdf^{(u)}$. This is a two-by-two matrix
of infinite-volume, divergence-free K matrices, defined below in Eq.~(\ref{eq:bKdfdef}).
The motivation for all these new quantities is described in more detail during the 
following derivation.

\bigskip

As above, our task is replace all summed loop momenta with integrals,
separating out the divergences due to both the
three-particle on-shell intermediate states and the poles in $\K_2$.
It turns out that, at first,
we do not need to decompose those factors of $\K_2$ that lie
directly adjacent to $\bKL^{(u)}$ components with a $3$ index.
This applies, for example, to the $\cK_2$s at both the left and right ends
of $\bKLth^{(n,u,u)}$ in Fig.~\ref{fig:KL}. 
These can remain as the full K matrices, despite containing poles,
since they do not appear in sums. 
Leaving these factors of $\K_2$ unseparated leads to shorter expressions at intermediate stages, 
at the cost of requiring an additional step to remove the final divergences.
We denote by $\bKi_{33}$, $\bKi_{3\ttil}$ and $\bKi_{\ttil 3}$
these intermediate infinite-volume quantities that still contain external divergences from the external $\K_2$.

The method we use here is simpler than the approach adopted in Ref.~\cite{\HSQCa},
where the result for any number of $\K_2$ factors was deduced by working out the
cases with $2$, $3$, and $4$ factors of $\K_2$ and then determining the pattern.
Here we use matrix equations that take care of all orders at once.
We find it convenient to keep track of finite-volume contributions 
in two stages: first those from $\bG$ cuts and second those from K-matrix poles,
the latter leading to $\bGr$ and $\bFrp$ cuts.

\bigskip

\begin{figure}
\begin{center}
\includegraphics[width=1\textwidth]{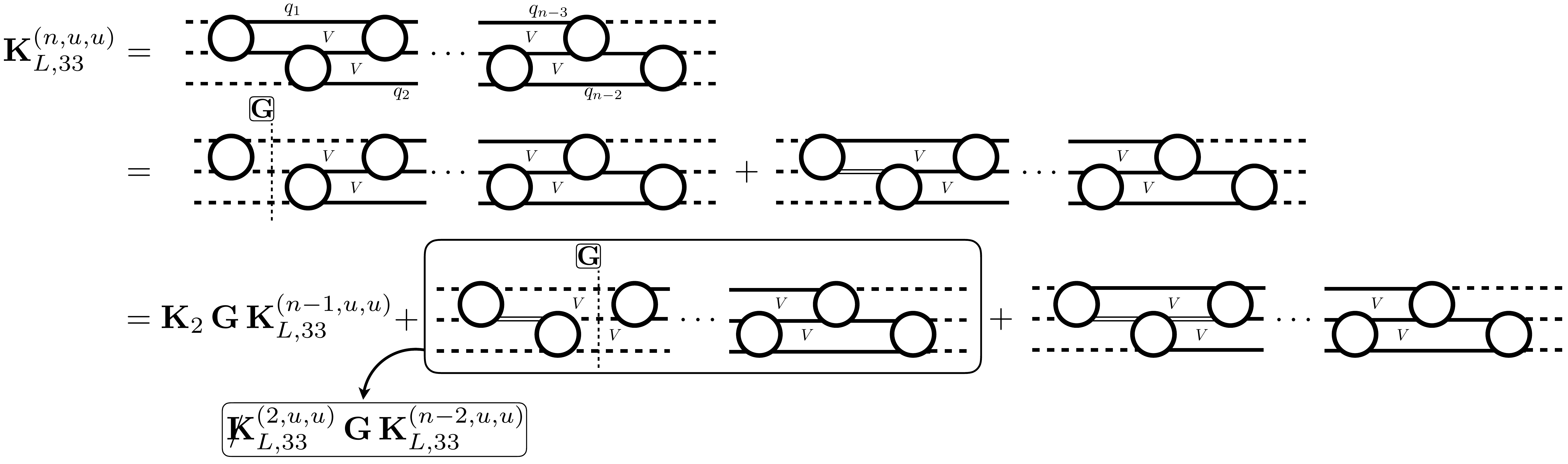}
\caption{
Diagrammatic description of the procedure described in the text leading to the Eq.~(\ref{eq:KL33uu}). 
The notation is the same as in Fig.~\ref{fig:CnL0F},
with the addition of a dashed line to indicate a propagator proportional to $\bG$ (if internal) or an amputated propagator (if external).
}
\label{fig:KL33_decomp}
\end{center}
\end{figure}

We begin by considering $\bKLth^{(u,u)}$. 
Moving from left to right, we consider each three-particle intermediate state in turn.
At each stage this consists of two fully-dressed propagators, e.g.~$\Delta(a)\Delta(b)$ with
$a$ the spectator momentum. We replace this with the product 
$2\pi \delta(a^0-\omega_a) (2\omega) \bG L^6$ together with the difference,
which is a smooth function of $\vec a$.
By construction, the insertion of $\bG$ sets the nonspectator pairs on either side on shell.
The details of how this works are unchanged from Ref.~\cite{\HSQCa} 
and we do not repeat them here.
After the substitution is made, in the term containing the factor of $\bG$ this first 
stage of decomposition is complete and a factor of $\bKLth^{(u,u)}$ appears to the right 
of $\bG$. 
In the term containing the smooth residue we proceed to the next intermediate state to 
the right and repeat the decomposition. See Fig.~\ref{fig:KL33_decomp} for a diagrammatic sketch of the first steps in this procedure.

This procedure leads to the equation
\begin{equation}
  \bKLth^{(u,u)} =  \bK  \bG \bK +
    \sbKLth  \left(1 + \bG   \bK \right) 
 +\left( \sbKLth    + \bK   \right) \bG \bKLth^{(u,u)} 
 \,,
 \label{eq:KL33uu}
 \end{equation}
where $\sbKLth$ is the same as $\bKLth$ except that all intermediate states have propagators replaced by the smooth difference described above.
For brevity, we have dropped the ``$u$" superscripts on $\sbKLth$.
We note also that the terms involving $\bK$ in this result arise from special cases
where, after the insertion of $\bG$, there is only a single $\bK$ on one or both sides.

If there were no poles in $\K_2$ we could replace the momentum sums in $\sbKLth$
with integrals and obtain the divergence-free K matrix. This was the procedure followed
in Ref.~\cite{\HSQCa}. However, 
here we need to extract the finite-volume effects that arise from the
K-matrix poles. To do so, we work through $\sbKLth$ from left to right, replacing
each full $\K_2$ with the $\bFrp$ cut and the difference, with the latter being
a smooth function of the spectator momentum. In the term with the $\bFrp$ cut the procedure stops, leaving a factor of $\sbKLtwth$ to the right. 
The remaining, $\bFrp$-independent terms build up quantities in which all loop sums can be replaced by
integrals because the integrands are divergence free. These are the quantities mentioned above that contain divergences only in the external K matrices and are denoted by
$\bKi_{33}$, $\bKi_{3\ttil}$ etc. (Again we drop the superscripts $(u)$ for brevity.)
The result is
\begin{equation}
\sbKLth = \bKi_{33} + \bKi_{3\ttil} \, \bFrp \sbKLtwth \,.
\label{eq:slashedKL33}
\end{equation}

Proceeding in the same way for $\bKLtwth^{(u)}$ and $\sbKLtwth $ we obtain
\begin{align}
 \bKLtwth^{(u)} & = 
 \sbKLtwth 
\left(1 + \bG \bK
+ \bG \bKLth^{(u,u)} \right)\,, 
\label{eq:KL23u} \\
\sbKLtwth  & = \bKi_{\ttil 3} + \bKi_{\ttil\ttil} \bFrp \sbKLtwth   \,.
\label{eq:slashedKL23}
\end{align}
We note here the appearance of $\bKi_{\ttil\ttil}$, which is the infinite-volume
version of $\bKLtw$ once all divergences have been removed. 
This quantity does not have factors of $\cK_2$ at its ends, so it is already divergence-free.  
These matrix equations can now be solved sequentially. The solution to
Eq.~(\ref{eq:slashedKL23}) is
\begin{equation}
\sbKLtwth    = \frac1{1 - \bKi_{\ttil\ttil}  \bFrp  }  \bKi_{\ttil 3}   \,,
\label{eq:slashedKL23soln}
\end{equation}
and inserting this in Eq.~(\ref{eq:slashedKL33})  yields
\begin{equation}
\sbKLth = \bKi_{33} + \bKi_{3 \ttil} \bFrp\frac1{1 - \bKi_{\ttil \ttil} \bFrp}  \bKi_{\ttil 3}\,.
\label{eq:slashedKL33soln}
\end{equation}

Taken together, Eqs.~(\ref{eq:KL33uu}) and (\ref{eq:slashedKL33soln})
give a complete prescription for writing $\bKLth^{(u,u)}$ in terms of
infinite-volume quantities and finite-volume cuts. In Appendix~\ref{app:KLdecom} 
we outline the remaining steps in this decomposition explicitly.
In the appendix we also work through the decompositions for the
remaining finite-volume K matrices, $\bKLthtw^{(u)}$ and $\bKLtw$, and
for the slashed objects, $\sbKLthtw$ and $\sbKLtw$. The procedure in
all cases is similar to that outlined above: One works through the
summed loops in a diagram from left to right, substituting singular
and smooth pieces for the propagators to reach matrix equations for
the various finite-volume objects entering the correlator. 
We find that the solutions to the resulting equations 
can be succinctly displayed in two key relations 
\begin{align}
\label{eq:KLdecom}
  \bKL^{(u)}  
& = 
\begin{pmatrix}
0 & 0 \\ 0 &  \bKLth^{(0)}
\end{pmatrix}
+
\bEL \frac1{      \sbKL^{-1}  - \bcGK}\bER
\,,  \\[5pt]
   \sbKL^{-1}    & =  {\pmb {\mathcal K}} ^{-1} - \bcFrp  \,,
   \label{eq:Kslashdecom}
\end{align}
where we have introduced two-by-two matrix generalizations of the various quantities appearing above
\begin{equation}
 \sbKL  \equiv
\begin{pmatrix}
\sbKLtw &   \sbKLtwth  \\
\sbKLthtw &  \sbKLth 
\end{pmatrix}\,,
\qquad
\pmb {\mathcal K} \equiv
\begin{pmatrix}
 {\bKi}_{\ttil\ttil} &  {\bKi}_{\ttil 3} \\
 {\bKi}_{3\ttil} &  {\bKi}_{33} 
\end{pmatrix}\,,
\qquad
\bcFrp \equiv
\begin{pmatrix}
\bFrp & 0 \\ 0 & 0
\end{pmatrix}\,,
\ \ \ {\rm and} \ \ \
\bcGK \equiv
\begin{pmatrix}
0 & 0 \\ 0 & \bGK
\end{pmatrix} \,.
\end{equation}
     
To complete the work of this section we need to remove the K-matrix poles contained
in ${\pmb {\mathcal K}}$. This is neccesary in order to symmetrize over choices of spectator,
as we see in the next subsection. To do so we introduce appropriate factors of
$\bGr$ corresponding to the poles in the external factors of $\K_2$. 
This leads to the result
\begin{equation}
\label{eq:bKdfdef}
{\pmb {\mathcal K}} = \bVD \,
\bKdf^{(u)} \, \bV
\,,
\end{equation}
where $\bVD$ and $\bV$ are defined in Eq.~(\ref{eq:Xdef}).
This relation defines a matrix of non-singular infinite-volume K-matrices 
already displayed in the result given at the beginning of the subsection\footnote{%
We note that $\bKi_{\df, \ttil \ttil}= \bKi_{\ttil \ttil}$, so the $\df$ subscript is not needed for this 
component. We include it anyway for uniformity of notation.}
\begin{equation}
 \bKdf^{(u)}  
\equiv
\begin{pmatrix}
\bKi_{\df,\ttil \ttil} & \bKi_{\df, \ttil 3}^{(u)} \\[8pt]
\bKi_{\df,3 \ttil}^{(u)} & \bKi_{\df,33}^{(u,u)} 
\end{pmatrix}\,.
\end{equation}

Combining Eqs.~(\ref{eq:KLdecom}), (\ref{eq:Kslashdecom}) and (\ref{eq:bKdfdef}) we reach the main result of this subsection given in Eq.~(\ref{eq:KLmatrix3}) above. The quantity $\bX$ appears as
\begin{equation}
\bX  \equiv  \bV  \Big ( \bcFrp + \bcGK \! \Big) \bVD \,,
\end{equation}
which can be rearranged into the form shown in Eq.~(\ref{eq:Xdef}). 

At this stage we have decomposed all objects appearing in $C_L^{[B_2]}$ into matrix products of finite- and infinite-volume quantities. In the following subsections we reshuffle these decompositions into a compact form for this partial finite-volume correlator. We then show how the three-particle Bethe-Salpeter kernls, $B_3$, can be reintroduced to derive the main result of the section, Eq.~(\ref{eq:CL}).


\subsection{$( \bKdf)^0$ contribution to $C^{[B_2]}_{L}$ }

We now have all the ingredients needed to determine the volume dependence of the
correlator $C_L^{[B_2]}$. The initial decomposition of this object is given in Eq.~(\ref{eq:Cno3MLres}). To derive the final form we work order by order in $\bKdf$, and begin by considering the contributions
that are independent of this local three-body interaction. In particular, in this subsection we demonstrate
\begin{align}
C^{[B_2]}_{L} - C^{[B_2]}_{\infty} - \delta  C^{[B_2]}_{\infty}
= \bApMat \bFMat \bAMat + \mathcal O(\bKdf) \,,
\label{eq:CLB2Kdf0}
\end{align}
where $\bF_{33}$, $\bF_{\ttil \ttil}$, $\bF_{\ttil 3}$, and $\bF_{3 \ttil}$ are 
defined in Eqs.~(\ref{eq:bFandbTdef})-(\ref{eq:bF32def}), respectively,
while $\delta  C^{[B_2]}_{\infty}$ is an additional volume-independent term, defined at leading order in $\bKdf$ in Eq.~(\ref{eq:CLnoKdf2_IV}) below.  As mentioned in the introduction, many of the steps in the derivation of Eq.~(\ref{eq:CLB2Kdf0}) presented here
have been checked using a {\em Mathematica} notebook implementing the package 
{\em The NCAlgebra Suite} \cite{NCA}. 
Equations verified in this way are preceded by the indicator ``\pNCA''.

\bigskip

If $\bKdf^{(u)}=0$, the only nonzero component of
$\bKL^{(u)}$ is $\bKLth^{(u,u)}$, which becomes $\bKLth^{(0)}$, defined in Eq.~(\ref{eq:KL330}).
Thus the infinite sum in Eq.~(\ref{eq:Cno3MLres}) becomes \pNCA
\begin{equation}
\bF_{33}^{(0)}
\sum_{n=0}^\infty 
\left( \bKLth^{(u,u)}  
       \bF_{33}^{(0)} \right)^n \ \ \  \longrightarrow \ \ \  \bZ \equiv 
\bF_{33}^{(0)}
\sum_{n=0}^\infty   \left( \bKLth^{(0)}  \bF_{33}^{(0)} \right)^n  
=    \bF \frac1{1- \bT \bF} \,,
\label{eq:Zdef}
\end{equation}
where the arrow indicates $\bKdf^{(u)} \to 0$. Here we have used $\bF_{33}^{(0)}$ and $\bT$, defined in Eqs.~(\ref{eq:F33zerodef}) and (\ref{eq:bT}) respectively. In this same limit the quantities $\bA'^{(u)}_{L,3}$, $\bA^{(u)}_{L,3}$ and $\slashed{\bA}_{L,\ttil}$ simplify to \pNCA
\begin{align}
\bA'^{(u)}_{L,3}
& \ \ \  \longrightarrow \ \ \  \bA'^{(u),\{0\}}_{L,3} =
\bA'^{(u)}_3 
+ 2 \bA'^{(s)}_3 \bF \bT 
+  \bA'_{\ttil}  \bGr \bGam 
(1 + \bG \bT)
\,, 
\label{eq:ApL3noKdf} \\
\bA^{(u)}_{L,3} &
\ \ \ \longrightarrow \ \ \ \bA^{(u),\{0\}}_{L,3} =
\bA^{(u)}_3
+\bT \bF \, 2 \bA^{(s)}_3 
+  (1 + \bT \bG) \bGam \bGrD \bA_{\ttil}
\,, \label{eq:AL3noKdf} \\
\slashed{\bA}_{L,\ttil} & \ \ \  \longrightarrow \ \ \  \slashed{\bA}_{\ttil} \,,
\end{align}
where the superscript $\{n\}$ indicates the contribution to the indicated object with $n$ factors of $\bKdf$.

At this stage we can use the following result from Ref.~\cite{\HSQCa}:
if $2 \bA'^{(s)}_3$ is adjacent to a factor of $\bF \bK $ then it can be replaced
by $\bA'^{(s)} + \bA'^{(\tilde s)}$, with $(\tilde s)$ indicating the third independent
permutation of the external momenta. This is the case in Eq.~(\ref{eq:ApL3noKdf})
because $\bT$ always has a factor $\bK$ on its left-hand end. 
The same holds for the factor of $2 \bA'^{(s)}_3$ in Eq.~(\ref{eq:CL0Ffinal}),
because $\bA_{L,3}^{(u)}$ and $\bSigD$ are symmetric under the
interchange of the nonspectator pair.
Similarly, the factor of $2 \bA^{(s)}_3$ in Eq.~(\ref{eq:AL3noKdf}) can be
replaced by $\bA^{(s)}_3 + \bA^{(\tilde s)}_3$.
These substitutions are important because the fully symmetrized endcaps are given by
\begin{equation}
\bA'_3 \equiv \bA'^\u_3+\bA'^{(s)}_3+\bA'^{(\tilde s)}_3\,,\ \ {\rm and }\ \ 
\bA_3 \equiv \bA^{(u)}_3+\bA^{(s)}_3+\bA^{(\tilde s)}_3\,.
\label{eq:Asymmdef}
\end{equation}
We expect the final result to depend only on symmetrized quantities.
In the following, for the sake of brevity, we use
$2 \bA^{(s)}$ as an abbreviation for $\bA^{(s)} + \bA^{(\tilde s)}$ and
$2 \bA'^{(s)}$ for $\bA'^{(s)} + \bA'^{(\tilde s)}$. Using this simplification, we can rewrite Eqs.~(\ref{eq:ApL3noKdf}) and (\ref{eq:AL3noKdf}) as
\begin{align}
 \bA'^{(u),\{0\}}_{L,3} & = 
\bA'_3 
-  2 \bA'^{(s)}_3 (1- \bF \bT )
+  \bA'_{\ttil}   \bGr \bGam 
(1 + \bG \bT)
\,,
\label{eq:ApL3noKdf_v2}
\\[5pt]
 \bA^{(u),\{0\}}_{L,3} & =
\bA_3
- (1 - \bT \bF ) 2 \bA^{(s)}_3 
+  (1 + \bT \bG) \bGam \bGrD \bA_{\ttil}
\,.
\label{eq:AL3noKdf_v2}
\end{align}

The final quantity we are missing is $C_{L,0F}$, whose decomposition is given in Eq.~(\ref{eq:CL0Ffinal}). Sending $\bKdf \to 0$ in this result, 
and using Eq.~(\ref{eq:AL3noKdf_v2}), gives \pNCA
\begin{multline}
C_{L,0F} \ \ \ 
\longrightarrow \ \ \ C_{L,0F}^{\{0\}}   = 
{C}^{[B_2],\{0\}}_{\infty }
 +
\left( 2 \bA'^{(s)}_3   {\bF} +
\bA'_{\ttil}  \bGr \bGam  \bG \right)
\\  \times \left ( 
\bA_3
-(1-\bT \bF) 2 \bA^{(s)}_3 
+  (1 + \bT \bG) \bGam \bGrD \bA_{\ttil}
- \bSigD \right)
+
\bA'_{\ttil} \bFrp   \slashed \bA_{\ttil} \,.
\end{multline}

We have now gathered all the pieces to evaluate the full correlation function, decomposed in Eq.~(\ref{eq:Cno3MLres}), at $\mathcal{O}([\bKdf]^0)$. 
This equation reduces to
\begin{equation}
C^{[B_2],\{0\}}_{L}   = 
 C^{\{0\}}_{L,0F} 
- \frac23 \bSig  \bF  \bSigD
+ 
\bA'^{(u),\{0\}}_{L,3} 
\mathbf Z  
\bA^{(u),\{0\}}_{L,3} \,.
\end{equation}
Substituting Eqs.~(\ref{eq:Zdef}), (\ref{eq:ApL3noKdf_v2}) and (\ref{eq:AL3noKdf_v2}) and significantly rearranging, we find \pNCA
\begin{equation}
C^{[B_2],\{0\}}_{L}  = {C}^{[B_2],\{0\}}_{\infty } +  \delta C^{[B_2],\{0\}}_{\infty } +     \bApMat \bFMat \bAMat \,,
\label{eq:CB2NoKdfResult}
\end{equation}
where
\begin{multline}
\delta C^{[B_2],\{0\}}_{\infty } 
\equiv - \tfrac23 \bSig \bF \bSigD 
- 2 \bA'^{(s)}_3 \bF  \bSigD - \bA'_3 \bF \, 2 \bA_3^{(s)}
+ \tfrac23 \bA'_3 \bF \bA_3
\\
+ \bA'_{\ttil} \bFrp     (\slashed \bA_{\ttil} - \bA_{\ttil})
+ \bA'_{\ttil}  \bGr \bGam    
\left[  \bG  (\bA_3^{(u)} - \bSigD)  - \bF \, 2 \bA_3^{(s)}\right] \,.
\label{eq:CLnoKdf2_IV}
\end{multline}
To obtain Eq.~(\ref{eq:CB2NoKdfResult}) we have made use of the following identities \pNCA:
\begin{align}
\left[ - \frac23 + \frac1{1- \bF \bT }\right] \bF & = \bF  \left[\frac13 +   \mathbf T_L  \bF \right ] \,,
\label{eq:iden0}
\\
\frac1{1 - \bT \bF} \bT &= \frac1{1- \bK (\bF+\bG)} \bK\,,
\label{eq:iden1}
\\
 \frac1{1-\bT \bF} (1+ \bT \bG) &= \frac1{1- \bK (\bF + \bG)}\,,
\label{eq:iden2}
\\
(1 + \bG \bT) \frac1{1- \bF \bT} &= \frac1{1- (\bF + \bG) \bK}\,,
\label{eq:iden3}
\\
\left[\bG + (1+ \bG \bT)\frac1{1- \bF \bT } \bF  \right]
(1 + \bT \bG) &=   \frac1{1- (\bF+\bG) \bK} (\bF + \bG)\,,
\label{eq:iden4}
\end{align}
which follow from straightforward manipulations using
the definitions Eqs.~(\ref{eq:bFandbTdef}) and (\ref{eq:bT}).

Equation~(\ref{eq:CB2NoKdfResult}) is equivalent to Eq.~(\ref{eq:CLB2Kdf0}), where $\delta C^{[B_2],\{0\}}_{\infty } $ is understood as the $\mathcal O[(\bKdf)^0]$ contribution to $\delta C^{[B_2]}_\infty$. At this stage it remains only to show that $\delta C^{[B_2],\{0\}}_{\infty } $ only has exponentially suppressed volume dependence. This is done in Appendix \ref{app:deltaCL}.


\subsection{$C^{[B_2]}_{L}$  to all orders in $ \bKdf^\u$: Unsymmetrized }

In this subsection we collect the terms
contributing to $C_L^{[B_2]}$,  Eq.~(\ref{eq:Cno3MLres}),
that contain at least one factor of $\bKdf^\u$. 
Throughout this subsection and the next, we use the superscript ${}^{[\bKdf]}$ to 
denote the contribution to a quantity with one or more factors of the unsymmetrized 
divergence-free K matrix.

Beginning with $C_{L,0F}$, decomposed in Eq.~(\ref{eq:CL0Ffinal}), we use the
results in Eqs.~(\ref{eq:AL3final}), (\ref{eq:slashedALfinal}), (\ref{eq:KLmatrix3})
and (\ref{eq:Xdef}) and find that the part containing at least one factor of $\bKdf$
can be written as \pNCA
\begin{align}
\begin{split}
C_{L,0F}^{[\bKdf]}
& =
\left( 2  \bA'^{(s)}_3 \bF +
\bA'_{\ttil}  \bGr \bGam \bG \right)
\left[ \bKLth^{(u,u)} 
\left(\bF 2 \bA^{(s)}_3 +\bG \bGam \bGrD \bA_{\ttil}\right) +  \bKLthtw^{(u)} \bFrp \bA_{\ttil} \right] \\
& \hspace{170pt} + \bA'_{\ttil} \bFrp \left [ \bKLtwth^{(u)}
\left( \bF 2 \bA^{(s)}_3 + \bG \bGam \bGrD \bA_{\ttil} \right) + \bKLtw \bFrp \bA_{\ttil} \right ] \,,
\end{split} 
\label{eq:CL0FKdf1partA}\\[5pt]
& =
\begin{pmatrix}  \bA'_{\ttil} & 2 \bA'^\s_3 \end{pmatrix}
 \begin{pmatrix}
\bFrp & \bGr \bGam \bG 
 \\
0 & \bF
 \end{pmatrix}
  \bKL^{(u)} 
\begin{pmatrix}
\bFrp & 0 \\
\bG \bGam  \bGrD &
 \bF 
 \end{pmatrix}
  \begin{pmatrix}
 \bA_{\ttil} \\[3pt] 2 \bA_3^\s
 \end{pmatrix} \,, 
 \label{eq:CL0FKdf1partB}\\[5pt]
& =
\begin{pmatrix}  \bA'_{\ttil} & 2 \bA'^\s_3 \end{pmatrix}
 \begin{pmatrix}
\bFrp & \bGr \bGam \bG 
 \\
0 & \bF
 \end{pmatrix}
\bEL \bVD   \bKdf^{(u)} \frac1{1 - \bX \bKdf^{(u)} } \bV \bER  
\begin{pmatrix}
\bFrp & 0 \\
\bG \bGam  \bGrD &
 \bF 
 \end{pmatrix}
  \begin{pmatrix}
 \bA_{\ttil} \\[3pt] 2 \bA_3^\s
 \end{pmatrix} \,,
 \label{eq:CL0FKdf1partC} \\[5pt]
\begin{split}
&=\bigg [
 \bA'_{\ttil}  \begin{pmatrix} 1 &  0 \end{pmatrix} \bX
+ 2 \bA'^\s_3 \bF (1+ \bT \bG) 
\begin{pmatrix}  \bGam  \bGrD &   1 \end{pmatrix}
\bigg ]
\cdot \bKdf^\u  \frac1{1- \bX  \bKdf^\u } \\ 
& \hspace{230pt} \cdot \bigg [
\bX \begin{pmatrix}1\\0\end{pmatrix} \bA_{\ttil}
+ \begin{pmatrix} \bGr \bGam  \\ 1 \end{pmatrix} 
(1 + \bG \bT) \bF \, 2 \bA_3^\s
\bigg ]\,.
\end{split}
\label{eq:CL0Ffinal_v2}
\end{align}
In Eq.~(\ref{eq:CL0FKdf1partA}) we have simply substituted Eqs.~(\ref{eq:AL3final}) and (\ref{eq:slashedALfinal}) into Eq.~(\ref{eq:CL0Ffinal}) and dropped terms that have no factors of $\bKdf^{(u)}$.
To obtain Eq.~(\ref{eq:CL0FKdf1partB}) we then rearrange terms into a matrix form
using the definition of $\bKL^{(u)}$, Eq.~(\ref{eq:KLmatrix}).
Next we substitute the result Eq.~(\ref{eq:KLmatrix3}) for $\bKL^{(u)}$, dropping
terms with no factors of $\bKdf^{(u)}$, leading to Eq.~(\ref{eq:CL0FKdf1partC}).
We then use the definition of $\bX$, Eq.~(\ref{eq:Xdef}), to bring the result to
the final form, Eq.~(\ref{eq:CL0Ffinal_v2}).

We next turn to the terms in Eq.~(\ref{eq:Cno3MLres}) that contain at least one factor of $\bF_{33}^{(0)}$. These terms always include the endcap factors $\bA'^\u_{L,3}$ and $\bA^\u_{L,3}$ so that we first require the full decomposition of these. Beginning with $\bA'^\u_{L,3}$, decomposed in Eq.~(\ref{eq:ApL3final}), we insert the expressions for the different components of $\bKL^{(u)} $ [Eq.~(\ref{eq:KLmatrix3})] to find \pNCA
\begin{align}
\bA'^\u_{L,3} 
&=
\bA'^{(u)}_3
+
\left( 2 \bA'^{(s)}_3 \bF
+  \bA'_{\ttil}   \bGr  \bGam \bG \right) 
\left(  \bKLth^{(u,u)} + \bK \right)
+
\bA'_{\ttil} \left( \bFrp \bKLtwth^{(u)} + \bGr \bGam \right) \,, \\[5pt]
\begin{split}
&= \bA'_3 - 2 \bA'^{(s)}_3 +
\left( 2 \bA'^{(s)}_3 \bF   +  \bA'_{\ttil}   \bGr \bGam \bG \right) 
\left(  \bKLth^{(0)} + \bK \right)
+ \bA'_{\ttil} \bGr \bGam  \\
& \hspace{20pt} + \left( 2 \bA'^{(s)}_3  \bF
+  \bA'_{\ttil}    \bGr  \bGam \bG \right) 
(1+ \bT \bG) 
\begin{pmatrix}  \bGam  \bGrD &  1 \end{pmatrix}
\cdot \bKdf^{(u)}  \frac1{1 - \bX \bKdf^{(u)} }  \cdot
\begin{pmatrix} \bGr \bGam  \\ 1 \end{pmatrix} (1 +\bG  \bT ) \\[-5pt]
& \hspace{20pt} + \bA'_{\ttil} \bFrp  
\begin{pmatrix}   1 & 0 \end{pmatrix}
\cdot \bKdf^{(u)}  \frac1{1 - \bX \bKdf^{(u)} }  \cdot
\begin{pmatrix} \bGr \bGam  \\1 \end{pmatrix} (1 + \bG  \bT ) \,, 
\end{split}\\[5pt]
 \begin{split}
 & = \bA'_3 - 2 \bA'^\s_3 (1 - \bF \bT )
+ \bA'_{\ttil} 
\begin{pmatrix} 1 & 0 \end{pmatrix} 
\cdot  \frac1{1 - \bX \bKdf^\u } \cdot
\begin{pmatrix} \bGr \bGam  \\ 1 \end{pmatrix} (1 + \bG   \bT ) \\
& \hspace{120pt} + 2 \bA'^\s_3 \bF
(1 + \bT \bG)  
\begin{pmatrix}   \bGam  \bGrD & 1 \end{pmatrix}
\cdot {\bKdf^\u} \frac1{1 - \bX \bKdf^\u } \cdot
\begin{pmatrix} \bGr \bGam  \\ 1 \end{pmatrix} (1 + \bG  \bT )
\,.
\end{split}
\label{eq:ApL3final_v2}
\end{align}
Here the first line is just a repeat of Eq.~(\ref{eq:ApL3final}) and in the remaining lines we have substituted the expressions for $\bKLth^\uu$ and $\bKLtwth^\u$ and simplified.

The expression for the mirror-imaged endcap is then given by \pNCA
\begin{multline}
\bA^\u_{L,3} = \bA_3 - (1 - \bT \bF) 2 \bA^\s_3
+
(1 + \bT \bG) 
\begin{pmatrix}  \bGam  \bGrD & 1 \end{pmatrix}
\cdot \frac1{1 - \bKdf^\u  \bX}  \cdot
\begin{pmatrix} 1 \\ 0 \end{pmatrix} \bA_{\ttil}
\\
+ (1 + \bT \bG) 
\begin{pmatrix}  \bGam  \bGrD & 1 \end{pmatrix}
\cdot {\bKdf^\u} \frac1{1 - \bX \bKdf^\u } \cdot
\begin{pmatrix} \bGr \bGam \\1 \end{pmatrix} (1 + \bG  \bT)
\bF 2 \bA^\s_3
\,.
\label{eq:AL3final_v2}
\end{multline}
In both cases we include the $\big (\bKdf^\u \big )^0$ part, 
since the factors of $\bKdf^\u$ can come from the sum appearing between the finite-volume endcaps in the expression for $C_L^{[B_2]}$.

Finally, to derive an expression for the sum appearing between $\bA'^\u_{L,3}$ and $\bA^\u_{L,3}$ in Eq.~(\ref{eq:Cno3MLres}), we make use of the following identity \pNCA
\begin{align}
\bKLth^{(u,u)}
&= 
\bKLth^{(0)} 
+
\begin{pmatrix} 0 & 1 \end{pmatrix} \cdot
\bEL \bVD    \bKdf^{(u)}  \frac1{1 - \bX  \bKdf^{(u)} } \bV \bER \cdot
\begin{pmatrix}
0 \\   1
\end{pmatrix}\,, \\
&= 
\bKLth^{(0)} 
+
(1 + \bT \bG) \begin{pmatrix}  \bGam  \bGrD & 1 \end{pmatrix}
\cdot   \bKdf^{(u)}  \frac1{1 - \bX  \bKdf^{(u)} } \cdot
 \begin{pmatrix} \bG_\rho \bGam  \\1 \end{pmatrix} (1 + \bG  \bT ) 
 \,,  
  \label{eq:KLthKdfdefA} \\
&\equiv 
\bKLth^{(0)} 
+
 \bKLth^{(u,u),[\bKdf]}\,,
 \label{eq:KLthKdfdefB}
\end{align}
where $\bKLth^{(u,u),[\bKdf]}$ is defined by comparing Eqs.~(\ref{eq:KLthKdfdefA}) and (\ref{eq:KLthKdfdefB}).

 Combining this with the expression for $\bZ$, defined in Eq.~(\ref{eq:Zdef}), we find \pNCA
\begin{align}
\bF_{33}^{(0)} \frac1{1- \bKLth^\uu   \bF_{33}^{(0)}}
&=
 \bZ  \,
 \frac1{1- \bKLth^{(u,u),[\bKdf]}  \, \bZ } \,, \\
&=
 \bZ 
+
 \bZ \bKLth^{(u,u),[\bKdf]}
\sum_{n=0}^\infty  \left  ( \bZ \, \bKLth^{(u,u),[\bKdf]} \right )^n \,  \bZ  \,, \\
&=
 \bZ 
+
 \bZ  (1 + \bT \bG)  
 \begin{pmatrix} \bGam  \bGrD & 1 \end{pmatrix}
 \cdot \bKdf^{(u)}  \frac1{1 - \bX \bKdf^{(u)} }  
\sum_{n=0}^\infty  \left( \! \bY \, \bKdf^{(u)}  \frac1{1 - \bX \bKdf^{(u)} } \right)^{\!\!n} \cdot
\begin{pmatrix} \bGr  \bGam  \\ 1 \end{pmatrix} (1 + \bG \bT )  \bZ  \,, \\
&=
 \bZ  +  \bZ  (1 + \bT \bG)  
 \begin{pmatrix} \bGam  \bGrD & 1 \end{pmatrix}
 \cdot \bKdf^{(u)} \frac1{1 - \big ( \bX + \bY \big ) \bKdf^{(u)}} \cdot 
\begin{pmatrix} \bGr \bGam  \\1 \end{pmatrix} (1 + \bG  \bT)  \bZ 
\,,
\label{eq:F3K3LF3}
\end{align}
where the new matrix $\bY$ is \pNCA
\begin{align}
\bY
&\equiv
\begin{pmatrix} \bGr \bGam  \\1 \end{pmatrix} \cdot (1 +  \bG  \bT )  \bZ 
(1 + \bT \bG)  \cdot \begin{pmatrix} \bGam  \bGrD & 1 \end{pmatrix} \,,\\
&=
\begin{pmatrix} \bGr \bGam  \\1 \end{pmatrix}  \cdot 
\left[
- \bGK 
 +
\frac1{1- ( \bF +  \bG) \bK}   
( \bF +  \bG) 
\right]  \cdot \begin{pmatrix} \bGam  \bGrD & 1 \end{pmatrix}
\,.
\label{eq:Ndef}
\end{align}

From this, together with the expression for $\bX$ [Eq.~(\ref{eq:Xdef})] and $\bF_{\ttil\ttil}$ [Eq.~(\ref{eq:F22def})], we find that the combined matrix appearing between factors of $\bKdf^\u$ is \pNCA
\begin{equation}
\bX + \bY = \begin{pmatrix}
\bF_{\ttil\ttil} & \bGr \bGam  \frac1{1- (\bF + \bG) \bK} (\bF + \bG) 
\\[5pt]
  (\bF +  \bG) \frac1{1- \bK (\bF + \bG) } \bGam  \bGrD
&
\frac1{1- (\bF  +  \bG) \bK } (\bF +  \bG) 
\end{pmatrix}
\,.
\label{eq:XplusY}
\end{equation}
We observe that the off-diagonal elements are close to $\bF_{\ttil 3}$ and $\bF_{3\ttil}$,
differing only by the presence of $\bF + \bG$ rather than $\bF$ on the ends.
Similarly, the $33$ element is close to $\bF_{33}$. These differences will be removed
when we change from the unsymmetrized $\bKdf^\u$ to the symmetrized version.


\bigskip

With these preliminaries, we begin the determination of $C_L^{[B_2],[\bKdf]}$
by collecting the terms involving factors of $\bA'_{\ttil}$ and $\bA_{\ttil}$ on the ends.
All terms appearing in Eq.~(\ref{eq:CL0Ffinal_v2}) as well as the appropriate combinations of Eqs.~(\ref{eq:ApL3final_v2}), (\ref{eq:AL3final_v2}) and (\ref{eq:F3K3LF3}) that contain these endcaps have the form 
\begin{equation}
C_{L}^{[B_2],[\bKdf]} \supset \bA'_{\ttil} \begin{pmatrix}  1 & 0 \end{pmatrix} \cdot \mathbf W \cdot  \begin{pmatrix} 1\\ 0\end{pmatrix} \bA_{\ttil}\,,
\label{eq:A2A2}
\end{equation}
and our task is to determine the matrix $ \mathbf W $.
Collecting terms, we find\footnote{%
On the right-hand side of the first equality,
the final $-\bY$ term is needed to remove the $(\bKdf^\u)^0$ contribution to the
previous term.}
\pNCA
\begin{align}
 \mathbf W  
&=  \bX  \bKdf^\u \frac1{1- \bX  \bKdf^\u }  \bX 
+ \frac1{1- \bX  \bKdf^\u } \left( \bY + \bY   \bKdf^\u 
   \frac1{1 - \big (\bX + \bY \big ) \bKdf^\u }  \bY  \right)
   \frac1{1- \bKdf^\u  \bX } 
   -  \bY \,, \\
&=\big (\bX + \bY \big )  \bKdf^\u  \frac1{1-\big (\bX + \bY \big )  \bKdf^\u } \big (\bX + \bY \big )
\,.
\end{align}

Next we consider the cases with either $\bA'^\s_3$ on the left-hand side,
or $\bA^\s_3$ on the right, or both. After some algebra, we find that all such terms
vanish identically. 

The remaining, non-vanishing terms are those involving the endcaps $\bA'_3$ and $\bA_3$. 
We find
\begin{multline}
C_{L}^{[B_2],[\bKdf]} \supset \bA'_3 \bZ (1+\bT \bG) \begin{pmatrix} \bGam \bGrD & 1 \end{pmatrix} \cdot
 \bKdf^\u \frac1{1- \big ( \bX + \bY \big)  \bKdf^\u }  \big ( \bX + \bY \big)  \cdot
\begin{pmatrix} 1\\ 0\end{pmatrix}  \bA_{\ttil} \
\\
+  \bA'_{\ttil}  \begin{pmatrix} 1 & 0 \end{pmatrix} \cdot  \big ( \bX + \bY \big)   \bKdf^\u 
\frac1{1- \big ( \bX + \bY \big)  \bKdf^\u } 
\cdot \begin{pmatrix} \bGr  \bGam \\ 1 \end{pmatrix}
 (1 + \bG  \bT)  \bZ \bA_3
\\
+ \bA'_3 \bZ  (1+ \bT \bG)  \begin{pmatrix}  \bGam \bGrD & 1 \end{pmatrix}  \cdot
 \bKdf^\u  \frac1{1- \big ( \bX + \bY \big)  \bKdf^\u } \cdot
\begin{pmatrix} \bGr  \bGam \\ 1 \end{pmatrix}
 (1 + \bG  \bT)  \bZ \bA_3 \,.
\label{eq:A3parts}
\end{multline}

Finally, we can combine the results in Eqs.~(\ref{eq:A2A2}) and (\ref{eq:A3parts}) 
into a compact matrix form
\begin{equation}
C_{L}^{[B_2],[\bKdf]}
=
\begin{pmatrix}  \bA'_{\ttil} & \bA'_3 \end{pmatrix}  \cdot
\bF_{\mathbf L}
 \bKdf^\u  \frac1{1- \big ( \bX + \bY \big)  \bKdf^\u } 
\bF_{\mathbf R} \cdot
\begin{pmatrix}
\bA_{\ttil}\\ \bA_3
\end{pmatrix}
\,,
\label{eq:CLK}
\end{equation}
where we have introduced 
\begin{equation}
 \bF_{\mathbf L} \equiv
\begin{pmatrix}
\bF_{\ttil\ttil} & \bX_{\ttil 3} + \bY_{\ttil 3}
\\
\bF_{3\ttil} &  \bF \frac1{1 - \bK(\bF+\bG)}
\end{pmatrix}
\ \ {\rm and}\ \ 
\bF_{\mathbf R} \equiv
\begin{pmatrix}
\bF_{\ttil\ttil} & \bF_{\ttil 3} 
 \\
\bX_{3\ttil} +\bY_{3\ttil} & \frac1{1-(\bF+\bG) \bK}  \bF
\end{pmatrix} \,.
\label{eq:ZLRdef}
\end{equation}
To obtain this form, we have used the identities (\ref{eq:iden2}) and (\ref{eq:iden3}),
as well as the definitions of $\bF_{\ttil 3}$ and $\bF_{3\ttil}$,
given in Eqs.~(\ref{eq:bF23def}) and (\ref{eq:bF32def}), respectively.

\subsection{Symmetrization of $\bKdf$ in $C_L^{[B_2]}$}
\label{sec:symmKdf}

A pleasing feature of the result of the previous section, Eq.~(\ref{eq:CLK}), is
that it contains only symmetrized endcaps, despite the presence of unsymmetrized endcaps
at earlier stages. It does, however, contain the unsymmetrized quantity $\bKdf^\u$,
and in this section we manipulate the result so that all infinite-volume quantities have the desired exchange symmetry. Here we build upon the work of Ref.~\cite{\HSQCa}, but again need additional
techniques to deal with the poles in $\K_2$. We also have found ways to shorten the
argumentation given in Ref.~\cite{\HSQCa}. Nevertheless, this section is the most 
algebraically involved in this work.

A key observation for doing the symmetrization is that, if Eq.~(\ref{eq:CLK}) is expanded in powers of $\bKdf^\u$,
then in all terms with more than one factor of this unsymmetrized three-particle quantity, it
always lies next to a factor of $\bF + \bG$,
due to the structure of $\bX+\bY$, Eq.~(\ref{eq:XplusY}).
This allows us to use a class of symmetrization results exemplified by
\begin{equation}
\bK (\bF + \bG) \, \bKdfth^\uu 
=
\bK \bF \, {\cal S} \big [ \bKdfth^\uu \big ]
+
\bGam  \bGrD \, \mathcal I_{\ttil 3} \otimes   \bKdfth^\uu
+
\mathcal I_{33} \otimes  \bKdfth^\uu
\,,
\label{eq:symmL}
\end{equation}
where ${\cal S}$ is the
symmetrization operator than converts a $(u)$ quantity into the symmetric
$(u+s+\tilde s)$ version,\footnote{%
Here $\mathcal S$ acts to the right, but, in the following, it will also act to the
left. Which is the case will be clear from the context.}
while $\mathcal I_{\ttil 3}$ and $\mathcal I_{33}$ are integral operators, 
to be explained below.
The result (\ref{eq:symmL}) holds with $\bKdfth^\uu$ replaced 
by any three-particle quantity with the $(u)$ superscript, e.g.~$\bKdfthtw^\u$.
It also assumes that there is at least one factor of $\bF$ or $\bG$ on the
left, as is true in general because $ \big ( \bX + \bY \big) $, $\bFL$ and $\bFR$ contain the geometric series $1/(1-(\bF+\bG) \bK)$.

\begin{figure}
\begin{center}
\includegraphics[width=\textwidth]{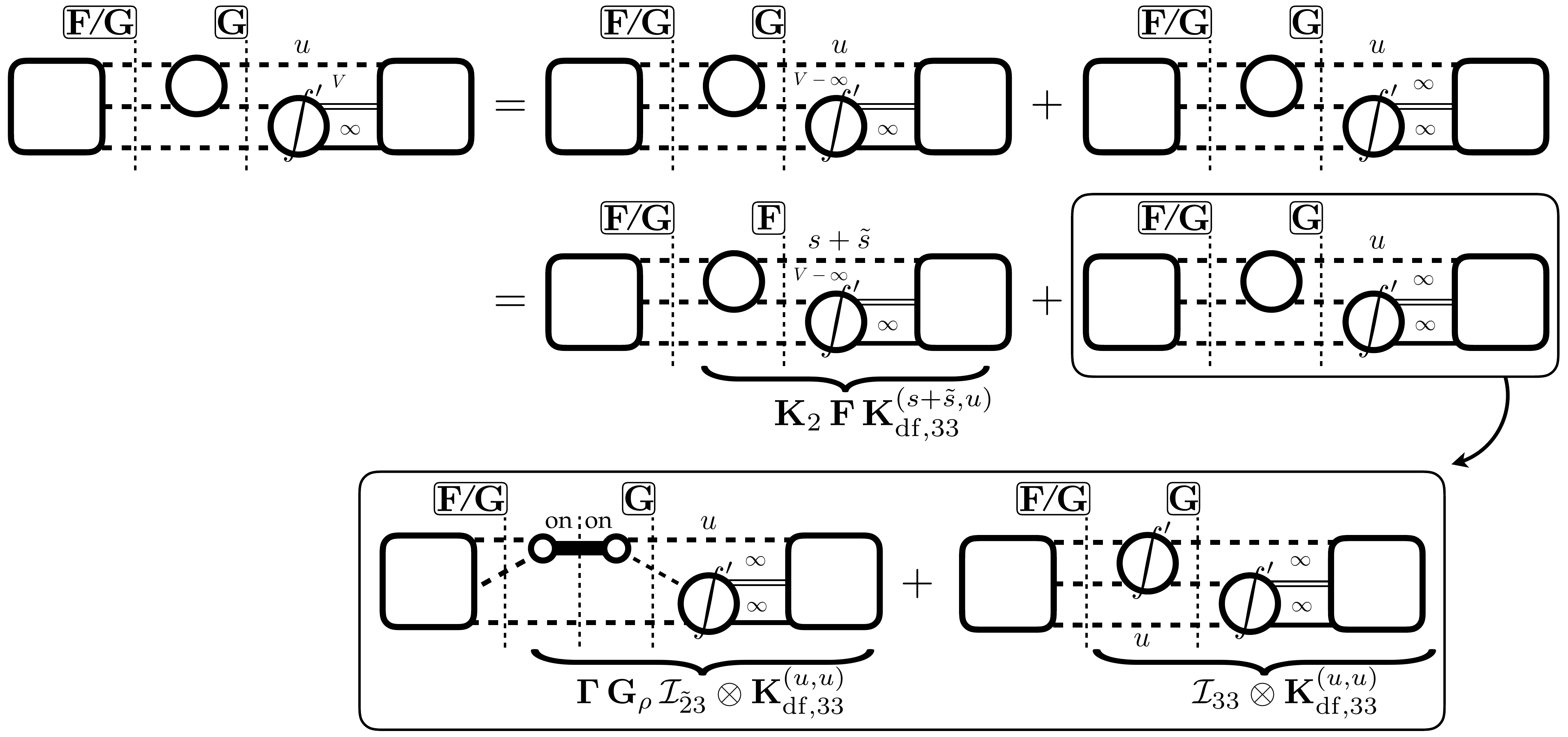}
\caption{
Derivation of Eq.~(\ref{eq:symmL0}), using the notation of Figs.~\ref{fig:CnL0F}
and \ref{fig:CL0Frecursion_v2}. On the left-hand side of the equality, the quantity
to the right of the $G$ cut is $\bKdfth^\uu$, with the $u$ above the upper-right dashed
line indicating that this is an unsymmetrized quantity. The left-hand cut in all diagrams must be
present, but can be either $\bF$ or $\bG$. The box at the left-hand end
of each diagram represents whatever lies
to the left of the $F/G$ cut, which depends on the context, but for which the details are irrelevant.
The first two equalities
show how $\bG$ is converted to $\bF$ by adding and subtracting an integral.
This method is used extensively in Ref.~\cite{\HSQCa} and is explained in
Eqs.~(163)-(165) of that work and the accompanying text.
It results in the $u$ superscript on $\bKdfth$ changing to $s+\tilde s$ in the $\bF$ term.
In the second step (indicated by the arrow connecting the two boxed diagrams), 
$\K_2$ is replaced by the pole term, with on-shell projection onto the K-matrix pole, 
and the smooth part. Since there is now an integral to the right of the $\bG$,
rather than a sum, the infinite volume quantity $\bKdfth^\uu$ is extended to
the left by the addition of either a $\Gamma$ or $\K'_2$, implicitly defining
the integral operators $\mathcal I_{23}$ and $\mathcal I_{33}$, respectively.
}
\label{fig:symm2}
\end{center}
\end{figure}

To demonstrate Eq.~(\ref{eq:symmL}) we derive the equivalent result
\begin{equation}
\bK \bG  \bKdfth^\uu 
=
\bK \bF   \bKdfth^{(s+\tilde s,u)}
+
\bGam  \bGrD  \, \mathcal I_{\ttil 3} \otimes  \bKdfth^\uu
+
\mathcal I_{33} \otimes \bKdfth^\uu \,,
\label{eq:symmL0}
\end{equation}
in Fig.~\ref{fig:symm2}.
As seen from the figure, the integral operator
$\mathcal I_{\ttil 3}$, attaches a factor of $\bGam$ to $\bKdfth^\uu$,
leading to an infinite-volume ``two-particle" quantity,
while $\mathcal I_{33}$ attaches $i \widetilde \K'_2$ to $\bKdfth^\uu$, 
creating another infinite-volume three-particle quantity with the $(u)$ superscript.

The reflected equation is derived similarly and is
\begin{equation}
\bKdfth^\uu  (\bF + \bG) \bK 
=
\big [ \bKdfth^\uu \big ] \mathcal S  \,  \bF \bK 
+
\bKdfth^\uu \otimes \mathcal I_{3\ttil} \, \bGr \bGam
+
\bKdfth^\uu \otimes \mathcal I_{33}^\dagger
\,,
\label{eq:symmR}
\end{equation}
where $\mathcal I_{3\ttil}$ and $\mathcal I_{33}^\dagger$ are integral operators acting to the left on three-particle unsymmetrized quantities.
The direction of action of the integral operators is indicated by the position of the
$\otimes$ symbol.

We can iterate Eq.~(\ref{eq:symmL}), assuming implicitly that it acts on an unsymmetrized
three-particle quantity on the right, and that there are additional implicit factors of
$\bF$ or $\bG$ on the left. We find
\begin{equation}
\frac1{1 - \bK(\bF+\bG)} 
=
\sum_{n=0}^\infty \left\{\bK (\bF+\bG)\right\}^n
=
\frac1{1 - \bK(\bF+\bG)} \left\{
\bK \bF \mathcal S 
+ \bGam \bGrD \, \mathcal I_{\ttil 3} \otimes  \right\}
\frac1{1- \mathcal I_{33} \otimes}
+  \frac1{1 - \mathcal I_{33} \otimes} \,.
\label{eq:symmLiterate}
\end{equation}
The first term in curly braces on the right-hand side leads to symmetrized quantities (since it contains the operator $\mathcal S$),
while the second, $\bGam$-dependent term does not require symmetrization.
The final term on the right-hand side of this result is an unsymmetrized residue that
will be dealt with subsequently.

We now apply this result to the quantity of interest, $C_L^{[B_2],[\bKdf]}$ in Eq.~(\ref{eq:CLK}).
We begin by considering the contribution
in which cuts appear between the endcaps $\bA'_3$ and $\bA_3$ and the 
outermost $\bKdf^\u$ insertions. 
Here the analysis is simplified by having a symmetrized quantity on one side. 
Focusing first on the right-side endcap, we find
\begin{align}
\bA'_3 [\bFL]_{33} \bKdfthtw^\u
&=
\bA'_3 \bF \frac1{1- \bK(\bF + \bG)} \bKdfthtw^\u
\label{eq:red1} \,,
\\ 
&=
\bA'_3   \bF \frac1{1 - \mathcal I_{33} \otimes }\bKdfthtw^\u
+
\bA'_3    \bF  \frac1{1 - \bK (\bF + \bG)} \left\{
\bK  \bF  \mathcal S 
+ \bGam \bGrD \mathcal I_{\ttil 3} \otimes  \right\} \frac1{1 - \mathcal I_{33} \otimes}\bKdfthtw^\u \,,
\\
&=
\tfrac13 \bA'_3    \bF  \bKdfthtw
+ \bA'_3    \bF  \frac1{1  -   \bK (\bF + \bG)} \bK  \bF  \bKdfthtw
+ \bA'_3  \bF _{3\ttil} \, \delta  \bKdftw
+ \delta_{33} \bA'_{\ttil} \,,
\\
&=
\bA'_3 \bF_{33} \bKdfthtw
+ \bA'_3  \bF_{3\ttil} \, \delta \bKdftw
+ \delta_{33} \bA'_{\ttil}
\,.
\label{eq:symm4}
\end{align}
{The first line recalls the definition of $\bFL$,
while the second substitutes Eq.~(\ref{eq:symmLiterate}).
To obtain the third line we use the definition of $\bF_{3\ttil}$ as well as
the following new definitions:}
\begin{align}
\bKdfthtw &\equiv \mathcal S \frac1{1- \mathcal I_{33} \otimes} \bKdfthtw^\u\,,
\label{eq:Kdf32def}
\\
\delta \bKdftw &\equiv \mathcal I_{\ttil 3} \otimes \frac1{1- \mathcal I_{33} \otimes} \bKdfthtw^\u\,,
\label{eq:deltaKdf22def}
\\
\delta_{33} \bA'_{\ttil} & \equiv \bA'_3 \frac{i\rho}{3\omega} \left(\frac1{1 - \mathcal I_{33} \otimes} \bKdfthtw^\u\right)^{(u-s)} \,.
\label{eq:deltaAp2}
\end{align}
{In addition we use the result from Ref.~\cite{\HSQCa} that a factor of $\bF$
sandwiched between a symmetric object (here $\bA'_3$) and a $(u-s)$ object
can be replaced by $i\rho/(2\omega)$, so that the resulting matrix sum can
be replaced by an integral.}
The final line follows immediately using the definition of $\bF_{33}$.
We see that the symmetrization has produced the desired factors of $\bF_{3\ttil}$ and
$\bF_{33}$, as well as an additional contribution to $\bKdftw$ and to the endcap $\bA'_{\ttil}$.
{An almost identical set of results holds with $\bKdfthtw^\u$ replaced with 
$\bKdfth^\uu$, except that the final index is changed from $\ttil$ to $3$, and an additional
$(u)$ superscript is added.
}

\begin{figure}
\begin{center}
\includegraphics[width=\textwidth]{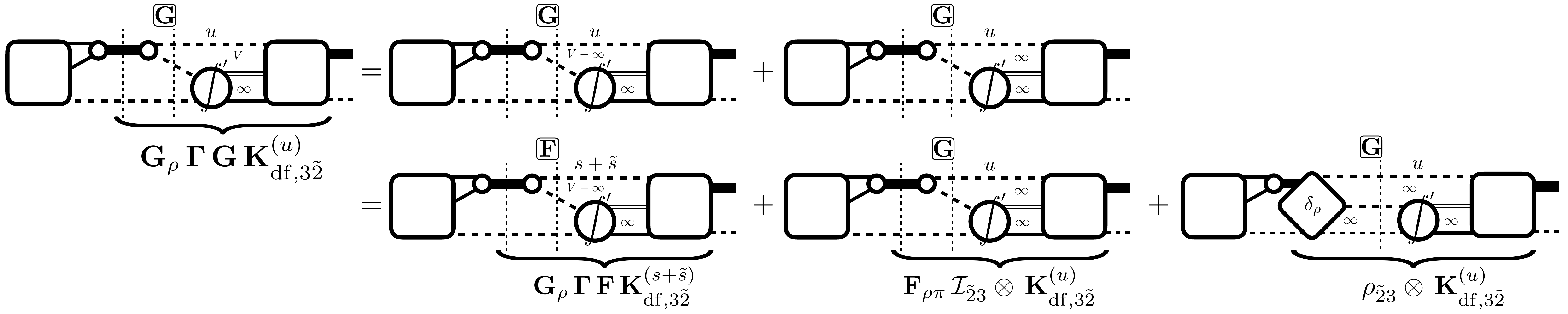}
 \caption{
Derivation of Eq.~(\ref{eq:symm5}), using the notation of Figs.~\ref{fig:CnL0F}
and \ref{fig:CL0Frecursion_v2}.
The left-hand boxes represent $\bA'_{\ttil}$, aside from the loop that is exposed
explicitly to the left of the left-most cut.
The steps are similar to those in Fig.~\ref{fig:symm2}: replacing the sum adjacent to the
$\bG$ with a sum-integral difference and an integral, the former giving rise to an $\bF$.
The difference from Fig.~\ref{fig:symm2} concerns the integral, in which the factor
of $\bGr$ can be converted into an $\bFrp$ cut by projecting the entire quantity
to the right onto the K-matrix pole onto the $\ttil$ mass shell, leading to the $\mathcal I_{\ttil 3}$ term. 
The residue (the $\delta_\rho$ term)
cancels the K-matrix pole, allowing the sum over the momentum $k$ to be replaced
by an integral, so that the implicit $\bA'_{\ttil}$ and the $\bKdfthtw^\u$ are connected
by an infinite-volume integral operator denoted $\rho_{\ttil 3}$.
}
\label{fig:symm3}
\end{center}
\end{figure}

We next consider terms where the endcap is $\bA'_{\ttil}$ or its reflection. 
In this case we need a slightly different symmetrization result,
\begin{equation}
\bA'_{\ttil}  \bGr  \bGam  (\bF + \bG) \bKdfthtw^\u
=
\bA'_{\ttil}  \bGr  \bGam  \bF  \mathcal S \big [ \bKdfthtw^\u \big ]
+ \bA'_{\ttil}  \bFrp \, \mathcal  I_{\ttil 3} \otimes  \bKdfthtw^\u
+ \bA'_{\ttil} \otimes \rho_{\ttil 3} \otimes \bKdfthtw^\u
\,.
\label{eq:symm6}
\end{equation}
This follows from
\begin{equation}
\bA'_{\ttil}  \bGr  \bGam   \bG  \bKdfthtw^\u
=
\bA'_{\ttil}  \bGr  \bGam   \bF  \bKdfthtw^{(s+\tilde s)}
+ \bA'_{\ttil}  \bFrp  \, \mathcal I_{\ttil 3} \otimes \bKdfthtw^\u
+ \bA'_{\ttil} \otimes \rho_{\ttil 3} \otimes \bKdfthtw^\u
\,,
\label{eq:symm5}
\end{equation}
the derivation of which is described in Fig.~\ref{fig:symm3}.
Here $\rho_{\ttil 3}$ is a second type of integral operator that acts both to the left and right,
and is defined in the figure.
It joins $\bA'_{\ttil}$ with $\bKdfthtw^\u$ into an expanded endcap. 
We stress that the results in Eqs.~(\ref{eq:symm6}) and (\ref{eq:symm5}) 
hold when $\bKdfthtw^\u$ is replaced by any unsymmetrized three-particle quantity.

Using the definition of $\bFL$ [Eq.~(\ref{eq:ZLRdef})] and Eqs.~(\ref{eq:symmLiterate}) and (\ref{eq:symm6}),
 we find
\begin{align}
\bA'_{\ttil} [\bFL]_{\ttil 3} \bKdfthtw^\u
&=
\bA'_{\ttil}  \bGr  \bGam  \frac1{1-(\bF + \bG)\bK}(\bF + \bG) \bKdfthtw^\u \,,
\\ &=
\bA'_{\ttil}  \bGr   \bGam   (\bF + \bG)
\left\{
\frac1{1-\bK(\bF + \bG)}\left( \bK  \bF  {\cal S}
+
 \bGam  \bGrD \, \mathcal I_{\ttil 3} \otimes \right)
+ 1
\right\} 
\frac1{1- \mathcal I_{33} \otimes} \bKdfthtw^\u\,,
\\ &=
\bA'_{\ttil} \bF_{\ttil 3} \bKdfthtw + \bA'_{\ttil} \bF_{\ttil\ttil} \delta \bKdftw + 
\delta_{\ttil 3} \bA'_{\ttil}
\,,
\label{eq:symm45}
\end{align}
where
\begin{equation}
\delta_{\ttil 3} \bA'_{\ttil} \equiv \bA'_{\ttil} \otimes \rho_{\ttil 3} \otimes \frac1{1- \mathcal I_{33} \otimes} \bKdfthtw^\u \,.
\end{equation} 
As above, an almost identical equation holds with $\bKdfthtw^\u$ replaced by 
$\bKdfth^\uu$.

Combining Eqs.~(\ref{eq:symm4}), (\ref{eq:symm45}) and their analogs with the
right-hand index changed to $3$, we find
\begin{equation}
\bA' \bFL \bKdf^\u
= 
\bA'
\pmb {\mathcal F} \, \bSL \! \big [ \bKdf^\u \big ]
+ \begin{pmatrix} \delta \bA'_{\ttil} & \delta \bA'^\u_3 \end{pmatrix}
\,,
\label{eq:symmZL}
\end{equation}
where $\pmb {\mathcal F}$ is defined in Eq.~(\ref{eq:bKdfbFMatdef}),
\begin{equation}
\bA' \equiv \begin{pmatrix} \bA'_{\ttil} & \bA'_3 \end{pmatrix}  \,, \qquad
\bSL \equiv \begin{pmatrix} 1 &  \mathcal I_{\ttil 3} \otimes \frac1{1- \mathcal I_{33} \otimes} \\ 
                         0 & \mathcal S \frac1{1 - \mathcal I_{33} \otimes} \end{pmatrix} 
\,,
\end{equation}
and we have introduced
\begin{equation}
\delta \bA'_{\ttil} = \delta_{33} \bA'_{\ttil} +  \delta_{\ttil 3} \bA'_{\ttil} \ \ \  {\rm and}\ \ \ 
\delta \bA'^\u_3 = \bA'_{\ttil} \frac{i\rho}{3\omega} \left(\frac1{1 - \mathcal I_{33}\otimes } \bKdfth^\uu\right)^{(u-s,u)}
+ \bA'_{\ttil} \otimes \rho_{\ttil 3} \otimes \frac1{1 - \mathcal I_{33} \otimes} \bKdfth^\uu\,.
\label{eq:deltaAL}
\end{equation}
Note that $\delta \bA'^\u_3$ inherits a superscript $(u)$ from the right-hand superscript
of $\bKdfth^\uu$. In the following it will be useful to rewrite the shifts in $\bA'$ as
\begin{equation}
\begin{pmatrix} \delta \bA'_{\ttil} & \delta \bA'^\u_3  \end{pmatrix}
= \bA' \otimes \IFL \otimes \bKdf^\u
\,,
\label{eq:IZL}
\end{equation}
where $\IFL$ is a matrix of integral operators.

The result for the $\bFR$ term is given by reflection and is
\begin{equation}
\bKdf^\u \bFR \bA
=
\big [ \bKdf^\u \big ] \bSR \, \pmb {\mathcal F} \bA
+
\begin{pmatrix} \delta \bA_{\ttil} \\ \delta \bA_3^\u \end{pmatrix}
\,,
\label{eq:symmZR}
\end{equation}
where
\begin{equation}
\bA= \begin{pmatrix} \bA_{\ttil} \\ \bA_3 \end{pmatrix}\,,\qquad
\bSR = 
\begin{pmatrix} 1 & 0\\
 \frac1{1- \otimes \mathcal I_{33}^\dagger } \otimes \mathcal I_{3\ttil} & \frac1{1- \otimes \mathcal I_{33}^\dagger} \mathcal S \end{pmatrix}
\,,
\end{equation} 
and $\delta \bA_{\ttil}$ and $\delta \bA_3^\u$ are reflections of the results in
Eq.~(\ref{eq:deltaAL}).
Again, we introduce the matrix of integral operators $\IFR$ such that
\begin{equation}
\begin{pmatrix} \delta \bA_{\ttil} \\ \delta \bA_3^\u \end{pmatrix}
=
\bKdf^\u \otimes \IFR \otimes \bA
\,.
\label{eq:IZR}
\end{equation}

Finally, we turn to the symmetrization {\em between} two factors of $\bKdf^\u$,
i.e. to the analysis of
\begin{equation}
\bKdf^\u \big ( \bX + \bY \big)  \bKdf^\u\,.
\end{equation}
Only the $33$ component of $ \big ( \bX + \bY \big)$ requires new work.
This is because $ [\bX + \bY] _{\ttil\ttil}= \bF_{\ttil\ttil}$ is already symmetrized,
while, since $[\bX + \bY]_{\ttil 3}=[\bF_{\bf L}]_{\ttil 3}$, the analysis for the $\ttil 3$
component is identical to that leading to Eq.~(\ref{eq:symm45}),
with the $3\ttil$ component given by reflection.

The contribution of the $33$ component is analyzed in Appendix \ref{app:w33sym}.
Combined with the results for the other components, we find that
\begin{align}
\bKdf^\u \big ( \bX + \bY \big)  \bKdf^\u
& = 
\bKdf^\u
\begin{pmatrix} 
1 & 0\\ 
\frac1{1- \otimes \mathcal I_{33}^\dagger } \otimes \mathcal I_{3\ttil} & \frac1{1- \otimes \mathcal I_{33}^\dagger} \mathcal S 
\end{pmatrix}
\bFMat
\begin{pmatrix} 
1 &  \mathcal I_{\ttil 3} \otimes \frac1{1- \mathcal I_{33} \otimes} \\ 
0 & \mathcal S \frac1{1 - \mathcal I_{33} \otimes} 
\end{pmatrix}
\bKdf^\u
+ \delta\bKdf^\u \,, 
\label{eq:symmXYA} \\
& = \bKdf^\u \bSR \, \pmb {\mathcal F} \, \bSL \bKdf^\u + \delta\bKdf^\u      \,.
\label{eq:symmXYB}
\end{align}
Many of the complications of the analysis are buried in the final term,
$\delta\bKdf^\u$.
This arises when the two factors of  $\bKdf^\u$ are joined by an integral. 
There are several  contributions to this term---those analogous to $\delta \bA'_{\ttil}$, $\delta \bA'^\u_{3}$, $\delta \bA_{\ttil}$ and $\delta \bA^\u_{3}$,
as well as additional terms discussed in the Appendix.
For this derivation we do not require the detailed form of $\delta \bKdf^\u$.
We only require that it is composed of
infinite-volume quantities, and that the symmetrization
structure of its external indices is the same as that of $\bKdf^\u$.
Again, it is useful to write this term using a matrix of integral operators
\begin{equation}
\delta\bKdf^\u \equiv \bKdf^\u \otimes \IXY \otimes \bKdf^\u
\,.
\label{eq:IXY}
\end{equation}
This emphasizes the fact that $\IXY$ is independent of the detailed form of the
quantities on either side.

\bigskip

We now have all the results to give a final form for the correlator.
Combining Eqs.~(\ref{eq:CLB2Kdf0}), (\ref{eq:CLK}), (\ref{eq:symmZL}),
(\ref{eq:symmZR}) and (\ref{eq:symmXYB}), and performing straightforward but
tedious algebra, we find
\begin{align}
C_L^{[B_2]} &= C_L^{[B_2],\{0\}}+ 
C_L^{[B_2],\bKdf} \,,
\\
&= C_\infty^{[B_2]} + \delta C_\infty^{[B_2]} 
+ \bA'^{[B_2]} \pmb {\mathcal F} \frac1{1-  \bKdf^{[B_2]} \pmb {\mathcal F}}  \bA^{[B_2]}  
\,,
\label{eq:CLB2}
\end{align}
where 
\begin{align}
 \bA'^{[B_2]} & \equiv \bA' + \bA' \otimes \IFL \otimes  \bKdf^u \frac1{1 - \otimes \, \IXY \otimes \bKdf^\u } \bSR \,,
\label{eq:deltaAp}
\\
 \bA^{[B_2]}    & \equiv \bA +  \bSL  \bKdf^\u \frac1{1- \otimes \, \IXY \otimes   \bKdf^\u } \otimes \IFR \otimes \bA \,,
\label{eq:deltaA}
\\
 \bKdf^{[B_2]}  & \equiv  \bSL \bKdf^\u \frac1{1 - \otimes \, \IXY \otimes \bKdf^\u } \bSR\,.
\label{eq:deltaKdf}
\\
\delta C_\infty^{[B_2]} &= 
\delta C_\infty^{[B_2],\{0\}}+ \bA' \otimes \IFL \otimes  \bKdf^\u  \frac1{1 - \otimes \, \IXY \otimes \bKdf^\u } \otimes \IFR \otimes \bA \,.
\end{align}
Equation~(\ref{eq:CLB2}) is the culmination of all the analysis contained in Secs.~\ref{sec:CLB2initaldecom}-\ref{sec:symmKdf}, together with the corresponding appendices, and is by far the most tedious result to derive in all our work on three-particle scattering. Having reached the very final form for all $B_2$-only diagrams, note that we introduce slightly more precise notation, labeling all infinite-volume quantities with
the $^{[B_2]}$ superscript to emphasize the missing $B_3$ kernels. In the next section we show that these are simple to incorporate.


\subsection{Including  three-to-three kernels, $B_3$}
\label{sec:B3}

In order to complete the derivation of Eq.~(\ref{eq:CL}) we must include the
contributions of the three-to-three kernel, $B_3$. This can be done by a straightforward
extension of the method used in Sec.~IVE of Ref.~\cite{\HSQCa}.
As in that work, the essential point is that the analysis described above, which takes
place between endcaps $\bSig$ and $\bSigD$, applies equally well if
one or both of the endcaps are replaced by factors of $i B_3 \equiv \bB$. This is because, like
$\bSig$ and $\bSigD$, $\bB$ is nonsingular in our kinematic regime.
The net result is that we can reuse all the work leading to
Eq.~(\ref{eq:CLB2}).

To do so we rewrite the components of Eq.~(\ref{eq:CLB2}) as
\begin{align}
C_\infty^{[B_2]} + \delta C_\infty^{[B_2]} 
&\equiv \bSig \otimes \DC \otimes \bSigD \,,
\\
\bA'^{[B_2]} &\equiv \bSig \otimes \DAp\,,
\\
\bA^{[B_2]} &\equiv \DA \otimes \bSigD\,,
\\
\pmb \cZ  &\equiv \pmb \cF \frac1{1-  \bKdf^{[B_2]} \pmb \cF}  \,,
\label{eq:calZdef}
\end{align}
in terms of which
\begin{equation}
C_L^{[B_2]} = \bSig \otimes  \left\{ \DC  + \DAp \pmb \cZ \DA \right\}  \otimes \bSigD \,.
\end{equation}
Here $\DC$, $\DAp$ and $\DA$ are infinite-volume decoration operators
that contain the complicated contributions worked out above.\footnote{%
In Ref.~\cite{\HSQCa} the corresponding decoration operators were given superscripts,
but here we drop these for the sake of brevity.}
Note that $\DA$ 
and $\DAp$ are, respectively, $2\times 1$ and $1\times 2$ matrices.
All we need to know in this section is that the decoration operators are well defined,
and apply just as well when the endcaps are replaced by factors of $\bB$.

The full finite-volume correlator, including all possible $B_2$ and $B_3$ insertions, can now be written
\begin{equation}
C_L =  \bSig \otimes \left\{ \DC  + \DAp \pmb \cZ \DA \right\}   \otimes
\sum_{n=0}^\infty \left(\bB \otimes \left\{ \DC  + \DAp \pmb \cZ \DA \right\}  \otimes \right)^n \bSigD
\,.
\end{equation}
Rearranging the series in powers of $\pmb \cZ$ we find
\begin{align}
C_L &= C_\infty + \sum_{n=0}^\infty \bA' \pmb \cZ  \left(  \bKdf^{[B_3]} \pmb \cZ  \right)^n \bA\,,
\end{align}
where we have defined the infinite-volume quantities
\begin{align}
C_\infty & \equiv \bSig \otimes \DC \otimes  \sum_{n=0}^\infty \left( \bB \otimes \DC \, \otimes \right)^n \bSigD \,,
\\
\bA' & \equiv \bSig \otimes \sum_{n=0}^\infty \left( \DC \otimes \bB \, \otimes \right)^n \DAp\,,
\label{eq:deltaApB3}
\\
\bA &= \DA \otimes \sum_{n=0}^\infty \left( \bB \otimes \DC \, \otimes \right)^n \bSig \,,
\label{eq:deltaAB3}
\\
\bKdf^{[B_3]} & \equiv  \DA \otimes \bB \otimes \sum_{n=0}^\infty \left( \DC \otimes \bB \, \otimes \right)^n \DAp\,.
\end{align}
Inserting the definition of $\pmb \cZ$, Eq.~(\ref{eq:calZdef}), into the result for $C_L$, and
rearranging, we reach the final form given in Eq.~(\ref{eq:CL}) above. In terms of our boldface quantities it reads
\begin{align}
C_L &= C_\infty + \bA' \pmb \cF  \frac1{1- \bKdf \pmb \cF } \bA\,,
\end{align}
where
\begin{equation}
 \bKdf \equiv  \bKdf^{[B_2]}  + \bKdf^{[B_3]} \,.
\label{eq:deltaKdfB3}
\end{equation}

\section{Relating $\cK_\df$ to the three-particle scattering amplitude \label{sec:KtoM}}


Having completed the derivation of the quantization condition, i.e.~the relation between the finite-volume spectrum and $\cK_\df$, we now turn to relating the latter to the physical three-to-three scattering amplitude, $\mathcal M_3$. Following Ref.~\cite{\HSQCb}, we derive equations relating $\cK_\df$ to $\mathcal M_3$ in two steps. First, in Sec.~\ref{sec:KtoML} we give a modified version of our main result, Eq.~(\ref{eq:CL}), in terms of a new finite-volume correlator, denoted $\mathcal M_{L,3}$. Second, in Sec.~\ref{sec:Linf}, we analytically study a carefully-defined $L \to \infty$ limit in which $\mathcal M_{L,3} \rightarrow \mathcal M_3$. The result is a series of integral equations relating the divergence-free K matrix to the scattering amplitude. In this section we return to the notation of Sec.~\ref{sec:summary} in which factors of $i$ and $1/(2 \omega L^3)$ are displayed explicitly.


\subsection{Relating $  \cK_{\df}$ to $  \mathcal M_{L,3}$}
\label{sec:KtoML}

In order to relate the components of $\cK_\df$ to physical quantities, we need to
determine the volume-dependence of $\cM_{L,3}$,  
first introduced in Ref.~\cite{\HSQCb}.
$\cM_{L,3}$ differs from $C_L$ in two ways. First, the diagrams have three on-shell,
amputated propagators on each end, rather than the generic operators $\mathcal O(x)$ and $\mathcal O^\dagger(x)$ included in Eq.~(\ref{eq:CLintermsofOO}). 
Second, we allow the momenta of these external particles to be arbitrary, and not
constrained to lie in the finite-volume set.
As discussed at length in Ref.~\cite{\HSQCb},
the latter property is necessary in order to take the infinite-volume limit.
Despite these differences we argue here that we can obtain the result for
$\cM_{L,3}$ from that for $C_L$, Eq.~(\ref{eq:CL}).

We rely on several key observations from Ref.~\cite{\HSQCb}, where, we recall,
$\cM_{L,3}$ was analyzed for systems without poles in $\cK_2$.
The first is that $C_L$ contains all the diagrams contributing to $\cM_{L,3}$.
The task is to separate these out. In particular, we need contributions in which 
three particles are on shell, rather than part of an unconstrained loop sum.
The second observation is that, in the final form for $C_L$, on-shell three-particle states
occur whenever there is a factor of $F$ or $G$.
In particular, if we take the expression for $C_L$ and restrict attention to terms with at least two
$F$ or $G$ ``cuts", then the expression lying between the outermost such cuts 
will contain all contributions to $\cM_{L,3}$.
It will turn out that the outermost cuts are always factors of $F$ rather than $G$.
The third observation is that amputation is effected by removing the external factors of $iF$
and multiplying by $2 \omega L^3$ on the left and right.
After doing so, the result is equal to $\cM_{L,3}$ aside from two final adjustments.
The first is to drop disconnected contributions, and the second is to
symmetrize. We discuss these two relatively minor steps in more detail below.

In fact, Ref.~\cite{\HSQCb} did not apply these observations to the final result for
$C_L$, but rather to an intermediate result. Additional analysis was then required to obtain
the final expression for $\cM_{L,3}$. It was noted that the result for $\cM_{L,3}$ could have been
obtained by applying the amputation procedure directly to the decomposition of $C_L$,
but it was argued that this was a mnemonic rather than a rigorous procedure
(see footnote 10 of Ref.~\cite{\HSQCb}).
We now think, however, that inferring the form of $\cM_{L,3}$ from $C_L$, by directly converting the final result is
justified, and indeed that the work of Ref.~\cite{\HSQCb} supports this claim.
We explain additional justification for this new approach below, once we have obtained the result for $\cM_{L,3}$.

\bigskip

Due to the presence of poles in $\K_2$, the procedure described above must be
amended. To understand the issue, we focus on the contribution to $C_L$
arising from a single insertion of the $3\ttil$ component of $\cF$, namely
\begin{align}
i A'_3 i F_{3\ttil} i A_{\ttil} &=
 i A'_3 \frac{1}{2 \omega L^3} i F \frac1{1 - i\K_2(iF\!+\!iG)} i\Gamma_J i G_\rho^\dagger i A_{\ttil} \,,
 \\
 \begin{split}
&= i A'_3 \frac{1}{2 \omega L^3} i F  i\Gamma_J i G_\rho^\dagger i A_{\ttil}
+
 i A'_3 \frac{1}{2 \omega L^3} i F i\K_2 (iF\!+\!iG) i\Gamma_J i G_\rho^\dagger i A_{\ttil} \\
& \hspace{170pt} + 
 i A'_3 \frac{1}{2 \omega L^3} i F i\K_2 (iF\!+\!iG) i\K_2 (iF\!+\!iG)i\Gamma_J i G_\rho^\dagger i A_{\ttil}
 + 
 \cdots
 \end{split}
 \,.
\label{eq:ex32}
\end{align}
The first term in Eq.~(\ref{eq:ex32}) can be dropped as it has only one $F$ or $G$ cut.
The second term has two such cuts, but only a single $\K_2$ lies between them,
so this corresponds to a disconnected contribution to $\cM_{L,3}$.
Thus this term is also dropped. 
The third term has two external cuts, and part of the contribution between them
is connected, namely the $i \K_2 i G i \K_2$ part.
However, such a contribution is already contained in the
$i A'_3 i F_{33} i A_3$ term, as is readily checked. A signal for this double counting is
that there is a $\rho\pi$ cut, $G_\rho^\dagger$, that
is external relative to the right-hand $F/G$-cut in each of the terms in Eq.~(\ref{eq:ex32}).
Indeed, one can show that  the complete set of contributions to $\cM_{L,3}$
are obtained by taking only terms in which the outermost cut contains three particles
rather than the $\rho\pi$ effective channel. This extra criterion implies that none of the terms
in Eq.~(\ref{eq:ex32}) should be kept. 

The same conclusion holds for single insertions of $F_{\ttil 3}$ or $F_{\ttil\ttil}$, 
which have, respectively, one and two external $\rho\pi$ cuts. 
Thus the only surviving contribution from a single insertion of $\cF$ is that from $F_{33}$.
This contribution is unaffected by the presence of poles in $\cK_2$, 
and so is unchanged from that obtained in Ref.~\cite{\HSQCb}.
We recall briefly how this term is obtained. 
Using the result for $F_{33}$, Eq.~(\ref{eq:F3def}), we find that the term with
at least two three-particle cuts is
\begin{align}
C_L &\supset i A'_3 iF_{33} i A_3 
 \supset i A'_3 \frac{1}{2 \omega L^3} i F \frac1{1 - i\K_2 (i F+i G)} i \K_2 i F i A_3
 \,.
 \label{eq:ex33}
\end{align}
Applying the recipe given above we obtain
\begin{align}
i\cM_{L,3} &\supset
\left\{ \frac1{1 - i\K_2 (i F+i G)} i\K_2  [2 \omega L^3] \right\}\Bigg|_{\rm connected,\ symmetrized}
\,.
\end{align}
The disconnected part is that obtained by setting $G\to 0$. We can remove this,
and at the same time make contact with the notation of Ref.~\cite{\HSQCb}, 
using the identity
\begin{align}
 \frac1{1 - i\K_2 (i F+i G)} i\K_2 [2 \omega L^3] &= 
 i\cM_{L,2} [2 \omega L^3]  + i\cD_L^\uu\,,
 \label{eq:idenML2}
 \end{align}
 where
 \begin{align}
 i\cM_{L,2} &= \frac1{1 - i \cK_2 i F} i \cK_2
 \,,
 \label{eq:ML2def}
 \\
i\cD_L^\uu &= \frac1{1-i\cM_{2,L} iG}  i\cM_{L,2} i G  i\cM_{L,2} [2 \omega L^3] \,.
\label{eq:DLuudef}
\end{align}
The first term in Eq.~(\ref{eq:idenML2}) contains no switches and thus leads to
a disconnected contribution. 
The second term contains at least one switch and thus is connected; it agrees
with the quantity of the same name given in Eq.~(25) of Ref.~\cite{\HSQCb}.
Thus we find
\begin{align}
i\cM_{L,3} &\supset
\left\{ i\cD_{L,2}^\uu\right\}\Bigg|_{\rm symmetrized} 
\equiv \cS  i \cD_{L,2}^\uu \cS
\,.
\label{eq:ML3part1}
\end{align}

We can now explain the need for symmetrization. In the original expression,
$i A'_3 i F_{33} i A_3$, the endcaps $A'_3$ and $A_3$ are fully symmetrized, 
as described earlier in the derivation. By this we mean that the quantities are invariant under interchange of any of the three particle momenta. The adjacent factors of $F$ that will be removed
are not, however, symmetric, since they single out one of the on-shell particles as the
spectator. Similarly, what lies between the two amputated $F$s is not symmetric.
Within $C_L$ this does not matter, because of the symmetry of the endcaps.
But $\cM_{L,3}$ is defined to be symmetric on the external lines, and to reproduce this we must
sum over all permutations of the three incoming and three outgoing particle momenta. However, it turns out that all quantities entering these expressions are already invariant under interchange of the two non-spectators, so that one need only sum over the remaining three distinct permutations. The precise action of the symmetrization operators is described
by the paragraph containing Eqs.~(35)-(37) in Ref.~\cite{\HSQCb}.\footnote{%
Note that, in that work, the symmetrization operators acting to the right and left
are packaged into a single overall symmetrization operator. }

Now we apply the updated rules to the terms in Eq.~(\ref{eq:CL}) having more
than one factor of $ \cF $, namely
\begin{equation}
C_L \supset i A'  i\cF \frac1{1 -  i\K_\df i\cF }  i\cK_\df i\cF iA
\,.
\end{equation}
Since we are requiring an external three-particle cut rather than a $\rho\pi$ cut, 
only $F_{33}$ and $F_{3\ttil}$ contribute from the left-hand $ \cF $,
and only $F_{33}$ and $F_{\ttil 3}$ contribute from the right-hand $ \cF $.
Thus we find that the contribution to $\cM_{L,3}$ is
\begin{align}
i \cM_{L,3} &\supset
\cS  \cL^\u_L \frac1{1 -  i\K_\df i\cF }  i\cK_\df    \cR^\u_L \cS\,,
\label{eq:ML3part2}
\\
\cL^\u_L &=\begin{pmatrix}  \frac1{1-i\K_2(iF+iG)} i\Gamma_J iG_\rho^\dagger  \ \ & \ \  \frac13   + \frac1{1-i\K_2 (iF+iG)} i\K_2 i F \end{pmatrix} \label{eq:cLdef}\,,
\\
\cR^\u_L &= \begin{pmatrix} 
iG_\rho i\Gamma_J \frac1{1-(iF+iG^\dagger) i\K_2} \\
\frac13+ iF i\K_2\frac1{1-(iF+iG^\dagger) i\K_2} 
\end{pmatrix}   \label{eq:cRdef}\,.
\end{align}
Here $G^\dagger =[2 \omega L^3]^{-1} G [2 \omega L^3]$, as follows from the
definition of the matrix $G$, Eq.~(\ref{eq:Gdef}).
Combining this result with that from Eq.~(\ref{eq:ML3part1}) leads to the
full expression for $\cM_{L,3}$
\begin{equation}
\mathcal M_{L,3} = \mathcal S \left\{ \mathcal D^{(u,u)}_L +
  \mathcal L^{(u)}_L   \mathcal K_{\mathrm{df}}  \frac{1}{1 +
     \mathcal F    \mathcal K_{\mathrm{df}}  } \mathcal R^{(u)}_L \right\}
\mathcal S \,.
\label{eq:ML3}
\end{equation}
Here we have multiplied various factors of $i$ together and divided both sides of the equation by $i$. We stress again that no factors of $i$ or $2\omega L^3$ have been absorbed here by redefinitions.

A consistency check on this derivation is that the external factors that
are ``amputated" to obtain $\cM_{L,3}$ from $C_L$ are the same for both
Eqs.~(\ref{eq:ML3part1}) and (\ref{eq:ML3part2}), namely $iA'_3 \frac{1}{2 \omega L^3} iF$ on
the left and $iF \frac{1}{2 \omega L^3} iA_3$ on the right.\footnote{Note that the matrices $\frac{1}{2 \omega L^3}$ and $F$ commute, though neither commutes with $G$.} 

Finally, we return to the issue of why we now think the above procedure for
obtaining $\cM_{L,3}$ from $C_L$ is valid.
We raised two concerns in Ref.~\cite{\HSQCb}.
The first was essentially that the infinite-volume quantities appearing in $C_L$
resulted from a sequence of redefinitions, obscuring the
relation to the underlying diagrams. 
Here we have been able to give a more explicit form for
these redefinitions, i.e.~those in Eqs.~(\ref{eq:deltaAp})-(\ref{eq:deltaKdf}),
(\ref{eq:deltaApB3}), (\ref{eq:deltaAB3}) and (\ref{eq:deltaKdfB3}).
This gives us confidence that there are no subtleties in picking out the parts of
the diagrams that contribute to $\cM_{L,3}$.
The second concern was that the symmetrization procedure after amputation was
not justified. We have now convinced ourselves, as described above, that it
is correct.


\subsection{Applying the formal $L \to \infty$ limit to relate $\cK_{\df}$ to the three-to-three scattering amplitude}
\label{sec:Linf}

We are now ready to apply the $L \to \infty$ limit to Eq.~(\ref{eq:ML3}), and thereby derive an integral equation relating $\cK_\df$ to the physical three-to-three scattering amplitude, $\cM_3$. We begin by recalling the expression for $\cF$ [Eqs.~(\ref{eq:Fmat_def}), (\ref{eq:F3def}), (\ref{eq:F22def}), (\ref{eq:F23def}) and (\ref{eq:F32def})]
\begin{align}
\cF  &=
\begin{pmatrix}
  F_{\ttil \ttil} &   F_{ \ttil 3} \\ 
 F_{ 3 \ttil } &   F_{33}
\end{pmatrix}
\,,
\\
 F_{\ttil\ttil} &\equiv  F_{\rho\pi} 
+  G_\rho \Gamma_J \frac{1}{2 \omega L^3} (F+G) \frac1{1+ \K_2(F+G)} \Gamma_J  G_\rho^\dagger\,,
\\
 F_{\ttil 3} &\equiv - G_\rho \Gamma_J \frac{1}{2 \omega L^3} \frac1{1+ (F+G) \K_2} F\,,
\\
 F_{3\ttil} &\equiv - \frac{1}{2 \omega L^3}  F \frac1{1 + \K_2 (F+ G)} \Gamma_J G_\rho^\dagger \,, \\
  F_{33} & = \frac{1}{2 \omega L^3} \bigg [ \frac{F}{3} - F  \frac1{1 + \K_2 (F+G)} \K_2  F \bigg ] \,.
\end{align}
Here we have again combined various factors of $i$ to simplify the expressions.

The method we use is that developed in Ref.~\cite{\HSQCb}.
We want to take $L\to \infty$ in such a way that $\cM_{L,3}$ goes over to $\cM_3$.
This requires that that we first regularize poles in integrands with the $i\epsilon$
prescription, and then take the $L \to \infty$ limit with $\epsilon$ held fixed.
As explained in Ref.~\cite{\HSQCb},
this limit sends $F\to \rho$ (since $F^{i\epsilon}\to 0$), and $\cM_{L,2}\to \cM_2$.
Matrix products, combined with factors of $1/L^3$, go over to integrals.
We also need to introduce $G^\infty$, defined by
\begin{equation}
 G^\infty_{\ell' m',\ell m}(\vec p, \vec k)
=
\cY_{3,\ell'm'}(\vec k^{\,*}_{2,p})  {\bf S}^{i\epsilon}_3(\vec p,\vec k) \cY^*_{3,\ell m}(\vec p^{\,*}_{2,k})
\,.
\end{equation}

In Ref.~\cite{\HSQCb}, the only poles present were the three-particle poles in ${\bf S}_3$.
Here we also have the possibility of K-matrix poles, 
which are present in $G_\rho$ and $F_{\rho\pi}$ as well as in $\K_2$ itself.
However, we know that K-matrix poles cannot be present in $\cM_3$, 
because poles on the real axis of scattering amplitudes would imply a violation of unitarity.
In fact, we will show that they are absent also in $\cM_{L,3}$, 
so that there is no need to regularize them.

To see the absence of K-matrix poles  we begin by rewriting Eq.~(\ref{eq:cRdef}) as
\begin{align}
 \cR^\u_L &= \begin{pmatrix} 
 - G_\rho  \Gamma_J \frac1{1 +F  \K_2}  \\
\frac13 -  F  \cM_{L,2} 
\end{pmatrix} \frac1{1 +  G^\dagger  \cM_{L,2}}\,.
\end{align}
Here we recall that
\begin{equation}
 \cM_{L,2} = \frac1{1 + \K_2  F}  \K_2 = \frac1{\K_2^{-1}  + F} 
\,,
\end{equation}
which shows explicitly that poles in $\K_2$ do not lead to poles in $\cM_{L,2}$.
The same cancellation occurs for the poles in $G_\rho$:
\begin{equation}
- G_\rho  \Gamma_J \frac1{1 +  F  \K_2}
= - G_\rho  \Gamma_J  \K_2^{-1}  \cM_{L,2}
\,.
\end{equation}
Since $\Gamma_J$ is a constant, $G_\rho \Gamma_J \K_2^{-1}$ is smooth at the pole position.
It is also a known quantity, assuming that we know $\K_2$ from the two-particle quantization condition, and has a well-defined infinite-volume limit.

We can similarly rewrite the other quantities involving $G_\rho$ (or its hermitian conjugate)
in such a way that they are manifestly free of K-matrix poles:
\begin{align}
\cL^\u_L &= \frac1{1 + \cM_{L,2}  G}
\begin{pmatrix} - \cM_{L,2}   \K_2^{-1}  \Gamma_J  G_\rho^\dagger \ \ & \ \ 
\frac13 -  \cM_{L,2}   F \end{pmatrix}\,,
\\
  F_{\ttil 3} &= 
- G_\rho  \Gamma_J  \K_2^{-1}  \cM_{L,2} \frac1{1 + G^\dagger  \cM_{L,2}} \frac{1}{2 \omega L^3} F
=
- G_\rho  \Gamma_J \K_2^{-1} \frac{1}{2 \omega L^3} \frac1{1 + \cM_{L,2} G}  \cM_{L,2}  F
\,,
\\
F_{3\ttil} &= -  \frac{1}{2 \omega L^3} F \frac1{1 + \cM_{L,2}  G}  \cM_{L,2} 
 \K_2^{-1}  \Gamma_J  G_\rho^\dagger
\,.
\end{align}
This leaves $F_{\ttil \ttil}$, which contains $F_{\rho\pi}$. This can be rewritten as
\begin{equation}
F_{\ttil\ttil} = 
- G_\rho  \Gamma_J  \K_2^{-1} 
 \frac{1}{2 \omega L^3} 
\frac1{1 +  \cM_{L,2} G} \cM_{L,2}  \cK_2^{-1} \Gamma_J G_\rho^\dagger
+
\left\{ F_{\rho\pi} + G_\rho  \Gamma_J \frac{1}{2 \omega L^3}  \K_2^{-1}  \Gamma_J G_\rho^\dagger\right\}
\,.
\end{equation}
The first term is manifestly free of K-matrix poles.
For the term in curly braces, the poles also cancel.
To see this we note that $F_{\rho\pi}$ contains a sum over spectator momenta,
which is matched in the $G_\rho [\cdots] G_\rho^\dagger$ part by the sum over matrix indices.
The infinite-volume limit of this term is known given knowledge of $\cK_2$.

The final quantity to be considered is $F_{33}$. Here the absence of K-matrix poles
is manifest, but it is still useful to rewrite it as
\begin{equation}
F_{33} = \frac{1}{2 \omega L^3} F \left[ \frac13
- \frac1{1 + \cM_{L,2}  G}  \cM_{L,2}  F\right]
\,.
\end{equation}

\bigskip
It is now a tedious but straightforward exercise to take the infinite volume limit of
Eq.~(\ref{eq:ML3}).
We first introduce useful infinite-volume quantities
\begin{align}
G_\rho \Gamma_J  \cK_2^{-1} 
&\xrightarrow{L\to\infty} 
\overline G_{\rho;M'_J\ell'm';\ell m}(\vec k)\,,
\\
\cK_2^{-1} \Gamma_J G_\rho^\dagger
&\xrightarrow{L\to\infty} 
\overline G^\dagger_{\rho;\ell m,M_J \ell' m'}(\vec k)\,,
\\
\left\{  F_{\rho\pi} +  G_\rho  \Gamma_J \frac{1}{2 \omega L^3} \K_2^{-1} \Gamma_J G_\rho^\dagger\right\}
&\xrightarrow{L\to\infty} 
\overline F_{\rho\pi;M'_J\ell'm';M_J\ell m}
\,.
\end{align}
We note that these quantities contain information about the spin of the resonance;
for example, $\overline G_\rho$ contains a factor of $\delta_{J \ell}$.
All three quantities are determined by $\K_2$.

The matrix $[1 +   \cM_{2,L} G ]^{-1}$ occurs repeatedly. In the $L\to\infty$ limit, multiplication
by this matrix is replaced by integration with the 
$\cU(\vec p, \vec k)_{\ell'm';\ell m}$, which solves the integral equation
\begin{equation}
\cU(\vec p,\vec k) = (2\pi)^3 \delta^3(\vec p-\vec k) -
\int_s   \cM_2(\vec p) G^\infty(\vec p, \vec s)\frac{1}{2\omega_s} \cU(\vec s, \vec k)
\,.
\end{equation}
Here $\int_s\equiv\int d^3s/(2\pi)^3$, 
and we are keeping the angular-momentum indices implicit.

We next construct the infinite-volume limits of the elements of $ \cF $.
Pulling out overall factors of $1/L^3$ that will turn sums into integrals, we find that
these limits give
\begin{align}
\overline F_{\ttil \ttil} &= - 
\int_s \int_t
 \overline G_\rho(\vec s) \frac1{2\omega_s} \;\cU(\vec s,\vec t)  \cM_2(\vec t) 
 \overline G_\rho^\dagger(\vec t)
+   \overline F_{\rho\pi} \,,
\\
 \overline F_{\ttil 3}(\vec k) &= -
\int_s \frac1{2\omega_s}
 \overline G_\rho(\vec s) \;\cU(\vec s, \vec k)  \cM_2(\vec k)  \rho(\vec k)
\,,
\\
 \overline F_{3 \ttil}(\vec p) &= -
\frac{ \rho(\vec p)}{2\omega_p}
\int_s 
 \;\cU(\vec p, \vec s)  \cM_2(\vec s)   \overline G^\dagger_\rho(\vec s)
\,,
\\
 \overline F_{33}(\vec p,\vec k) &= 
\frac{\rho(\vec p)}{6\omega_p} (2\pi)^3\delta^3(\vec p-\vec k)
-
\frac{ \rho(\vec p)}{2\omega_p} \cU(\vec p,\vec k) \cM_2(\vec k)  \rho(\vec k)
\,.
\end{align}
All these quantities can be determined given knowledge of $\cM_2$.
We also recall that $\rho(\vec k)$ contains the cutoff function $H(\vec k)$, so that all
integrals have finite range.

The next stage is to determine the limit of $ \K_\df (1+ \cF \K_\df )^{-1}$,
which we call $ \cT $. This leads to two pairs of coupled matrix-integral equations 
for the components of $ \cT $. The first pair is
\begin{align}
 \cT_{\ttil\ttil} & =  \K_{\df,\ttil\ttil}
 -   \K_{\df,\ttil\ttil}  \overline F_{\ttil\ttil}  \cT_{\ttil\ttil}
 -  \int_t   \K_{\df,\ttil\ttil}  \overline F_{\ttil 3}(\vec t)  \cT_{3\ttil}(\vec t)
 -  \int_s   \K_{\df,\ttil 3}(\vec s)  \overline F_{3\ttil}(\vec s)  \cT_{\ttil\ttil}
 -  \int_{s,t}   \K_{\df,\ttil 3}(\vec s)  \overline F_{33}(\vec s,\vec t)  \cT_{3\ttil}(\vec t)\,, \\[5pt]
 \begin{split}
 \cT_{3\ttil}(\vec p) & =  \K_{\df,3\ttil}(\vec p)
 -   \K_{\df,3\ttil}(\vec p)  \overline F_{\ttil\ttil}  \cT_{\ttil\ttil}
 -  \int_t   \K_{\df,3\ttil}(\vec p)  \overline F_{\ttil 3}(\vec t)  \cT_{3\ttil}(\vec t)
 -  \int_s   \K_{\df,3 3}(\vec p,\vec s)  \overline F_{3\ttil}(\vec s)  \cT_{\ttil\ttil}\\
 & \hspace{300pt} -  \int_{s,t}   \K_{\df,3 3}(\vec p,\vec s)  \overline F_{33}(\vec s,\vec t)  \cT_{3\ttil}(\vec t)\,.
 \end{split}
\end{align}
The second pair is a straightforward generalization given by replacing all rightmost $\ttil$ indices with $3$ indices and including the appropriate additional momentum dependencies. 

Finally, given $ \cT $ we can obtain $\cM_3$ by doing integrals.
The contribution of $\cD_L^{(u,u)}$ is unchanged from Ref.~\cite{\HSQCb}.
We obtain it using
\begin{equation}
 \cD^{(u,u)}(\vec p, \vec k) = -
\int_s \cU(\vec p,\vec s)  \cM_2(\vec s)  G^\infty(\vec s,\vec k)  \cM_2(\vec k)
\,.
\end{equation}
For the remaining term we multiply $ \cT $ on the left with
\begin{equation}
\overline {\mathcal L}_\infty^\u(\vec p, \vec s) \equiv \begin{pmatrix} \ \  -    \cU(\vec p,  \vec s)  \cM_2(\vec s)  \overline G_\rho(\vec s) \ \ & \ \ 
  \cU(\vec p, \vec s) [\tfrac13 -   \cM_2(\vec s)  \rho(\vec s)] \ \ \end{pmatrix}
\,,
\end{equation}
and integrate the $\vec s$ coordinate. Similarly we multiply with the conjugate, $\overline {\mathcal R}_\infty^\u(\vec  t, \vec k)$, on the right and integrate again to reach
\begin{equation}
\mathcal M^\uu_{3}(\vec p, \vec k) 
= \cD^{(u,u)}(\vec p, \vec k) +  \int_{s,t} \overline {\mathcal L}_\infty^\u(\vec p, \vec s) \mathcal T(\vec s, \vec t)  \overline {\mathcal R}_\infty^\u(\vec t, \vec k)  \,.
\end{equation}

This result can then be converted to a function of the three incoming and three outgoing momenta via
\begin{equation}
\mathcal M^\uu_{3}(\vec p , \vec a', \vec b'; \vec k, \vec a, \vec b)  \equiv 4 \pi Y_{\ell' m'}(\hat a'^*_{2,p}) \mathcal M^\uu_{3; \ell' m'; \ell m}(\vec p, \vec k) Y^*_{\ell m}(\hat a^*_{2,k})  \,,
\end{equation}
where $\vec b' \equiv \vec P - \vec p - \vec a'$ and $\vec b \equiv \vec P - \vec k - \vec a$,
and we have restored the angular momentum indices on $\cM_3^\uu$ on the right-hand side.
Finally, the physical scattering amplitude is reached by symmetrizing
\begin{equation}
\mathcal M_{3}(\vec p , \vec a', \vec b'; \vec k, \vec a, \vec b)  = \mathcal S \big [    \mathcal M^\uu_{3}     \big ] \mathcal S \equiv \sum_{\vec p_1, \vec p_2, \vec p_3 \in \mathcal P_{\vec p}} \sum_{\vec k_1, \vec k_2, \vec k_3 \in \mathcal P_{\vec k}} \mathcal M^\uu_{3}(\vec p_1, \vec p_2, \vec p_3 ; \vec k_1, \vec k_2, \vec k_3) \,,
\end{equation}
where
\begin{align}
 \mathcal P_{\vec p} \equiv \big \{  \{ \vec p, \vec a', \vec b'\},  \{ \vec b', \vec p, \vec a'\},  \{ \vec a', \vec b', \vec p\} \big \} \,, \ \ \text{and} \ \  \mathcal P_{\vec k} \equiv \big \{  \{ \vec k, \vec a, \vec b\},  \{ \vec b, \vec k, \vec a\},  \{ \vec a, \vec b, \vec k\} \big \} \,.
\end{align}

\section{Conclusion \label{sec:conclusion}}


In this work we have lifted the final major restriction on our formalism relating finite-volume energies to relativistic two- and three-particle scattering amplitudes. To summarize, at this stage we have the building blocks to treat any system of identical scalar particles. Our results fall into three classes: 
\begin{enumerate}
\item $\textbf 3 \to \textbf 3$ scattering assuming a $\mathbb Z_2$ symmetry and no subchannel resonances (i.e.~no poles in $\mathcal K_2$, see Refs.~\cite{\HSQCa,\HSQCb}), 
\item $\{\textbf 2, \textbf 3 \} \to \{\textbf 2, \textbf 3 \}$ scattering in the case of no $\mathbb Z_2$ symmetry and, again, no subchannel resonances (see Ref.~\cite{\BHSQC}), 
\item $\textbf 3 \to \textbf 3$ scattering for systems with a pole in $\mathcal K_2$ (this work).
\end{enumerate}

To complete the formalism for all two- and three-particle systems of identical scalars, it remains only to extend item 3 to any number of $\mathcal K_2$ poles in any angular momentum channels, and then to combine items 2 and 3 to describe $\textbf 2 \to \textbf 3$ systems with resonant subprocesses. Beyond this, the remaining extensions to general two- and three-particle systems require incorporating non-identical and non-degenerate particles, multiple two- and three-particle channels and, finally, particles with spin. Based on the structure of the results derived so far and on our experience with two-particle quantization conditions, we expect that all of these extensions will be significantly easier than the derivation presented here.

The approach detailed in this article requires treating the pole in $\mathcal K_2$ as a pseudoparticle and constructing an effective two-particle state, labeled $\ttil$, built from the pole together with the remaining spectator. From this set-up we have derived a quantization condition in the usual form of a determinant involving a finite-volume matrix, $\mathcal F$, and a divergence-free K matrix, $\K_\df$, both of which carry matrix indices on the $\ttil + 3$ effective channel space. The final aspect of the result presented here is the relation between $\K_\df$ and the physical scattering amplitude, denoted $\mathcal M_3$. The latter has the usual degrees of freedom and in particular carries no memory of the unphysical $\ttil$ channel.

One of the central questions raised by this derivation, to be further explored in future work, is whether it is really neccesary or natural to explicitly treat the $\mathcal K_2$ poles as we have done. One motivation for this approach follows from considering, e.g., isospin two $\pi \pi \pi$ scattering for varying quark masses. For physical-mass pions, in the allowed energy range of $3 M_\pi < \sqrt{s} < 5 M_\pi$, the energy of the $\pi \pi$ subsystem is well below the $\rho$ mass and therefore well below any poles in $\mathcal K_2$. Thus, for this system, the formalism of item 1 above is appropriate. By contrast, for sufficiently heavy pions the $\rho$ is stable so that one requires the formalism of item 2 to describe the $\rho \pi \to \pi \pi \pi$ scattering amplitude. The latter depends on a two-channel version of $\mathcal K_\df$ represented by a two-by-two matrix with indices $2$ and $3$.\footnote{Strictly speaking the only available $\textbf 2 \to \textbf 3$ formalism requires that all particles in the two- and three-particle states are identical. However based on the nature of the derivation, and the corresponding results in the two-particle sector, it is quite clear that the basic structure of the quantization conditions, in particular the appearance of channel indices, will persist in the case of non-identical particles.} Since one can, at least in principle, vary the quark masses continuously between these two scenarios, it is necessary to understand how the quantization conditions transition between the two different matrix spaces. 

The result of this work provides a natural answer to this question. As the quark mass increases from the physical point, the $\rho$ pole moves into the sampled energy range and the corresponding pole in $\mathcal K_2$ is treated by opening an effective $\ttil$ channel. If the quark mass is further increased, this pole location moves closer to the two-particle threshold until it drops below, leading to a stable $\rho$. Note that, even for the case of $M_\rho < 2 M_\pi$, if the mass hierarchy is such that $\kappa^2 \equiv  M_\pi^2 - M_\rho^2/4  \ll M_\pi^2$, i.e.~the state is shallow, then the quantization condition derived here should be used to properly incorporate potentially large volume effects of the form $e^{- \kappa L}$, arising from the large size of the weakly bound state. If the quark masses are instead chosen very large, such that $\kappa > M_\pi$, then the finite size of the $\rho$ can be neglected and the two-to-three formalism may be applied. 

We further remark that the key difference between the case of the unphysical $\ttil$ and the physical $2$ channels is that the off-diagonal elements of $\mathcal F$ vanish only in the latter case. Future work is needed to understand exactly how the result derived here can be used to recover to the case of physical $\textbf 2 \to \textbf 3$ scattering considered previously. Conversely, we recall that the elimination of off-diagonal elements in the $\textbf 2 \to \textbf 3$ formalism of Ref.~\cite{\BHSQC} required construction of the cutoff function $H(\vec k)$ such that the finite-volume cuts of one- and two-particle subspaces (within the two- and three-particle states respectively) did not overlap. The results of this work could also allow one to explore more freedom in the definition of $H(\vec k)$, at the cost of allowing unsuppressed off-diagonal entries in the finite-volume matrix.

\bigskip

Although these observations give some motivation for the $\ttil$ effective channel, it is nonetheless possible that one might reformulate the results without this unphysical aspect. We are motivated to consider this in more detail especially following the demonstration in Sec.~\ref{sec:Linf} that all entries of $\mathcal F$ do not contain $\mathcal K_2$ poles. We note, in addition, that our result requires special treatment of $\mathcal K_2$ poles regardless of the sign of the residue. Thus also poles with no connection to a resonance state must be separated out. In this case we can provide no physical motivation for this mathematical necessity. 

Having removed the largest limitation of our previous formalism, we think it is now feasible to arrive at a quantization condition for completely general two- and three-particle systems. Even after this is achieved, several open issues still remain to be addressed. First, we hope to understand simplifications in both the derivation and the final result that can be made without adding any approximations. We have a sense that these can be identified by better understanding the relation of this work to Refs.~\cite{Hammer:2017uqm,Hammer:2017kms,Mai:2017bge},
and by studying the pole structures of the final quantities appearing in our results. Second, we plan to understand systematic approximations and truncations. This will likely involve subducing the quantization condition to irreducible representations of the finite-volume symmetry groups and truncating the angular momentum basis as is done in all two-particle studies. 
Third, we intend to continue our numerical investigations of these results, along the lines of Ref.~\cite{\BHSnum}. Fourth, and finally, we aim to implement this formalism in numerical LQCD calculations. To do so, it is necessary to identify a set of possible functional forms for the scattering amplitudes. Input from the dispersive and amplitude analysis communities is likely to play a key role in this next step. [See Refs.~\cite{Mai:2017vot, Jackura:2018xnx} for significant progress on this front.]

\section{Acknowledgements}


The work of SRS was supported in part by the United States Department of Energy grant No.~DE-SC0011637. RAB acknowledges support from U.S. Department of Energy contract DE-AC05-06OR23177, under which Jefferson Science Associates, LLC, manages and operates Jefferson Lab. 
The authors would like to thank A. Szczepaniak, A. Pilloni, J. Dudek and the late M. Pennington for useful discussions

\appendix
\section{Factorization of the off-shell two-particle K matrix at the pole\label{app:factorize}}



The aim of this appendix is to present a derivation of Eq.~(\ref{eq:Kofffromon})
and its consequences. 
We first consider K matrix poles above threshold and turn at the end to the case of subthreshold
poles.

We begin by reviewing the constraints that unitarity places on two-particle scattering amplitudes. 
The S-matrix is related to the on-shell scattering amplitude in the standard way
\begin{align}
S_2^{(\ell)}&=
1+2i\tilde{\rho}_2 \cM^{(\ell)}_{2;\rm on;on},
\end{align}
where we have introduced $\tilde{\rho}_2=i\tilde{\rho}$ with $\tilde \rho$ defined in Eq.~(\ref{eq:rhotdef}). For the purpose of this appendix, $\tilde{\rho}_2$ is more convenient, in particular because 
it is real above threshold. On the physical scattering axis, 
i.e. for real energies above threshold on the physical sheet, the S-matrix is unitary, 
implying
\begin{align}
 \text{Im}(\cM^{(\ell)}_{2;\rm on;on})&= 
\cM^{(\ell)}_{2;\rm on;on}\,\tilde{\rho}_2 \,\cM^{(\ell)\,\dag}_{2;\rm on;on}.
\end{align}
Given that $\tilde{\rho}_2$ is finite, this result prohibits $\cM^{(\ell)}_{2;\rm on;on}$
from having poles on the physical axis (since the left-hand side would then have a single-pole and
the right-hand side a double pole). Of course, poles below threshold for real $s$, 
corresponding to bound states, are allowed, 
since this constraint applies only for real energies above threshold. 

Unitarity alone cannot put constraints on the analytic structure of 
off-shell scattering amplitudes. Instead, as described in
the main text,  we consider this system as a generic effective field theory, with quantities calculated to all orders in perturbation theory. In this context we 
can connect the off- and on-shell scattering amplitudes, as we now show.

The ingredients we need are, first,
the product of two fully-dressed propagators with the appropriate symmetry factor,
\begin{align}
\Delta_2\equiv \frac{1}{2}\Delta(P-k)\,\Delta(k)\,,
\label{eq:Delta2def}
\end{align}
and, second, the fully off shell Bethe-Salpeter kernel $B^{(\ell)}_{2;\rm off;off}$.
We recall that the latter is defined as the sum of all amputated two-to-two diagrams
that are two-particle irreducible in the s channel.\footnote{%
As noted in the main text, we are implicitly making a choice of single-particle 
interpolating operator when defining this kernel. None of the subsequent considerations
depend on this choice.}
The pair of subscripts indicates that both initial and final
states are off shell. The on-shell versions have the same
definitions except that the 4-momenta in either one or both states are set to the physical values, $p^2 \to m^2$. 
In terms of these building blocks, the off-shell amplitude can be written as an iteration
of s-channel two-particle loops,
\begin{align}
i\cM^{(\ell)}_{2;\rm off;off}
&=i{ B}^{(\ell)}_{2;\rm off;off}
+\int\, i{ B}^{(\ell)}_{2;\rm off;off}\,\Delta_2\,i{ B}^{(\ell)}_{2;\rm off;off}
+\iint\, i{ B}^{(\ell)}_{2;\rm off;off}\,\Delta_2\,i{ B}^{(\ell)}_{2;\rm off;off}\,\Delta_2\,i{ B}^{(\ell)}_{2;\rm off;off}+\cdots\,,
\end{align}
where the integrals are over the loop momenta, e.g. over $k$ in Eq.~(\ref{eq:Delta2def}).

Fully or partially on-shell amplitudes are then given by appropropriate changes to the subscripts, e.g.
\begin{align}
i\cM^{(\ell)}_{2;\rm on;on}
&=i{ B}^{(\ell)}_{2;\rm on;on}
+\int\, i{ B}^{(\ell)}_{2;\rm on;off}\,\Delta_2\,i{ B}^{(\ell)}_{2;\rm off;on}
+\iint\, i{ B}^{(\ell)}_{2;\rm on;off}\,\Delta_2\,i{ B}^{(\ell)}_{2;\rm off;off}\,\Delta_2\,i{ B}^{(\ell)}_{2;\rm off;on}+\cdots\,.
\label{eq:Monon}
\end{align}
These results can be used to rewrite the on-shell amplitude in three useful forms
\begin{align}
i\cM^{(\ell)}_{2;\rm on;on}
&=i{ B}^{(\ell)}_{2;\rm on;on}
+\int\, i{ B}^{(\ell)}_{2;\rm on;off}\,\Delta_2\,i{\cM}^{(\ell)}_{2;\rm off;on}
\label{eq:Kon_to_off_v2}
\\
&=i{ B}^{(\ell)}_{2;\rm on;on}
+\int\, i{\cM}^{(\ell)}_{2;\rm on;off}\,\Delta_2\,i{B}^{(\ell)}_{2;\rm off;on}
\label{eq:Kon_to_off_v3}
\\
&=i{ B}^{(\ell)}_{2;\rm on;on}+\int\, i{ B}^{(\ell)}_{2;\rm on;off}\,\Delta_2\,i{ B}^{(\ell)}_{2;\rm off;on}
+\iint\, i{ B}^{(\ell)}_{2;\rm on;off}\,\Delta_2\,i{\cM}^{(\ell)}_{2;\rm off;off}\,\Delta_2\,i{ B}^{(\ell)}_{2;\rm off;on}\,.
\label{eq:Kon_to_off_v1}
\end{align}

To proceed, we assume that resonances in $\cM^{(\ell)}_{2;\rm on;on}$ arise by
the iteration of the two particle loops in Eq.~(\ref{eq:Monon}) and are not present in 
the kernel $B^{(\ell)}_2$ itself (whether on or off shell). 
In other words, since $B^{(\ell)}_2$ has no intermediate states
that are on shell in the kinematic range of interest, $4 m^2 < s < 16 m^2$
(or $4 m^2 < s < 9 m^2$ if there is no G-parity-like symmetry),
it can be treated as a nearly local two-particle interaction, 
and it is the iteration of this interaction that leads to resonances. 
Given this assumption, $B^{(\ell)}_2$ has no s-channel singularities on the physical axis.
This will be a key input into the following arguments.
We note that $B^{(\ell)}$ can  have t- and u-channel singularities
(e.g. the left-hand cut) but these occur for $s\le 0$
and are thus outside of the kinematic range of interest. 

Given the inputs that neither $\cM^{(\ell)}_{2;\rm on;on}$ nor $B^{(\ell)}_2$ have
poles on the physical axis, it follow from Eqs.~(\ref{eq:Kon_to_off_v2}),
~(\ref{eq:Kon_to_off_v3}) and ~(\ref{eq:Kon_to_off_v1}), respectively,
that $\cM^{(\ell)}_{2;\rm off;on}$, $\cM^{(\ell)}_{2;\rm on;off}$
and $\cM^{(\ell)}_{2;\rm off;off}$ cannot have such poles either.
Of course, all these quantities can have poles in the complex plane corresponding
to resonances, but the key point here is that the off-shell amplitudes inherit from
$\cM^{(\ell)}_{2;\rm on;on}$ the absence of poles on the real axis above threshold.

With this in hand, we can finally turn our attention to the K matrix. Again,
unitarity alone places no constraints on the K matrix, but we can use its
all orders effective field theory definition to relate it to the scattering amplitude.
Indeed, whether on or off shell,
the two quantities differ only by the replacement of the $i\epsilon$ prescription
in two-particle loops with the principal-value prescription.
The difference in these definitions is proportional to 
$\tilde{\rho}_2$ and a $\delta$-function that places the states on-shell.
From this we find that the fully off shell K matrix can be written as [see also Eq.~(\ref{eq:ourK})]
\begin{align}
i\K^{(\ell)}_{2;\rm off;off}
&=
i\cM^{(\ell)}_{2;\rm off;off}
-i\cM^{(\ell)}_{2;\rm off;on}\,\tilde{\rho}_2\,i\cM^{(\ell)}_{2;\rm on;off}
+i\cM^{(\ell)}_{2;\rm off;on}\,\tilde{\rho}_2\,i\cM^{(\ell)}_{2;\rm on;on}\,\tilde{\rho}_2\,i\cM^{(\ell)}_{2;\rm on;off}+\cdots
\nn\\
&=
i\cM^{(\ell)}_{2;\rm off;off}
-i\cM^{(\ell)}_{2;\rm off;on}\,\tilde{\rho}_2\,
\left[\frac{1}{1+i\cM^{(\ell)}_{2;\rm on;on}\tilde{\rho}_2}\right]\,
i\cM^{(\ell)}_{2;\rm on;off}
\nn\\
&=
i\cM^{(\ell)}_{2;\rm off;off}
-i\cM^{(\ell)}_{2;\rm off;on}\,\tilde{\rho}_2\,
\left[\frac{1}{\cM^{(\ell)\,-1}_{2;\rm on;on}+i\tilde{\rho}_2}\right]\,\cM^{(\ell)\,-1}_{2;\rm on;on}\,i\cM^{(\ell)}_{2;\rm on;off}
\nn\\
&=
i\cM^{(\ell)}_{2;\rm off;off}
-i\cM^{(\ell)}_{2;\rm off;on}\,\tilde{\rho}_2\,
\K^{(\ell)}_{2;\rm on;on}\,\cM^{(\ell)\,-1}_{2;\rm on;on}\,i\cM^{(\ell)}_{2;\rm on;off}.
\label{eq:Kff}
\end{align}
In the last step, we have expressed the off-shell $\K_2$ in terms of its on shell form.
This gives the desired result, Eq.~(\ref{eq:Kofffromon}), when working above threshold
so that $H=1$.
The key point is that, on the right-hand side of Eq.~(\ref{eq:Kff}), the only quantity
that has poles on the physical axis is $\K^{(\ell)}_{2;\rm on;\on}$.
Thus we conclude that $\K^{(\ell)}_{2;\rm off;off}$ must share these poles 
with $\K^{(\ell)}_{2;\rm on;on}$ in order for the equality to hold.

The second result we wish to show is the factorization of the residues of
poles in $\K^{(\ell)}_{2;\rm off;off}$.
To do this we note that the on-shell scattering amplitude is purely imaginary
at the poles of $\K^{(\ell)}_{2;\rm on;on}$,
\begin{align}
\cM^{(\ell)\,-1}_{2;\rm on;on} \longrightarrow -i\tilde{\rho}_2\,.
\end{align}
Therefore, near the poles, the off- and on-shell K matrices are related by
\begin{align}
i\K^{(\ell)}_{2;\rm off;off}
&\sim
Z^{(\ell)}_{2;\rm off;on}
\,
i\K^{(\ell)}_{2;\rm on;on}\,
Z^{(\ell)}_{2;\rm on;off},
\end{align}
where $Z^{(\ell)}_{2;\rm off;on}= -i\cM^{(\ell)}_{2;\rm off;on}\tilde{\rho}_2$ 
and $Z^{(\ell)}_{2;\rm on;off}= - i\tilde{\rho}_2\,\cM^{(\ell)}_{2;\rm on;off}$.
These two quantities depend, respectively, only on the final (initial) momenta,
thus demonstrating the claimed factorization of momentum dependence.
Both quantities equal unity when the corresponding external legs are placed on shell. 
Comparing the definitions of the residues of poles in 
on- and off-shell K matrices, given in Eqs.~(\ref{eq:K2poles})
and Eq.~(\ref{eq:K2off_v2}), respectively,
we see that
\begin{align}
 \Gamma_J(M^2, a'^2, b'^2) (a'^*_{2,k})^{J} = \Gamma_J
(q^*_{2,k})^{J}\, Z^{(J)}_{2;\rm on;off}\,,
\end{align} 
with a similar relation for $Z^{(J)}_{2;\rm off;on}$.

Before concluding this appendix, we return to the situation in which the K matrix has poles for real values of
the energy lying below threshold.
In this case, Eq.~(\ref{eq:Kff}) continues to hold---since it is based on a diagrammatic
analysis---except that $\tilde{\rho}_2$ becomes $i \tilde\rho H$, with the
factor of $H$ required by our definition of $\K_2$ [see Eq.~(\ref{eq:ourK})].
Thus we obtain Eq.~(\ref{eq:Kofffromon}) also when working below threshold,
and consequently it remains true that poles in the on-shell K matrix appear in
its off-shell extension, in the same locations.

There is, however, an additional issue that must be considered.
This arises because the scattering amplitude itself can have poles for real, subthreshold energies,
corresponding to bound states. At such poles, the on-shell K matrix becomes
$\K^{(\ell)}_{2,\rm on;on}\to -1/(\tilde\rho H)$, which is real and finite. 
The issue is whether the off-shell K matrix is also finite. 
To see that this is in fact the case, we make use of the factorization of $\cM_2$
at the pole, allowing us to write
\begin{equation}
i{\cM}^{(\ell)}_{2;\rm on;on}\sim\frac{i(ig_{\rm on})^2}{(s-E_{b}^2)}\,,\quad
i{\cM}^{(\ell)}_{2;\rm off;on}\sim\frac{i(ig_{\rm off})\,(ig_{\rm on})}{(s-E_{b}^2)}\,,
\ \ {\rm and}\ \ 
i{\cM}^{(\ell)}_{2;\rm off;off}\sim\frac{i(ig_{\rm off})^2}{(s-E_{b}^2)}\,,
\end{equation}
where $E_b$ is the energy of the bound state pole and $s=P^2_{2,k}$ is the
two-particle c.m.~energy. Substituting these results into Eq.~(\ref{eq:Kff})
and using the value of $\K^{(\ell)}_{2;\rm on;on}$ at the pole, we find that $\K^{(\ell)}_{2; \rm off; off}$ is indeed finite at $s=E_b^2$.

\section{Details of the derivation of results presented 
in Sec.~\ref{sec:derivation} \label{app:derapp}}


In this appendix we present technical details of the derivations
outlined in Sec.~\ref{sec:derivation}.


\subsection{Derivation of the recursion formula for $C_{L,0F}^{(m,n)}$
 [Eq.~(\ref{eq:CL0Fmaster})]}
\label{app:decomCLmn}

Here we derive Eq.~(\ref{eq:CL0Fmaster}) and, in doing so, 
give complete definitions of the quantities defined therein.

\begin{figure}
\begin{center}
\includegraphics[width=0.7\textwidth]{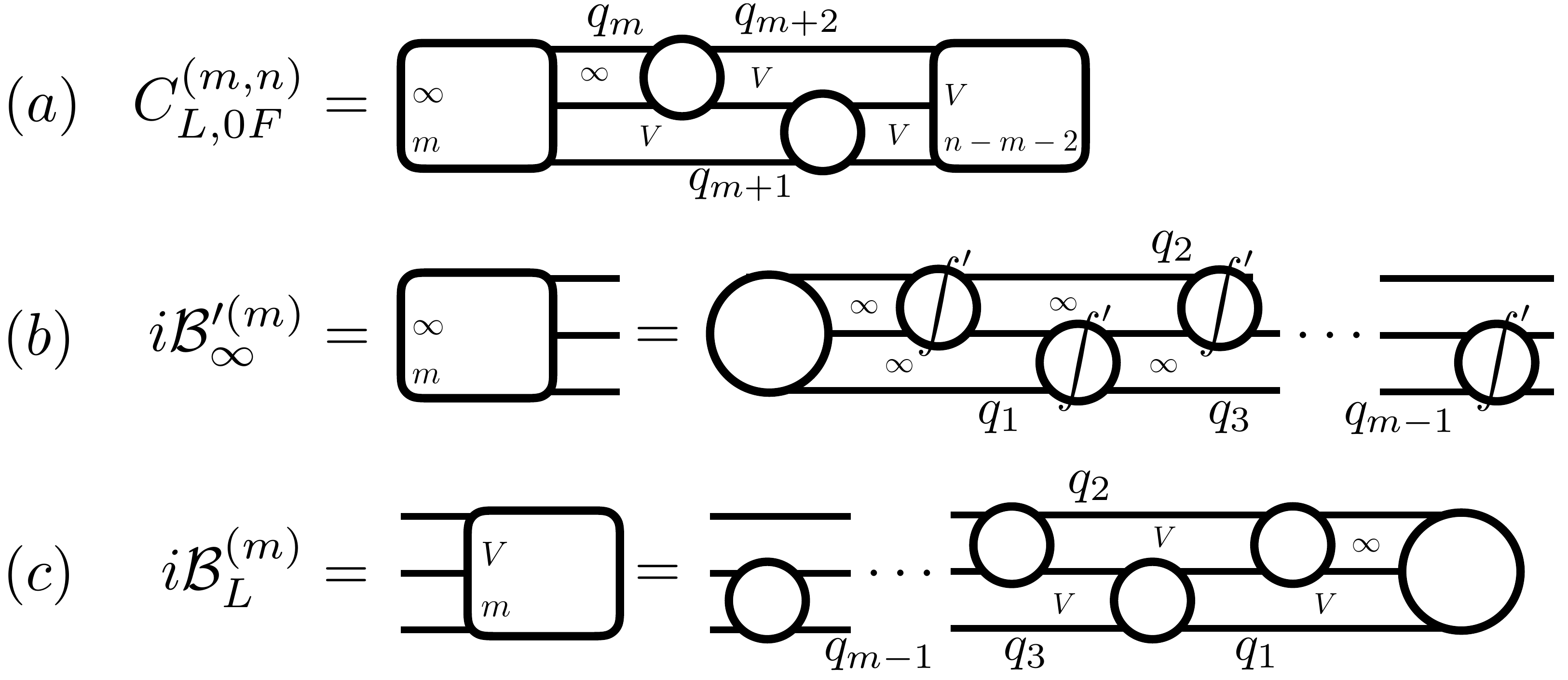}
\caption{
Diagrammatic definitions for the objects appearing in the
initial decomposition of $C_{L,0F}^{(m,n)}$, Eq.~(\ref{eq:CL0Frelevant}).
The square boxes with rounded corners represent the endcaps $i\cB$, with the
entries inside the box corresponding to the superscripts and subscripts.
The infinite-volume and finite-volume versions of these endcaps are shown,
respectively, in (b) and (c). Remaining notation is as in Fig.~\ref{fig:CnL0F}.
}
\label{fig:Bdef}
\end{center}
\end{figure}

\begin{figure}
\begin{center}
\includegraphics[width=\textwidth]{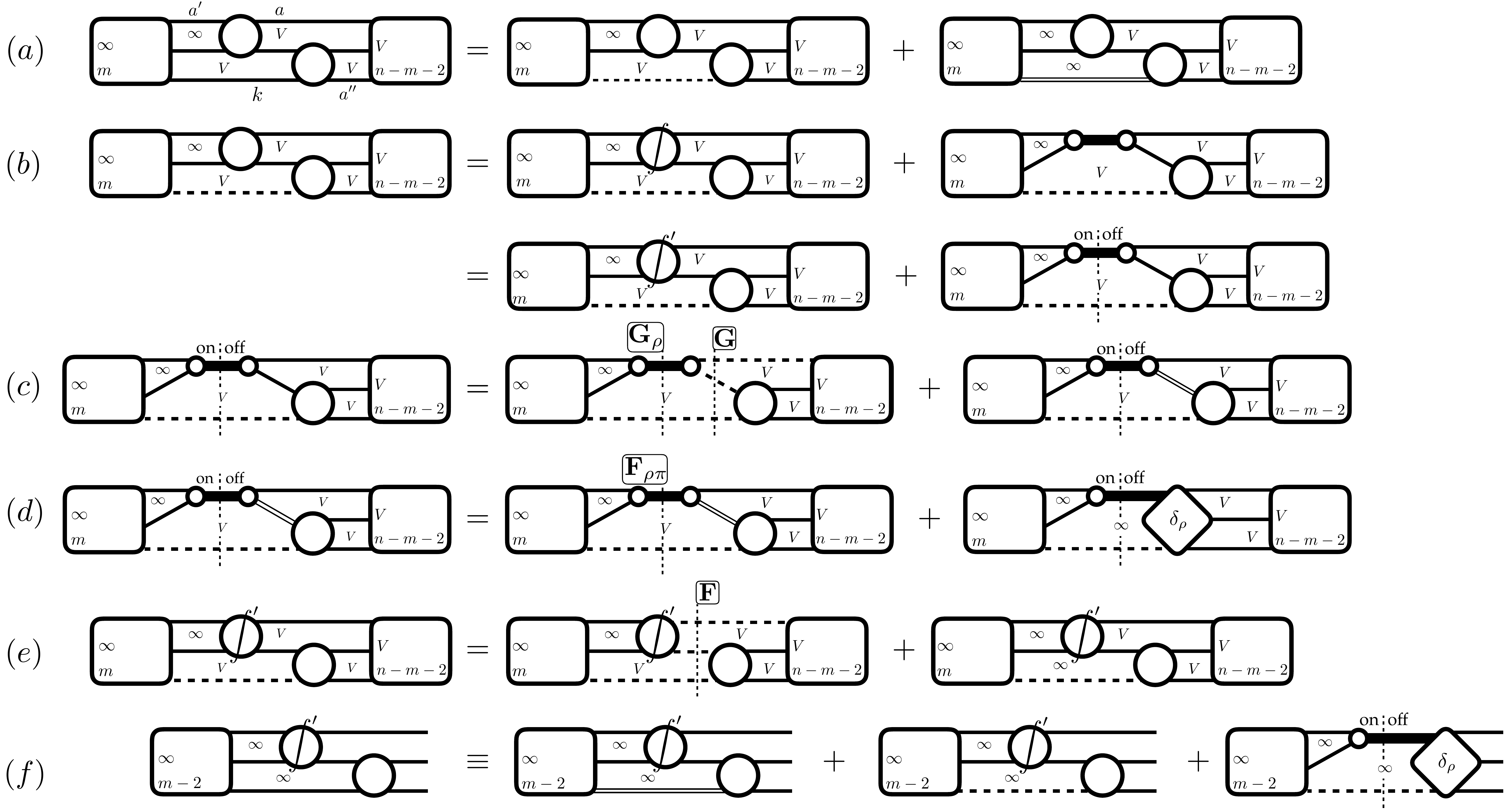}
\caption{
        Summary of the various steps used to derive Eq.~(\ref{eq:CL0Fmaster}), the recursion relation for decomposing $C_{L,0F}^{(m,n)}$. (a) First the leftmost spectator is split into its pole contribution and a second term that is smooth at the pole and thus contributes to $C_{L,0F}^{(m+1,n)}$. (b) Next, in the first term from (a) the leftmost K matrix is decomposed into a smooth and singular part and the latter is projected partially on shell, as explained in the text. (c) The complete on-shell projection at the K-matrix pole is performed simultaneously with the projection of the exchange propagator, leading to factors of $\bGr$ and $\bG$. (d) This leads to a remainder term that is smooth everywhere except at the K-matrix pole and can be separated into an on-shell term and another contribution to $C_{L,0F}^{(m+1,n)}$. (e) The remaining piece to decompose, in which the rightmost K matrix is smooth, leads to an $F$ cut in the middle of the diagram and yet another contribution to $C_{L,0F}^{(m+1,n)}$. (f) Finally we summarize all terms entering $\cB^{(m+1)}_{\infty}$, collected from (a), (d) and (e) above. 
The notation is as in Figs.~\ref{fig:CnL0F} and \ref{fig:Bdef}, with the addition that
dashed lines represent on shell propagators. See text for further discussion.
}
\label{fig:CL0Frecursion_v2}
\end{center}
\end{figure}

$C_{L,0F}^{(m,n)}$ is shown diagrammatically in Fig.~\ref{fig:CnL0F}(b).
Here we focus on the next momentum to be converted from a sum to an integral,
labeled $q_{m+1}$ in Fig.~\ref{fig:CnL0F}(b). Thus it is convenient to
absorb the integrated loops to the left of $q_{m+1}$ into a new endcap $i\cB_\infty^{(m)}$,
and similarly to absorb the summed loops to the right into $i\cB_L^{(n-m-2)}$,
since these new endcaps will maintain their forms throughout the derivation.
This new notation is shown in Fig.~\ref{fig:Bdef}(a), with the diagrammatic
definitions of the endcaps $\cB_\infty^{(m)}$ and $\cB_L^{(m)}$ sketched,
respectively, in Figs.~\ref{fig:Bdef}(b) and (c).\footnote{%
The definition of $\cB_\infty^{(m)}$ is imprecise, since additional terms are included
for each extra factor of $i\K_2$ that is added. 
This is explained in Fig.~\ref{fig:CL0Frecursion_v2} and the accompanying text.
}
The superscripts on the
$\cB$s denote the number of factors of $i\K_2$ or its smooth counterpart that they contain.
Note that these endcaps are closely related to $A'^{(m,u)}_3$ and
$A^{(m,u)}_{L,3}$, respectively, quantities discussed in the main text. 
The differences are that the
$\cB$ endcaps are not projected on shell, and also not decomposed into spherical harmonics.
The $\cB$ endcaps are shown in the figures as open squares with rounded corners.

Using this notation, we can explicitly display the relevant part of $C^{(m,n)}_{L,0F}$,
\begin{multline}
C_{L,0F}^{(m,n)} = \frac14
\sum_{k,a,a''} \int_{a'} 
i \cB_\infty^{(m)}(k,a') \Delta(a')\Delta(b') i\K_{2,\off}(a',b',-a) \\ 
\times  \Delta(k) \Delta(a)\Delta(b)
i\K_{2,\off}(k,b,-a'') \Delta(a'')\Delta(b'') i \cB_L^{(n-m-2)}(a,a'')
\,.
\label{eq:CL0Frelevant}
\end{multline}
Here we have changed the labels to the more manageable choices shown in
Fig.~\ref{fig:CL0Frecursion_v2}(a),
and used the definitions $b' \equiv P-k-a'$, $b \equiv P-a-k$, $b'' \equiv P-a-a''$ and
\begin{equation}
\int_a \equiv  \PV \int \frac{d^4a}{(2 \pi)^4} \,, \ \ \ \ \  \sum_k \equiv \int \frac{dk^0}{2 \pi}   \frac{1}{L^3}  \sum_{\vec k} \,.
\end{equation}
The quantity $\K_{2,\off}$ is the fully off-shell two-particle
K matrix defined as in Ref.~\cite{\HSQCa},
with the first two labels denoting the outgoing momenta, and the third one of the
incoming momenta. Note that in Appendix~\ref{app:factorize} we refer to this
K matrix as $\K_{2;\off,\off}$; here the double subscript is not necessary.
The expression (\ref{eq:CL0Frelevant})
holds for $n-m > 2$ and $n>2$, which is the case shown
in Fig.~\ref{fig:CL0Frecursion_v2}(a). For $n-m=2$ and $n \ge 2$ the 
sum over $a''$ is replaced by an integral.
Other cases are simpler and will be discussed at the end.

To derive Eq.~(\ref{eq:CL0Fmaster}), we begin by making the substitution
\begin{equation}
\Delta(k) = (2\pi) \delta(k^0-\omega_k) \frac1{2\omega_k} + \cR(k)
\,,
\end{equation}
 thereby separating the particle-pole contribution to the propagator (which is the only part that can lead to singularities as a function of $k$)
from the remainder, $\cR(k)$.
This is shown in Fig.~\ref{fig:CL0Frecursion_v2}(a) where, just as in Fig.~\ref{fig:KL33_decomp}, the pole is shown by a dashed line 
and the remainder by double solid lines. 
In the contribution of $\cR(k)$ to $C_{L,0F}^{(m,n)}$, 
we can replace the sum over $k$ with an integral, leading to a contribution
to $C_{L,0F}^{(m+1,n)}$, shown {as the second term on the right-hand side of Fig.~\ref{fig:CL0Frecursion_v2}(a).}
Thus we focus only on the particle-pole contribution, {the first term on the right side of \ref{fig:CL0Frecursion_v2}(a)}, in the following.

The next step is to insert a variant of Eq.~(\ref{eq:K2off_v2}) for the left-hand K matrix
\begin{multline}
i\K_{2,\off}(a',b',-a) = 
4\pi Y^*_{J M_J}(\hat a'^*_{2,k}) (a'^*_{2,k})^J i\Gamma_J(M^2,a'^2,b'^2) 
\frac{i\eta_J H_\rho(\vec k)}{(P_{2,k}^2 - M^2)}
i \Gamma_J(M^2, a^2,b^2) (a_{2,k}^*)^J Y_{J M_J}(\hat a^*_{2,k})
\\ + i \widetilde \K_{2,\off}(a',b',-a)
\label{eq:modK2poleDecom}
\,.
\end{multline} 
Here we have added back in the spherical harmonics needed to recreate the full
K matrix. Note that, by assumption, the pole appears only in the $J$th partial wave, 
while the second, smooth term includes contributions from all partial waves.
In the first term of Eq.~(\ref{eq:modK2poleDecom}), $J$ is fixed, 
while $M_J$ is summed from $-J$ to $J$.
Note also, as compared to Eq.~(\ref{eq:K2off_v2}),
we have included the UV regulator $H_\rho$ in the pole term. 
This can be added since, by construction, $1-H_\rho$ cancels the K-matrix pole,
and thus leads to a smooth contribution that can be absorbed into $\widetilde \K_{2,\off}$.

The result of this insertion is shown in Fig.~\ref{fig:CL0Frecursion_v2}(b).
We first consider the K-matrix pole contribution, which is represented by the {second
term on the right-hand side of Fig.~\ref{fig:CL0Frecursion_v2}(b)}, and has the explicit expression
\begin{multline}
C_{L,0F}^{(m,n)} \supset \frac12
\frac{1}{L^3} \sum_{\vec k} \sum_{a,a''} \int_{a'} 
i A'^{(m)}_{\ttil;M_J}(\vec k)
\frac{i\eta_J H_\rho(\vec k)}{(P_{2,k}^2 - M^2)}
i \Gamma_J(M^2, a^2,b^2) (a_{2,k}^*)^J
\frac{1}{2 \omega_k  }  {  \sqrt{4\pi}} Y_{J M_J}(\hat a^*_{2,k})
\\
\times \Delta(a)\Delta(b) i\K_{2,\off}(k,b,-a'') \Delta(a'')\Delta(b'') i \mathcal B_L^{(n-1)}(a,a'')
\,,
\label{eq:CL0FmnKpole}
\end{multline}
where 
\begin{equation}
i A'^{(m)}_{\ttil;M_J}(\vec k) \equiv \frac12 \int_{a'} i B_\infty^{(m-1)}(k,a') \Delta(a')\Delta(b') 
\sqrt{4\pi} Y^*_{J M_J}(\hat a'^*_{2,k})
(a'^*_{2,k})^J i\Gamma_J(M^2,a'^2,b'^2) 
\,.
\end{equation}
In both of these equations $k$ is on shell, $k^\mu=(\omega_k,\vec k)$.
In the figures, we represent the factors of $i \Gamma_J$ by small closed circles, and the
K-matrix pole by a thick horizontal line.

At this stage $A'^{(m)}_{\ttil;M_J}(\vec k)$ is not evaluated at the K pole, 
i.e. $P_k^2 \ne M^2$.
We can pick out the ``on shell'' part (where on shell here refers to the $\ttil$ state consisting of a particle plus the K-matrix pole)
by hand, by introducing a $\delta$ operator analogous to those used in Ref.~\cite{\HSQCa}
\begin{equation}
A'^{(m)}_{\ttil;M_J}(\vec k) \equiv 
A'^{(m)}_{\ttil;M_J \ell' m'} 
\Y_{2;\ell'm'}(\vec k^*)
+ \delta_\rho A'^{(m)}_{\ttil;M_J}(\vec k) 
\,,
\label{eq:deltarho}
\end{equation}
where the on-shell value of $A'_\ttil$ is
\begin{equation}
A'^{(m)}_{\ttil;M_J \ell' m'} \sqrt{4\pi}\, Y_{\ell' m'}(\hat k^*) 
\equiv A'^{(m)}_{\ttil;M_J}(q^*_\rho \hat k^*)
\,.
\label{eq:A2_vf}
\end{equation}
Here we are using the definitions of $\cY_2$ from Eq.~(\ref{eq:harmonic2})
and of $q^*_\rho$ from Eq.~(\ref{eq:krho}).
This step is represented by the second line in Fig.~\ref{fig:CL0Frecursion_v2}(b),
where in the second term we use the label ``on'' to indicate those quantities for which the
$\rho\pi$ relative momentum has been set to its on-shell value, 
$\vec k^* \to q_\rho^* \hat k^*$. If $\vec k$ is left at its original value, then
we use the label ``off''.
The $\delta_\rho$ operator cancels the K-matrix pole, and thus its contribution
can be absorbed into that from $\Kt_{2}$  to $C_{L,0F}^{(m,n)}$.
This is indicated by the prime on the integrated K matrix symbol 
in the first term on the right-hand side of Fig.~\ref{fig:CL0Frecursion_v2}(b).
We return to this contribution later. 
Equation~(\ref{eq:A2_vf}) completes the definition of the  infinite-volume endcap 
$A'^{(m)}_{\ttil;M_J \ell m}$. 

Substituting the on-shell term from Eq.~(\ref{eq:deltarho}) into
Eq.~(\ref{eq:CL0FmnKpole}), we obtain the second term on the second line of Fig.~\ref{fig:CL0Frecursion_v2}(b). The explicit expression is
\begin{multline}
C_{L,0F}^{(m,n)} \supset \frac12
\sum_{\vec k} \sum_{a,a''} \int_{a'} 
iA'^{(m)}_{\ttil;M_J \ell' m'} 
{  \Y_{\ttil,\ell' m'}(\vec k^*) }
\frac{i\eta_J H_\rho(\vec k)}{(P_{2,k}^2 - M^2)}
i \Gamma_J(M^2, a^2,b^2) (a_{2,k}^*)^J \frac{1}{2 \omega_k L^3} \sqrt{4\pi} Y_{J M_J}(\hat a^*_{2,k})
\\
\times \Delta(a)\Delta(b) i\K_{2,\off}(k,b,-a'') \Delta(a'')\Delta(b'') i \cB_L^{(n-1)}(a,a'')
\,.
\label{eq:CL0FmnKpole2}
\end{multline}
The final step for this term is to introduce a ``$G$ cut'' through the $a$, $b$ and $k$
propagators, following the approach of Ref.~\cite{\HSQCa}.
This cut places all three particles on shell, but in a different manner to
the left and the right of the cut. In both cases, the spectator momentum
is unchanged ($\vec k$ to the left, and $\vec a$ to the right), 
while the interacting pair have their momenta rescaled in their c.m.~frame. 
The $G$-cut term thus replaces $\Gamma_J(M^2, a^2, b^2)$ with the fully on shell 
$\Gamma_J$, and $a_{2,k}^*$ with $q_{2,k}^*$.
{This is shown in Fig.~\ref{fig:CL0Frecursion_v2}(c)}.
Using the definitions given in Sec.~\ref{sec:summary}, 
we find that the $G$-cut contribution 
[the first term on the right-hand side of Fig.~\ref{fig:CL0Frecursion_v2}(c)] is
\begin{equation}
C_{L,0F}^{(m,n)} \supset
iA'^{(m)}_{\ttil;M_J \ell' m'} iG_{\rho;M_J \ell' m';k \ell'' m''}
i \Gamma_J \frac{1}{2 \omega_k L^3} iG_{k \ell''m''; a \ell'''m'''}
{  iA^{(n-m,u)}_{L,3;a \ell''' m'''}} = \bA'^{(m)}_{\ttil} \bGr \bGam \bG \bA^{(n-m,u)}_{L,3}
\,,
\label{eq:CL0FmnKpoleG}
\end{equation}
where all repeated indices are summed in the middle quality and left implicit in the last. 
This is the third term on the right-hand side of first line of Eq.~(\ref{eq:CL0Fmaster}).

The cut we have just discussed is the most singular that arises, 
having the $\K_2$ pole ($\bGr$) and the three-particle pole ($\bG$) separated only by the
constant $\bGam$. It is possible for both poles go on shell simultaneously, for special
values of $\vec k$ and $\vec a$. 
We stress that these potential double poles appear only in sums over the spectator momenta, and not in sum-integral differences.
Thus we do not need to introduce a generalized zeta-function to describe them, 
unlike, for example, in the analysis of finite-volume effects in
two-particle matrix elements~\cite{Briceno:2015tza}.

The difference between Eqs.~(\ref{eq:CL0FmnKpole2}) and (\ref{eq:CL0FmnKpoleG}), represented by the last term of Fig.~\ref{fig:CL0Frecursion_v2}(c),
has no three-particle singularity, but still retains the K-matrix pole.
The absence of this singularity is shown in Fig.~\ref{fig:CL0Frecursion_v2}(c) by
the double line for the $b$ propagator in the last term.
We now project the quantity to the right of this pole on shell using the
$\delta_\rho$ operator introduced above in Eq.~(\ref{eq:deltarho}), 
but now acting to the right.
This is shown in Fig.~\ref{fig:CL0Frecursion_v2}(d),
leading to the final term on the first line of Eq.~(\ref{eq:CL0Fmaster}),
\begin{equation}
\bA'^{(m+1)}_2 \bFrp \slashed{\bA}^{(n-m)}_{L,\ttil}
\,,
\end{equation}
in which $\bFrp$ acts like a cut, and provides an implicit definition of $\slashed{A}_{L,\ttil}$.
The term involving $\delta_\rho$ removes the K-matrix pole, and is thus free of
singularities. For this term the sum over $k$ can be replaced by an integral,
providing an additional contribution to  $C_{L,0F}^{(m+1,n)}$.

Finally we consider the part involving the smooth part of the left-hand $\K_2$, i.e the first term in the second line of Fig.~\ref{fig:CL0Frecursion_v2}(b),
whose explicit expression is
\begin{multline}
C_{L,0F}^{(m,n)} \supset \frac14 \frac{1}{L^3}
\sum_{\vec k}\sum_{a,a''} \int_{a'} 
i B_\infty^{(m)}(k,a') \Delta(a')\Delta(b') i\Kt'_{2,\off}(a',b',-a) \frac{1}{2 \omega_k  }
\\ \times \Delta(a)\Delta(b)
i\K_{2,\off}(k,b,-a'') \Delta(a'')\Delta(b'') i \cB_L^{(n-m-2)}(a,a'')
\,.
\end{multline}
As noted above, another term with the same pole structure  has
been implicitly absorbed into this expression. We represent this 
by adding a prime to $\Kt'_{2,\off}$. 
The situation is now just as in Ref.~\cite{\HSQCa}, since the K-matrix pole is absent.
Thus we can replace the sum over $\vec k$ with an integral plus the difference,
the latter giving rise to an ``$F$ cut". 
We do not present the details as they have been presented in Ref.~\cite{\HSQCa}.
This step is shown in Fig.~\ref{fig:CL0Frecursion_v2}(e).
The $F$ cut gives the second term on the right-hand side of the first line
of Eq.~(\ref{eq:CL0Fmaster}),
which has the form
\begin{equation}
2 \bA'^{(m+1,s)}_3 \bF \bA_{L,3}^{(n-m-1,u)}
\,,
\end{equation}
while the integral leads to the final contribution to $C_{L,0F}^{(m+1,n)}$.

As we have progressed through this derivation, we have picked up three contributions
that can be absorbed into $C_{L,0F}^{(m+1,n)}$.
In fact, given our definition $C_{L,0F}$ in terms of the $\cB$ endcaps,
Eq.~(\ref{eq:CL0Frelevant}), the contributions are specifically absorbed into
$\cB_\infty^{(m+1)}$. 
This is shown in Fig.~\ref{fig:CL0Frecursion_v2}(f).
In this way $\cB_\infty$ 
and the  meaning of the smooth $\K_2$ symbol, $\widetilde \K'_2$,
are defined recursively, and this feeds into the definitions of
the other infinite-volume endcaps.

The above discussion holds for $n-m \ge 2$, so that the two $ \K_2$ factors can
be pulled out and dealt with explicitly.
The case $n-m=1$ is special, since there is only a single summed loop and the
only singularity arises from the pole in  $\K_2$.
The analysis is simpler for this case and leads to the second line in
Eq.~(\ref{eq:CL0Fmaster}).


\subsection{Details on decomposition of $\bKL^{(u)}$ described in Sec.~\ref{sec:KLdecom} \label{app:KLdecom}}

In this appendix we provide various details in the derivation of Eq.~(\ref{eq:KLmatrix3}) described in Sec.~\ref{sec:KLdecom}. As in the main text, many of these results have been checked using a {\em Mathematica} notebook together with the package {\em The NCAlgebra Suite} \cite{NCA}. Equations verified in this way are preceded by the indicator ``\pNCA''.

We begin by solving Eq.~(\ref{eq:KL33uu}). By isolating $\bKLth^{(u,u)}$ in the matrix equation one finds
\begin{align}
\bKLth^{(u,u)} &= 
\frac1{1-  \sbKLth   \bG  -  \bK  \bG }
\left[  \bK   \bG   \bK  +
 \sbKLth \left(1 +  \bG  \bK \right) \right]
\,,
\end{align}
which can be rearranged into a compact, symmetric expression \pNCA 
\begin{align}
\bKLth^{(u,u)}  = \bKLth^{(0)} 
+ \left(1 +  \bT  \bG\right) 
\sbKLth \frac1{1- \bGK \sbKLth }
\left(1 + \bG   \bT  \right)
\,,
\label{eq:APPKL33uufinal}
\end{align}
where $\bKLth^{(0)}$ is defined in Eq.~(\ref{eq:KL330}).
This is identically the $33$ component of Eq.~(\ref{eq:KLdecom}). To see this, 
we rewrite the latter equation as
\begin{align}
  \bKL^{(u)}  
& = 
\begin{pmatrix}
0 & 0 \\ 0 &  \bKLth^{(0)}
\end{pmatrix}
+
\bEL  \sbKL \frac1{1- \bcGK \sbKL} \bER \,,
\\
&=
\begin{pmatrix}
0 & 0 \\ 0 &  \bKLth^{(0)}
\end{pmatrix}
+
\begin{pmatrix}
1 & 0 \\ 0 & 1+ \bT \bG
\end{pmatrix}
\begin{pmatrix}
\sbKLtw &   \sbKLtwth  \\
\sbKLthtw &  \sbKLth 
\end{pmatrix} 
 \left[1 -  
 \begin{pmatrix}
0 & 0 \\ 0 & \bGK
\end{pmatrix}  
 \begin{pmatrix}
\sbKLtw &   \sbKLtwth  \\
\sbKLthtw &  \sbKLth 
\end{pmatrix}  
 \right ]^{-1} 
\begin{pmatrix}
1 & 0 \\ 0 & 1 + \bG \bT
\end{pmatrix} \,.
\label{eq:KLdecom2}
\end{align} 
As the two-by-two matrix containing $\bGK$, as well as matrices $\bEL$ and
$\bER$, project onto the $3$ component of their neighbors, 
it is straightforward to determine the $33$ component of this relation 
and see that it indeed matches Eq.~(\ref{eq:APPKL33uufinal}).

We now turn to $\bKLtwth^{(u)}$.
Substituting the result for $\bKLth^{(u,u)}$ into Eq.~(\ref{eq:KL23u}), 
and simplifying yields \pNCA  
\begin{align}
\bKLtwth^{(u)} &= 
 \sbKLtwth 
 \frac1{1-  \bGK  \sbKLth } (1 +  \bG   \bT)
 \,.
 \label{eq:APPKL23ufinal}
 \end{align}
The expression for $\bKLthtw^{(u)}$ can be obtained similarly, with the result being
essentially the left-right reflection of Eq.~(\ref{eq:APPKL23ufinal}) \pNCA 
\begin{align}
\bKLthtw^{(u)} &= 
 (1 +  \bT  \bG)
 \frac1{1-  \sbKLth \bGK} 
\sbKLthtw
 \,.
 \label{eq:APPKL32ufinal}
 \end{align}
Together these results give the $\ttil 3$ and $3 \ttil$ components of Eq.~(\ref{eq:KLdecom}) [equivalently Eq.~(\ref{eq:KLdecom2})].

 The final quantity we need is $\bKLtw$. Using the method detailed in the main text for $\bKLth^{(u,u)}$, we find
 \begin{equation}
\bKLtw =  \sbKLtwth ^{(u)}    \bG \bKLthtw^{(u)} + \sbKLtw
\,.
\label{eq:APPKL22}
\end{equation}
Substituting 
(\ref{eq:APPKL32ufinal})
and rearranging leads to \pNCA 
\begin{equation}
 \bKLtw =  \sbKLtw + \sbKLtwth  
\bGK \frac1{1 - \sbKLth \bGK} \sbKLthtw
\,.
\label{eq:APPKL22final}
\end{equation}
This gives the $\ttil \ttil$ component of Eqs.~(\ref{eq:KLdecom}) and (\ref{eq:KLdecom2}) and completes the demonstration of this result.

It remains to verify Eq.~(\ref{eq:Kslashdecom}), the relation between slashed objects and the infinite-volume matrix, $\pmb {\mathcal K}$. 
In the main text we derived the relations for $\sbKLtwth$ and $\sbKLth$,
Eqs.~(\ref{eq:slashedKL23soln})  and (\ref{eq:slashedKL33soln}), respectively.
We find the result for $\sbKLthtw$ is essentially the reflection of that for $\sbKLtwth$,
 \begin{equation}
 \sbKLthtw = \bKi_{3 \ttil} \frac1{1-\bFrp \bKi_{\ttil \ttil}}\,.
 \label{eq:APPslashedKL32soln}
 \end{equation}
To complete the discussion we must address $\sbKLtw$.
Following the same decomposition strategy one last time we reach 
\begin{equation}
\sbKLtw = \bKi_{\ttil \ttil} \bFrp \sbKLtw + \bKi_{\ttil \ttil}
\,,
\label{eq:APPslashedKL22}
\end{equation}
whose solution is 
\begin{equation}
 \sbKLtw = \frac1{1 - \bKi_{\ttil \ttil} \bFrp} \bKi_{\ttil \ttil}\,.
\label{eq:APPslashedKL22soln}
\end{equation}

Our claim is that the four results  (\ref{eq:slashedKL23soln}), (\ref{eq:slashedKL33soln}),
(\ref{eq:APPslashedKL32soln}) and (\ref{eq:APPslashedKL22soln}) are
equivalent to the matrix result, Eq.~(\ref{eq:Kslashdecom}).
To show this, we rearrange the latter, and insert
the definitions for $\pmb {\mathcal K}$ and $\bcFrp$, yielding
\begin{align}
   \sbKL    = \left [ \begin{pmatrix} 1 & 0 \\ 0 & 1 \end{pmatrix} - \begin{pmatrix}
 {\bKi}_{\ttil\ttil} &  {\bKi}_{\ttil 3} \\
 {\bKi}_{3\ttil} &  {\bKi}_{33} 
\end{pmatrix} 
\begin{pmatrix}
\bFrp & 0 \\ 0 & 0
\end{pmatrix}  \right ]^{-1}  \begin{pmatrix}
 {\bKi}_{\ttil\ttil} &  {\bKi}_{\ttil 3} \\
 {\bKi}_{3\ttil} &  {\bKi}_{33} 
\end{pmatrix}   \,.
\label{eq:KLslashdecom2}
\end{align}
It is then straightforward to pick out various components of the equation by expanding the square-bracketed quantity, identifying a given component and then resumming.
The manipulations are simplified by the fact that the matrix containing
$\bFrp$ is a projector. The most complicated example is the $33$ component,
for which we find
\begin{equation}
\sbKLth = \bKi_{33} +  \bKi_{3 \ttil} \sum_{n=0}^\infty \bFrp \big ( \bKi_{\ttil \ttil}  \bFrp  \big )^n      \bKi_{\ttil 3} \,,
\end{equation} 
which sums into Eq.~(\ref{eq:slashedKL33soln}). 
Similarly one can show that the $\ttil 3$ component of the matrix relation matches Eq.~(\ref{eq:slashedKL23soln}), the $3\ttil$ component yields 
Eq.~(\ref{eq:APPslashedKL32soln}), and the $\ttil\ttil$ component gives
Eq.~(\ref{eq:APPslashedKL22soln}).

At this stage we have derived all relations summarized in Eqs.~(\ref{eq:KLdecom}) and (\ref{eq:Kslashdecom}) of Sec.~\ref{sec:KLdecom}. From this point the discussion in the main text completes the derivation, yielding a decomposition of all entries the matrix $\bKL^{(u)}$ in terms of infinite-volume divergence-free K matrices.


\subsection{Volume independence of $\delta C^{[B_2],\{0\}}_{\infty } $}
\label{app:deltaCL}

In this appendix we explain why $\delta C^{[B_2],\{0\}}_{\infty }$, defined in Eq.~(\ref{eq:CLnoKdf2_IV}) of the main text, has only exponentially suppressed volume dependence and can thus be taken as an infinite-volume quantitiy. 

To show this we begin by focusing on the first four terms, 
and noting that these can be rewritten as
\begin{align}
- \tfrac23 \bSig \bF \bSigD 
- 2 \bA'^{(s)}_3 \bF  \bSigD - \bA'_3 \bF \, 2 \bA_3^{(s)}
+ \tfrac23 \bA'_3 \bF \bA_3
&=
-\tfrac23\bA'^{(s-u)}_3 \bF \bSig
-\tfrac23 \bA'_3 \bF \bA_3^{(s-u)} \,,
\label{eq:sminusu}
\end{align}
where\footnote{%
The factors of $\bSig$ and $\bSigD$ appear here because $\bA'^{(u)}_3$
and $\bA^{(u)}_3$ are defined to include the $n=0$ terms [see Eq.~(\ref{eq:endcaps})] 
while $\bA'^{(s)}_3$ and $\bA^{(s)}_3$ do not [see Eq.~(\ref{eq:allorders2})].}
\begin{equation}
\bA'^{(s-u)}_3 \equiv \bA'^{(s)}_3 - \bA'^{(u)}_3  + \bSig \,,  \ \ {\rm and}\ \
\bA^{(s-u)}_3 \equiv \bA_3^{(s)} - \bA_3^{(u)} +\bSigD\,.
\end{equation}
As explained in Ref.~\cite{\HSQCa} 
[see Eqs.~(196)-(198) of that work, and the surrounding discussion], 
the $s-u$ differences in Eq.~(\ref{eq:sminusu}) can be written as
\begin{align}
-\tfrac23\bA'^{(s-u)}_3 \bF \bSig
-\tfrac23 \bA'_3 \bF \bA_3^{(s-u)}
 &= 
- \bA'^{(s-u)}_3 \frac{i\rho}{3\omega} \bSig
- \bA'_3 \frac{i\rho}{3\omega} \bA_3^{(s-u)}
+\mathcal O(e^{- m L})
\,.
\label{eq:sminusua}
\end{align}
The phase-space factor $\rho$ [defined in Eq.~(\ref{eq:rhodef})] is smooth, allowing
the implicit sums in our matrix notation to be replaced by integrals, up to
exponentially suppressed corrections.
Thus, within the framework of dropping exponentially-suppressed volume dependence, the right-hand side is an infinite-volume quantity.

\begin{figure}
\begin{center}
\includegraphics[width=\textwidth]{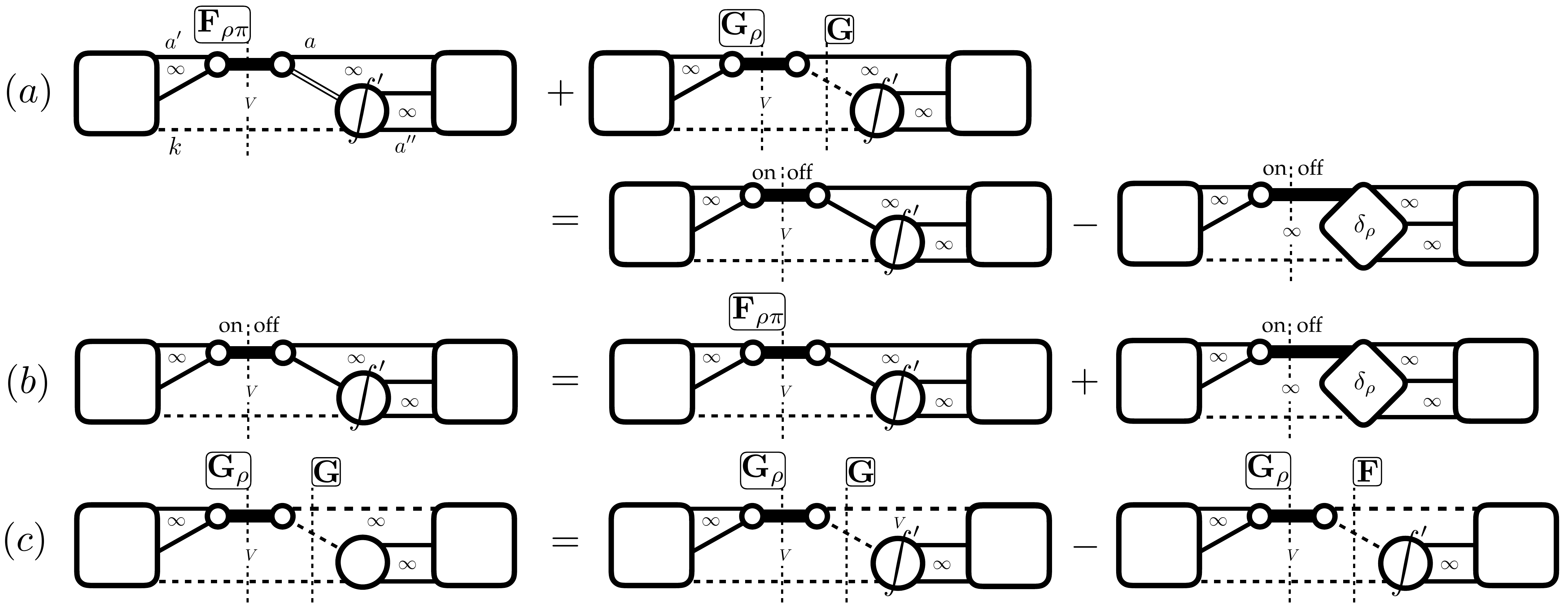}
 \caption{
Derivation of Eq.~(\ref{eq:symm1}), using the notation of Fig.~\ref{fig:CL0Frecursion_v2}.
\label{fig:symm1}}
\end{center}
\end{figure}

The remaining two terms in Eq.~(\ref{eq:CLnoKdf2_IV}) are
\begin{equation}
\bA'_{\ttil} \bFrp     (\slashed \bA_{\ttil} - \bA_{\ttil})
+ \bA'_{\ttil}  \bGr \bGam    
\left[  \bG  (\bA_3^{(u)} - \bSigD)  - \bF \, 2 \bA_3^{(s)}\right] \,.
\end{equation}
To show that this is also an infinite-volume quantity, 
 we need a new argument, since this quantity involves
K-matrix poles and thus was not encountered in Ref.~\cite{\HSQCa}.
We make the argument diagrammatically in Fig.~\ref{fig:symm1}, 
based in part on the derivation illustrated in Fig.~\ref{fig:CL0Frecursion_v2}. 
We do not give the corresponding analytic expressions, as
our earlier discussion explains the precise relation
between diagrams and equations.

We begin by substituting Fig.~\ref{fig:CL0Frecursion_v2}(d) into the final term in
Fig.~\ref{fig:CL0Frecursion_v2}(c) and rearranging so that the terms involving
$\bFrp$ and $\bGr $ are on the left-hand side.
We also make several changes to the parts of the diagrams away from the cuts
in order to apply the result to the present quantities of interest.
These changes do not impact the derivation.
They are (a) the box on the left end now represents $\bA'_{\ttil}$, with the
final loop explicitly exposed;
(b) the loops to the right of the cut are changed from sums to integrals;
(c) the kernel on the right is changed from $\K_2$ to $\widetilde {\cK}'_2$;
and, finally, (d), the box on the right end represents the remainder of
a full infinite-volume endcap.
These steps lead to the equality in Fig.~\ref{fig:symm1}(a).
At this stage, the first term on the left-hand side represents 
$\bA'_{\ttil}\bFrp \slashed \bA_{\ttil}$, 
the second term on the right-hand side is manifestly an infinite volume quantity,
while the other two terms need further manipulations to bring them to a useful form.

The first term on the right-hand side of Fig.~\ref{fig:symm1}(a) is rewritten in
Fig.~\ref{fig:symm1}(b). The approach here is to to expand the off-shell 
$\bA_{\ttil}$ factor lying to the right of the cut about the position of the K-matrix pole,
using the $\delta_\rho$ operator introduced above.
The leading term gives $\bA'_{\ttil}\bFrp \bA_{\ttil}$,
while the $\delta_\rho$-dependent term is smooth at the K-matrix pole 
allowing the sum over the spectator momentum to be replaced by
an integral (up to exponentially suppressed corrections).
This is shown in the second term on the right-hand side by the ``$\infty$" symbol
within the vertical dashed line. This term is manifestly an infinite-volume quantity.

The final step is shown in Fig.~\ref{fig:symm1}(c),
where we derive an equality for the second term on the right-hand side of
Fig.~\ref{fig:symm1}(a). 
On the left-hand side
we have a $\bG$ cut with the momentum $\vec a$ integrated. To obtain the
right-hand side we replace this integral with a sum minus a sum-integral difference.
The sum gives $\bA'_{\ttil}  \bGr \bGam  \bG \bA_3^{(u)}$,
shown by the first term on the right-hand side of Fig.~\ref{fig:symm1}(c).
The sum-integral difference gives rise to a factor of $\bF$, and, following the arguments of Ref.~\cite{\HSQCa}, switches $\bA_3^{(u)} - \bSigD$ to $\bA_3^{(s)}$,
leading to
$- \bA'_{\ttil}  \bGr \bGam   \bF  \, 2 \bA_3^{(s)}$.

The overall result of these steps is\footnote{%
Since the difference $\slashed{\bA}_{\ttil}-\bA_{\ttil}$ has at least one $\widetilde{\mathbf K}'_2$ insertion,
as shown in the figure, the $\bSigD$ part of $\bA_3^{(u)}$ does not
contribute.}
\begin{equation}
\bA'_{\ttil} \bFrp    \slashed \bA_{\ttil}
+ \bA'_{\ttil}  \bGr \bGam  
\left[  \bG (\bA_3^{(u)}  - \bSigD)  - \bF  \, 2 \bA_3^{(s)}\right]
= 
 \bA'_{\ttil}  \bFrp  \bA_{\ttil}
+ \delta' C^{[B_2],\{0\}}_{\infty } + \mathcal O(e^{-m L})\,,
\label{eq:symm1}
\end{equation}
where $ \delta' C^{[B_2],\{0\}}_{\infty }$ is a particular infinite-volume contribution, to be absorbed into $ \delta C^{[B_2],\{0\}}_{\infty }$ and ultimately into $C_\infty^{[B_2]}$.
After rearrangement, this demonstrates the desired result.


\subsection{Symmetrization of factors adjacent to $\big ( \bX + \bY \big) _{33}$}
\label{app:w33sym}

In this final appendix, we demonstrate that the contribution of the $33$
component of $\bX+\bY$ to 
$\bKdf^\u \big ( \bX + \bY \big)  \bKdf^\u$ is consistent with the
claimed general result, Eq.~(\ref{eq:symmXYA}).
For definiteness, we consider the term containing the $23$ component of $\bKdf^\u$
and the $32$ component of $\bKdf^\u$, although the derivation works for
any $(u)$-type three-particle quantities on the ends.
To match with Eq.~(\ref{eq:symmXYA}) we need to show that
\begin{align}
\chi &\equiv \bKdftwth^\u \big ( \bX + \bY \big)_{33}  \bKdfthtw^\u 
\\
& = 
\bKdftwth^\u \frac1{1 - \otimes \mathcal I_{33}^\dagger} \begin{pmatrix} \otimes \mathcal I_{3\ttil} & \mathcal S \end{pmatrix} \cdot
\bFMat
\cdot
\begin{pmatrix} \mathcal I_{\ttil 3} \otimes \\ \mathcal S \end{pmatrix}
\frac1{1 - \mathcal I_{33} \otimes } \bKdfthtw^\u
+
\delta'\bKdftw^\u
\,, 
\label{eq:symm10app}
\end{align}
where $\delta' \bKdftw^\u$ is an infinite-volume quantity that can be absorbed
into the quantity $\delta \bKdftw^\u$ defined in Eq.~(\ref{eq:symmXYA}).
Its explicit expression is determined below.

Using the definition of $\bX+\bY$ in Eq.~(\ref{eq:XplusY}) we find
\begin{equation}
\chi = 
\bKdftwth^\u (\bF+\bG) \frac1{1- \bK(\bF  +   \bG )} \bKdfthtw^\u \,,
\end{equation}
which can be divided into two contributions using Eq.~(\ref{eq:symmLiterate}),
\begin{align}
\chi &= \chi_{a} + \chi_{b}\,,
\\
\chi_{a} &= \bKdftwth^\u
 (\bF + \bG) 
\frac1{1 -  \bK (\bF+\bG)} \left\{
\bK \bF \mathcal S 
+ \bGam  \bGrD \mathcal I_{\ttil 3}  \otimes \right\} 
\frac1{1 - \mathcal I_{33} \otimes}  \bKdfthtw^\u\,,
\\
\chi_{b} &=
\bKdftwth^\u
 (\bF + \bG) 
 \frac1{1 - \mathcal I_{33} \, \otimes } \bKdfthtw^\u \,.
\end{align}
$\chi_a$ can be rewritten using the reflected form of Eq.~(\ref{eq:symmLiterate}),
\begin{align}
\chi_a &= \chi_{a1}+ \chi_{a2}+\chi_{a3}\,,
\\
\chi_{a1} &=
\bKdftwth^\u \frac1{1 - \otimes \mathcal I_{33}^\dagger}
\left\{ \otimes
\mathcal I_{3\ttil}  \bGr  \bGam 
+ \mathcal S  \bF \bK \right\}
\frac1{1  -  (\bF + \bG)\bK} 
(\bF + \bG)
\left\{
\bK \bF \mathcal S 
+ \bGam  \bGrD  \mathcal I_{\ttil 3} \otimes  \right\} 
\frac1{1 - \mathcal  I_{33} \otimes} \bKdfthtw^\u\,,
\\
\chi_{a2} &=
\bKdftwth^\u \frac1{1 - \otimes \mathcal I_{33}^\dagger} 
 (\bF + \bG)
\bK \bF \mathcal S 
\frac1{1 - \mathcal I_{33} \otimes} \bKdfthtw^\u\,,
\\
\chi_{a3} &=
\bKdftwth^\u \frac1{1 - \otimes \mathcal I_{33}^\dagger} 
 (\bF + \bG)
\bGam  \bGrD  
\mathcal I_{\ttil 3}  \otimes \frac1{1 - \mathcal I_{33} \otimes } \bKdfthtw^\u\,.
\end{align}
$\chi_{a1}$ includes only symmetrized quantities, but $\chi_{a2}$ and $\chi_{a3}$ need
further work. Noting the presence of $\bF+\bG$, we can apply Eq.~(\ref{eq:symmR})
to both quantities. For $\chi_a$ this leads to
\begin{align}
\chi_{a2} &= \chi_{a2A}+\chi_{a2B}\,,
\\
\chi_{a2A}&= \bKdftwth^\u \frac1{1 - \otimes \mathcal I_{33}^\dagger} 
\mathcal S 
  \bF \bK \bF \mathcal S 
\frac1{1 - \mathcal I_{33} \otimes } \bKdfthtw^\u
+
\bKdftwth^\u \frac1{1 - \otimes \mathcal I_{33}^\dagger}  \otimes \mathcal I_{3\ttil}  \bGr  \bGam
  \bF \mathcal S 
\frac1{1 - \mathcal I_{33} \otimes } \bKdfthtw^\u \,,
\\
\chi_{a2B} &=
 \bKdftwth^\u \frac1{1 - \otimes \mathcal I_{33}^\dagger} \otimes \mathcal  I_{33}^\dagger
  \bF \mathcal S 
\frac1{1- \mathcal I_{33} \otimes } \bKdfthtw^\u
\,.
\end{align}
Only $\chi_{a2B}$ contains an unsymmetrized quantity. It can be rewritten as
\begin{equation}
\chi_{a2B} = \frac13 \bKdftwth^\u \frac1{1 - \otimes \mathcal I_{33}^\dagger} \otimes \mathcal   I_{33}^\dagger
\mathcal S   \bF \mathcal S 
\frac1{1 - \mathcal I_{33} \otimes } \bKdfthtw^\u
+
\left(\bKdftwth^\u \frac1{1 - \otimes \mathcal I_{33}^\dagger}  \otimes \mathcal I_{33}^\dagger\right)^{(u-s)}
\frac{ i \rho}{3\omega} \mathcal S 
\frac1{1 - \mathcal I_{33} \otimes} \bKdfthtw^\u
\,,
\end{equation}
where the first term is symmetrized, while in the second the
two factors of $\bKdf^\u$ are bound together by an integral operator,
giving a contribution to $\delta' \bKdftw^\u$.

Returning to $\chi_{a3}$, we can apply the reflected form of Eq.~(\ref{eq:symm6}), yielding
\begin{multline}
\chi_{a3} =
\bKdftwth^\u \frac1{1 - \otimes \mathcal I_{33}^\dagger} \mathcal S
 \bF
\bGam  \bGrD  \mathcal I_{\ttil 3}  \otimes
\frac1{1 - \mathcal I_{33} \otimes}  \bKdfthtw^\u
+
\bKdftwth^\u \frac1{1 - \otimes \mathcal I_{33}^\dagger} \otimes \mathcal I_{3\ttil} \bFrp
\mathcal I_{\ttil 3} \otimes \frac1{1 - \mathcal I_{33} \otimes} \bKdfthtw^\u
\\
+
\bKdftwth^\u \frac1{1 - \otimes \mathcal I_{33}^\dagger} 
\otimes \rho_{3\ttil} \otimes
\mathcal I_{\ttil 3} \otimes  \frac1{1 - \mathcal I_{33} \otimes} \bKdfthtw^\u\,.
\end{multline}
The final term in this expression
gives an additional contribution to $\delta' \bKdftw^\u$.

\begin{figure}
\begin{center}
\includegraphics[width=\textwidth]{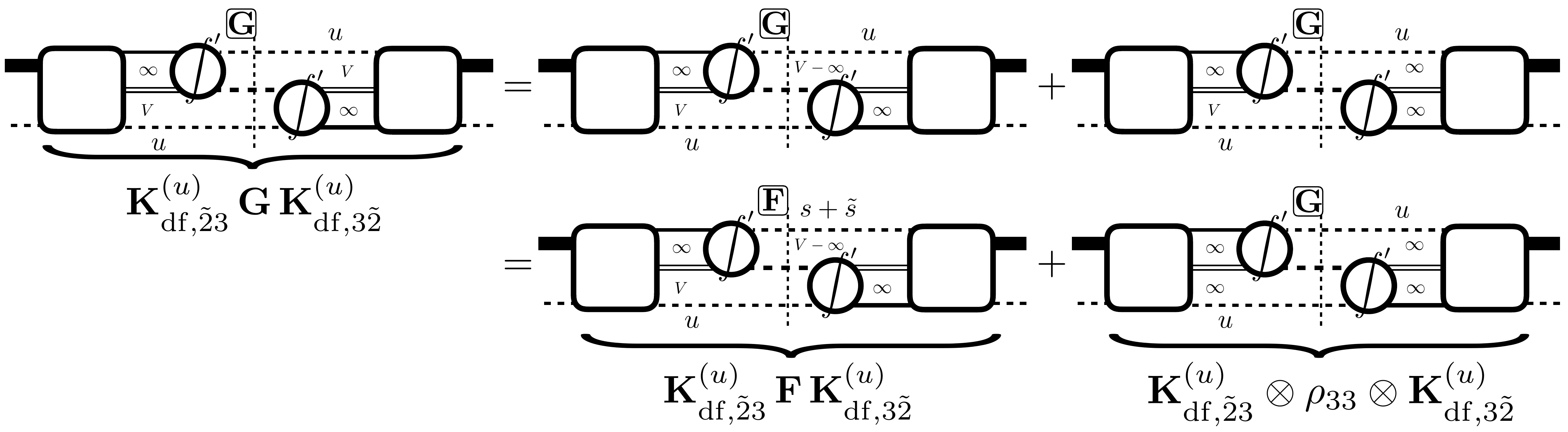}
\caption{
Derivation of Eq.~(\ref{eq:symm7}), using the notation of Figs.~\ref{fig:CnL0F},
\ref{fig:symm2} and  \ref{fig:CL0Frecursion_v2}.
The first step is identical to that in Fig.~\ref{fig:symm2}: replacing the sum adjacent to the
$G$ cut with a sum-integral difference and an integral, the former giving rise to an $F$ cut.
In this case, however, there are no K-matrix poles, so the integral in the upper-right loop
of the final term on the first line removes the divergence in the summand for 
the spectator momentum, allowing it to be replaced
by an integral, as shown by the final term on the second line.
Since all loops are integrated in this term, it can be represented as
the action of a new integral operator, denoted $\rho_{33}$, tying the 
$\bKdftwth^\u$ and $\bKdfthtw^\u$ together.
}
\label{fig:symm4}
\end{center}
\end{figure}

The final step is to analyze $\chi_b$. This requires the result
\begin{equation}
\bKdftwth^\u  \bG \bKdfthtw^\u
=
\bKdftwth^\u  \bF \bKdfthtw^{(s+\tilde s)}
+
\bKdftwth^\u \otimes \rho_{33} \otimes \bKdfthtw^\u
\,,
\label{eq:symm7}
\end{equation}
which implies
\begin{align}
\bKdftwth^\u  (\bF + \bG) \bKdfthtw^\u
&=
\bKdftwth^\u  \bF \mathcal S \bKdfthtw^\u
+
\bKdftwth^\u \otimes \rho_{33} \otimes \bKdfthtw^\u \,,
\\&=
\frac13
\bKdftwth^\u \mathcal S  \bF \mathcal S \bKdfthtw^\u
+ \bKdftwth^\u \otimes \rho_{33}  \otimes \bKdfthtw^\u
+ \bKdftwth^{(u-s)} \frac{i\rho}{3\omega} \bKdfthtw^\u
\,.
\label{eq:symm8}
\end{align}
The derivation of Eq.~(\ref{eq:symm7}) is shown diagrammatically in Fig.~\ref{fig:symm4}.
The result holds, as usual, for any choice of unsymmetrized $(u)$-like three-particle quantities
on the ends, and thus can be applied to $\chi_b$, yielding
\begin{equation}
\chi_b = \frac13 \bKdftwth^\u \mathcal S  \bF \mathcal S 
\frac1{1 - \mathcal I_{33} \otimes}\bKdfthtw^\u
+ \bKdftwth^\u \otimes \rho_{33} \otimes \frac1{1 - \mathcal I_{33} \otimes}\bKdfthtw^\u
+ \bKdftwth^{(u-s)} \frac{i \rho}{3\omega} \frac1{1 - \mathcal I_{33} \otimes}\bKdfthtw^\u
\,.
\end{equation}
The last two terms give additional contributions to $\delta' \bKdftw^\u$.

Combining all these results we find the desired result, Eq.~(\ref{eq:symm10app}),
with
\begin{multline}
\delta' \bKdftw^\u  = 
\left(\bKdftwth^\u \frac1{1 - \otimes \mathcal I_{33}^\dagger} \otimes \mathcal I_{33}^\dagger\right)^{(u-s)}
\frac{i\rho}{3\omega} \mathcal S 
\frac1{1 - \mathcal I_{33} \otimes} \bKdfthtw^\u
+
\bKdftwth^\u \frac1{1 - \otimes \mathcal I^\dagger_{33} } \otimes\rho_{3\ttil} \otimes
\mathcal I_{\ttil 3} \otimes \frac1{1 - \mathcal I_{33} \otimes} \bKdfthtw^\u
\\
+
\bKdftwth^\u \otimes \rho_{33} \otimes \bKdfthtw^\u
+ \bKdftwth^{(u-s)} \frac{i\rho}{3\omega} \bKdfthtw^\u
\,.
\label{eq:deltapKdfu}
\end{multline}

\bibliography{ref} 

\end{document}